\documentclass[9pt,article,twoside]{rmaa-rho-class/rmaa-rho}
\RMxAAtemplatetype{\RMxAA} 

\usepackage{rotating}
\usepackage{amsmath}
\usepackage{hyperref}
\usepackage{lscape}
\usepackage{multirow}

\def\deg{$^\circ$}

\vol{0}
\pages{0-0}
\thisyear{2025}
\doi{\href{https://www.astroscu.unam.mx/rmaa/RMxAA..XX-X}{https://www.astroscu.unam.mx/rmaa/RMxAA..XX-X}}

\title{Hadronic Clues in Quasars Caught by Fermi-LAT}

\author[1]{A.~ Galván \orcidlink{0000-0001-5193-3693}}
\author[2]{N.~ Fraija \orcidlink{0000-0002-0173-6453}}
\author[3]{E.~ Aguilar-Ruiz}
\author[1]{H.~ León Vargas \orcidlink{0000-0001-5516-4975}}
\author[4,5]{M. G.~ Dainotti}
\author[2]{J.A.~ de Diego \orcidlink{0000-0001-7040-069X}}

\affil[1]{ Instituto de Física. Universidad Nacional Autónoma de México, AP 70-264, CDMX 04510, México}
\affil[2]{Instituto de Astronomía. Universidad Nacional Autónoma de México,AP 70-264, CDMX 04510, México}
\affil[3]{Instituto de Radioastronomía y Astrofísica. Universidad Nacional Autónoma de México, Antigua Carretera a Pátzcuaro 8701, Ex-Hda. San José de la Huerta, 58089, Morelia, Michoacán, México}
\affil[4]{Division of Science, National Astronomical Observatory of Japan, 2-21-1 Osawa, Mitaka, Tokyo 181-8588, Japan}
\affil[5]{The Graduate University for Advanced Studies (SOKENDAI), 2-21-1 Osawa, Mitaka, Tokyo 181-8588, Japan}

\leadauthor{Galván et al.}
\smalltitle{LaTex Macro RMxAA}

\corres{Antonio Galván}
\email{edwin@fisica.unam.mx}

\license{Texto de la licencia aquí}

\setbool{rho-abstract}{true} 
\setbool{rho-resumen}{true} 

\begin{abstract}
    This work explores whether hadronic processes could be responsible for the high-energy emission seen in quasars identified by the Large Area Telescope (LAT) instrument aboard the Fermi satellite. In contrast to purely leptonic models, this work investigates whether hadronic mechanisms can explain the observed gamma-ray spectra by analyzing the spectral energy distributions (SEDs) of a chosen sample of FSRQs (Flat-Spectrum Radio Quasars). By incorporating both hadronic and leptonic components into their multi-wavelength modeling, we evaluate the model's feasibility to simultaneously describe the data collected by Fermi-LAT and neutrinos detected by IceCube. According to the results, a hadronic contribution would be required to explain the SED of quasars detected by Fermi-LAT. However, their contribution to the neutrino flux detected by IceCube remains understated.
\end{abstract}


\begin{resumen}
    Este trabajo explora si los procesos hadr\'onicos podr\'ian ser responsables de la emisi\'on de alta energ\'ia observada en los cu\'asares identificados por el {instrumento} Large Area Telescope (LAT) abordo del Sat\'elite Fermi. En contraste con los modelos puramente lept\'onicos, el trabajo investiga si los mecanismos hadr\'onicos pueden explicar los espectros de rayos gamma observados analizando las distribuciones espectrales de energ\'ia (SEDs) de una muestra escogida de FSRQs (Flat Spectrum Radio Quasars). Al incorporar componentes hadr\'onicos y lept\'onicos en su modelizaci\'on multi-longitud de onda, evaluamos la viabilidad del modelo para describir simult\'aneamente los datos recogidos por Fermi-LAT y los neutrinos detectados por IceCube. Seg\'un los resultados, ser\'ia necesaria una contribuci\'on hadr\'onica para explicar la SED de los cu\'asares detectados por Fermi-LAT. Sin embargo, su contribuci\'on al flujo de neutrinos detectado por IceCube sigue siendo subestimada.
\end{resumen}

\begin{document}

\maketitle
\pagestyle{fancy}\thispagestyle{firststyle}


\section{INTRODUCTION}

Victor Hess discovered high-energy Cosmic Rays (HECRs) over a century ago, but their origin is still a mystery. The interaction between cosmic rays (CRs) and magnetic fields makes it impossible to determine where a CR was created \citep{2013FrPhy...8..748G}. Interactions between $\gamma$-ray photons and Cosmic Microwave Background (CMB) photons can reduce the intensity of $\gamma$-rays by a factor that depends both on the redshift and the energy of $\gamma$-rays \citep[e.g.][]{2012MNRAS.422.3189G, 2017A&A...603A..34F}. Since neutrinos have no net electric charge, they pass through magnetic fields unimpeded from their point of origin toward Earth. Because of their weak force and gravitational interactions with matter, neutrinos are not attenuated as they travel across the universe. The presence of high-energy neutrinos would give hints for the search for astronomical UHCR accelerators \citep{2002RPPh...65.1025H, 2018MNRAS.481.4461F}.\\

So far, only one steady neutrino source has been identified, the Active Nuclei Galaxy (AGN) Seyfert type 2 NGC 1068, which is consistent with a neutrino cluster with a significance of 4.2 $\sigma$ \citep{2022Sci...378..538I}. Furthermore, on 22 September 2017, IceCube detected the neutrino IceCube-170922A with an energy above 0.1 PeV. The blazar TXS 0506 + 056 is located within the uncertainty of the neutrino arrival position; it was in a flare state at the time of the detection. The probability of this happening is $\sim$ 3$\sigma$ \citep{2018Sci...361.1378I}. IceCube searched 9.5 years of data in the direction of TXS 0506+056 and found an excess respect to the expected neutrino background spanning from September 2014 to March 2015. This variable flux represents evidence of 3.5$\sigma$ of the neutrino emission from this blazar \citep{2018Sci...361..147I} without an increase in its electromagnetic counterpart.\\

IceCube does not observe neutrinos directly but rather detects with their 5160 digital optical modules (DOMs) \citep{2013Sci...342E...1I} the Cherenkov light produced by secondary particles created by the interaction of the neutrino-nucleon interactions. Neutrinos produce charged-current (CC) interactions with nuclei. The nucleus is broken apart, releasing a shower of hadrons, and a percentage (80\%) of the energy is transferred to the charged lepton created. For the case of $\nu_{\mu}$ (as well as $\bar{\nu}_{\mu}$), secondary muons are created along the hadronic shower; those muons can travel several kilometers before decay. As a result, the light detected exhibits a track-like signature. In contrast, an electron (positron) neutrino $\nu_{e}$ ($\bar{\nu}_{e}$) causes an immediate electromagnetic cascade. This shower overlaps the hadronic shower generated at the vertex point. In the detector, this topology has a elliptical shape. With the light deposited in the DOMs, the neutrino's arrival position and energy can be reconstructed. In the case of CC-$\nu_{\mu}$, the long light paths allow for a better reconstruction of the neutrino position in the sky, with errors $\mathcal{O}(1) \rm{deg}$ but with a worse energy reconstruction, because only a fraction of the track events lies in the instrument. On the other hand, for CC-$\nu_{e}$ a better energy reconstruction is archived, but a worse arrival resolution is obtained, with median angular errors at the position of $\mathcal{O}(10-15) \rm{deg}$\citep{2013Sci...342E...1I}. Neutral-current (NC) cascades involve neutrinos transferring only a small portion of their energy to the nucleus before leaving. Therefore, we can expect only cascade topologies for all three types of neutrinos, with production at the vertex point limited to shower events \citep{2021hean.book.....B}. \\

The Fermi Large Area Telescope (LAT) is a pair-conversion telescope that detects $\gamma$ rays between 20 MeV and 300 GeV \citep{2009ApJ...697.1071A}. The fourth Fermi-LAT $\gamma$-ray source catalog (4FGL) \citep{2020ApJS..247...33A} uses Pass 8 data, which improved angular resolution above 3 GeV, increased acceptance by $\sim20\%$, and expanded the effective area to 2.5 $\mathrm{m^{2},sr}$ (2–300 GeV). Galactic diffuse emission modeling was refined, and energy dispersion effects were included for more accurate reconstructions. The catalog is based on eight years of data from 50 MeV to 1 TeV, with later updates: 4FGL-DR2 (10 years; \citealt{2020arXiv200511208B}) and 4FGL-DR3 (12 years; \citealt{2022ApJS..260...53A}). A dedicated AGN subset, 4LAC \citep{2020ApJ...892..105A}, covers sources in the 50 MeV–1 TeV range.   In this paper, we describe the observations in Section \ref{Sec:Observations}, outline the procedure for finding correlations with spatial arguments in Section \ref{sec:Analysis}, present the physical formulation for describing the spectra of the candidates in correlation with neutrinos in Section \ref{sec:model}, and discuss our findings in Section \ref{sec:discussion}.


\section{Observations.}\label{Sec:Observations}

\subsection{IceCube Neutrinos}\label{subsec:neutrino_data}

Significant angular uncertainties in cascade-like topologies motivated us to rule them out from this work. We are looking to conciliate point-like sources detected by Fermi-LAT with the neutrinos detected by IceCube. This study examines the IceCube catalog IceCat-1 \citep{2023ApJS..269...25A}. This catalog contains 275 high-energy neutrinos that exhibit a track-like topology, recorded within a time frame from 2011 to 2023 in its initial release. The data is publicly accessible and regularly updated\footnote{\url{https://dataverse.harvard.edu/dataset.xhtml?persistentId=doi:10.7910/DVN/SCRUCD}}.
\\

\subsection{Fermi-LAT $\gamma$-rays}\label{subsec:photons}


This work is based on the Second Data Release of the Fourth LAT AGN Catalog (4LAC-DR2; \citealt{2020arXiv201008406L}), which extends the 4LAC catalog \citep{2020ApJ...892..105A} to ten years of Fermi-LAT observations. The original 4LAC catalog comprises 2,863 AGNs detected in $\gamma$~rays, including 665 flat-spectrum radio quasars (FSRQs), 1067 BL Lacertae objects (BL Lacs), 1077 blazar candidates of uncertain type (BCUs), and 64 other AGNs, such as radio galaxies and narrow-line Seyfert~1s. The 4LAC-DR2 release adds 285 new AGNs, consisting of 39 FSRQs, 59 BL~Lacs, 185 BCUs, and 2 radio galaxies. In this work, we focus on the FSRQ population from the high-latitude subset of 4LAC-DR2 to ensure reliable source associations and well-constrained $\gamma$-ray spectra.\footnote{The \texttt{FITS} tables for these catalogs are publicly available at \url{https://fermi.gsfc.nasa.gov/ssc/data/access/lat/4LACDR2/}.}\\

\noindent The spectra for each source were obtained through binned likelihood analysis following the standard steps of the Fermi-LAT Collaboration (as described below).\footnote{\url{https://fermi.gsfc.nasa.gov/ssc/data/analysis/scitools/binned\_likelihood\_tutorial.html}} The data and the spacecraft files were obtained from the Fermi-LAT Data Server with pass 8.\footnote{\url{https://fermi.gsfc.nasa.gov/cgi-bin/ssc/LAT/LATDataQuery.cgi}} The data set was selected at the neutrino position in the energy range of 100 $\mathrm{MeV}$ to 1 $\mathrm{TeV}$ and a mask radius of 15$^{\circ}$. The likelihood analysis was performed using \texttt{Fermi Science Tools} \citep{2019ascl.soft05011F} in the conda-based python interface with version \texttt{2.2.0} of Fermi-tools.\\



\noindent A selection of the data set was applied at the neutrino position in the energy range from 100 $\mathrm{MeV}$ to 1 $\mathrm{TeV}$ and a search radius of 15$^{\circ}$. From this, only source events were retained (\texttt{evclass} = 128), and the types of events from the front+back section of the tracker were converted to events within all subclasses PSF and Energy subclasses (\texttt{evtype} = 3). We also take into account, according to the Fermi-LAT team\footnote{\url{https://fermi.gsfc.nasa.gov/ssc/data/analysis/documentation/Cicerone/Cicerone_Data_Exploration/Data_preparation.html}} a maximum zenith angle of 90\deg  recommended for point like sources and a filter \texttt{(DATA\_QUAL>0)\&\&(LAT\_CONFIG==1)} on the Good Time Intervals centered at the position of the neutrino arrival position. Since a Region Of Interest (ROI) based zenith cut was not applied. The angle of inclination between a source and the LAT normal determines the response functions of the LAT instrument. The count of a source should depend on the time it spent at a given inclination angle during an observation.  The LAT's livetime, or time observed at a given sky position and inclination angle, depends only on its orientation history, not the source model, the tool \texttt{gtltcube} performs these corrections and then is part of our pipeline. Finally, we compute the exposure map with the response function \texttt{P8R3\_SOURCE\_V3} within a radius of 40\deg of the ROI and taking into account ten energy bins.\\

\noindent In order to perform the likelihood analysis, ROI was modeled into an \texttt{xml} file, in which the spectral properties of the gamma ray sources reported in the 4FGL catalog are considered, freeing the sources within  5\deg from the neutrino arrival position. In addition to this, we searched for a potential new gamma ray source at the neutrino arrival position; then a new source was added assuming a power-law spectral shape. The Galactic diffuse emission model used is \texttt{gll\_iem\_v07}, while the extragalactic isotropic diffuse emission model chosen is \texttt{iso\_P8R3\_SOURCE\_V3\_v1}. Taking this into account, the likelihood  was calculated using the python interface \texttt{pyLikelihood} which is part of the fermi-tools package \citep{2019ascl.soft05011F}. We iterate over 50 times to find that the likelihood converges, the first ten times, we chose the \texttt{DRMNFB} minimizer if this converges on these steps we use the output as input for at least another interaction, but using \texttt{the NewMinuit} minimizer\footnote{\url{https://fermi.gsfc.nasa.gov/ssc/data/analysis/documentation/Cicerone/Cicerone\_Likelihood/Fitting\_Models.html}}. \\

We build the light curves shown in Figures \ref{fig:LC_IC110807A}-\ref{fig:IC230914A}. Specifically, we analyzed the area surrounding the detected neutrino for a period of 30 days before and after its arrival, focusing our analysis on the time reported by IceCube and using one-day intervals. We searched for any emissions from a potential new gamma-ray source and examined the behavior of the 4FGL sources located within the error region of the reported position.


\section{Analysis}\label{sec:Analysis}

To determine whether the FSRQ identified by Fermi-LAT are progenitors of neutrinos, we connect them to the IceCat-1 neutrino database from IceCube. In many situations, neutrinos and $\gamma$-ray sources are not coincident. However, the $\gamma$-ray sources are located inside the error region of the reconstructed neutrino position \citep[e.g.][]{2018Sci...361.1378I, 2019ApJ...880..103G, 2023MNRAS.519.1396S, 2020ApJ...899..113P}, the methodology employed in this study to determine spatial coincidence is as follows:

Table \ref{table:Correlations} presents the correlation data along with the angular separation and localization of neutrinos and 4FGL sources. The table presented displays neutrino events in the first column, followed by the (RA, Dec) best arrival direction reconstructed with a 90\% confidence level uncertainty, and the best-fit energy reconstruction in the third column. The fourth column presents the 4FLG source associated with each neutrino, along with the celestial coordinates (RA, Dec) of the $\gamma$-ray source and the angular separation between the $\gamma$-ray source and the neutrino arrival position. This source's counterpart and the redshift are presented. The counterpart is sourced from the 4LAC data, and the redshift is acquired from astronomical databases including NED\footnote{\url{https://ned.ipac.caltech.edu/}} and CDS Portal\footnote{\url{http://cdsportal.u-strasbg.fr/}}, with queries performed using the 4FGL designation. With the exception of the $\gamma$-ray source 4FGL J1834.2+3136, the investigation was carried out using the name of the corresponding source. \\

\begin{table*}[htbp]
  \centering\small
  \caption{Spatial correlation found between high energy neutrinos reported by IceCube and $\gamma$-ray sources with a FSRQ counterpart reported by Fermi-LAT. \label{table:Correlations}} 
    \begin{tabular}{c|c|c|c|c|c|c|c}
 \toprule
 
 \begin{tabular}[c]{@{}c@{}}IceCube \\ Neutrino\end{tabular} & \begin{tabular}[c]{@{}c@{}}Position \\ (RA$^{\circ}\substack{+\Delta {\rm RA} \\ -\Delta {\rm RA}}$, Dec$^{\circ}\substack{+\Delta {\rm Dec} \\ -\Delta {\rm Dec}}$)\end{tabular} & \begin{tabular}[c]{@{}c@{}}Energy \\ (TeV)\end{tabular} & \begin{tabular}[c]{@{}c@{}}4FGL \\ Source\end{tabular} & \begin{tabular}[c]{@{}c@{}}Position \\ (RA$^{\circ}$, Dec$^{\circ}$)\end{tabular} & \begin{tabular}[c]{@{}c@{}}Angular \\ Separation ($^{\circ}$) \end{tabular} & Counterpart & Redshift \\
 \midrule
 110807A & (336.80$\substack{+1.36 \\ -1.98}$, 1.53$\substack{+0.93 \\ -0.78}$)  & 108.0 & J2226.8+0051 & (336.71, 0.86) & 0.67 & PKS B2224+006 & 2.25 \\
110930A & (267.01$\substack{+1.19 \\ -1.14}$, -4.44$\substack{+0.60 \\ -0.79}$) & 160.0 & J1744.2-0353 & (266.05, -3.88) & 1.10 & PKS 1741-03  & 1.05 \\
120515A & (198.94$\substack{+1.71 \\ -1.41}$, 32.00$\substack{+0.97 \\ -1.09}$) & 194.0 & J1310.5+3221 & (197.63, 32.35) & 1.16 & OP 313  & 1.67 \\
  - &                                                               -       & -     & J1311.0+3233 & (197.75, 32.55) & 1.14 & RX J131058.8+323335 & 1.63 \\
120916A & (182.24$\substack{+1.36 \\ -1.71}$, 3.88$\substack{+0.67 \\ -0.82}$)  & 174.0 & J1204.8+0407 & (181.20, 4.12) & 1.06 & MG1 J120448+0408 & 1.94 \\
130127A & (352.97$\substack{+1.32 \\ -1.01}$, -1.98$\substack{+0.97 \\ -0.90}$) & 235.0 & J2333.4-0133 & (353.36, -1.55) & 0.58 & PKS B2330-017 & 1.06 \\
  - &                                                               -       & -     & J2335.4-0128 & (353.86, -1.47) & 1.03 & PKS 2332-017 & 1.18 \\
140114A & (337.59$\substack{+0.57 \\ -0.92}$, 0.71$\substack{+0.97 \\ -0.86}$)  & 54.0  & J2226.8+0051 & (336.71, 0.86) & 0.89 & PKS B2224+006 & 2.25 \\
141012A & (63.85$\substack{+2.24 \\ -1.36}$, 3.21$\substack{+0.90 \\ -1.08}$)   & 173.0 & J0422.8+0225 & (65.70, 2.42) & 2.01 & PKS 0420+022 & 2.27 \\
141210A & (318.12$\substack{+2.33 \\ -1.93}$, 1.57$\substack{+1.57 \\ -1.72}$)  & 154.0 & J2118.0+0019 & (319.50, 0.32) & 1.86 & PMN J2118+0013 & 0.46 \\
150104A & (272.11$\substack{+1.71 \\ -1.54}$, 28.76$\substack{+2.41 \\ -1.86}$) & 133.0 & J1814.4+2953 & (273.61, 29.89) & 1.73 & B2 1811+29  & 1.35 \\
150904A & (133.77$\substack{+0.53 \\ -0.88}$, 28.08$\substack{+0.51 \\ -0.55}$) & 302.0 & J0852.2+2834 & (133.06, 28.57) & 0.79 & B2 0849+28  & 1.28 \\
150919A & (279.54$\substack{+1.76 \\ -2.29}$, 30.35$\substack{+2.19 \\ -1.50}$) & 228.0 & J1834.2+3136 & (278.56, 31.60) & 1.51 & 4C +31.51  & 0.59 \\
170308A & (155.35$\substack{+2.02 \\ -1.19}$, 5.53$\substack{+0.98 \\ -0.90}$)  & 107.0 & J1018.4+0528 & (154.61, 5.47) & 0.73 & TXS 1015+057 & 1.94 \\
181212A & (316.41$\substack{+1.85 \\ -2.02}$, -31.0$\substack{+1.68 \\ -1.58}$) & 162.0 & J2101.4-2935 & (315.36, -29.59)& 1.67 & PKS 2058-297 & 1.50 \\
201130A & (30.54$\substack{+1.10 \\ -1.27}$, -12.10$\substack{+1.14 \\ -1.11}$) & 203.0 & J0206.4-1151 & (31.60, -11.85) & 1.07 & PMN J0206-1150 & 1.66 \\
211216A & (316.05$\substack{+2.55 \\ -1.93}$, 15.79$\substack{+1.62 \\ -1.24}$) & 113.0 & J2108.5+1434 & (317.14, 14.58) & 1.61 & OX 110  & 2.02 \\
220509A & (334.25$\substack{+1.93 \\ -1.41}$, 5.38$\substack{+1.65 \\ -1.58}$)  & 177.0 & J2212.8+0647 & (333.21, 6.78) & 1.74 & TXS 2210+065 & 1.12 \\
220928A & (207.42$\substack{+1.41 \\ -2.46}$, 10.43$\substack{+0.91 \\ -0.91}$) & 143.0 & J1342.6+0944 & (205.67, 9.73) & 1.86 & NVSS J134240+094752 & 0.28 \\
230914A & (163.83$\substack{+2.55 \\ -2.02}$, 31.83$\substack{+2.08 \\ -1.77}$) & 168.0 & J1102.9+3014 & (165.74, 30.24) & 2.28 & B2 1100+30B  & 0.38 \\
 
 \bottomrule
    \end{tabular}%
  \label{tab:addlabel}%
\end{table*}


For completeness, we calculated the probability that a neutrino and a gamma-ray source are spatially correlated using statistical arguments motivated by the strategy followed by \citet{2020MNRAS.492.5011D}. To do this, we considered the neutrinos detected by IceCube listed in Table \ref{table:Correlations} together with the uncertainty of the reconstructed position of the neutrino. This motivates us to consider only events with track-type topology, as these are the events that have less uncertainty when locating the neutrino. Once this population was taken, we simulated a synthetic distribution of gamma-ray sources distributed uniformly across the celestial sphere, except for galactic longitude values less than 10º. This last argument is because we sought to recreate the FSRQs reported in the 4LAC \citep{2022ApJS..263...24A} on the objects with $|b| > 10$. Thus, a coincidence will be taken into account if a synthetic gamma-ray source is located within the error region of the neutrino position. We repeated this action 1,000 times for each neutrino, calculating the probability as the ratio of favorable outcomes to total attempts throughout the process. The probability of serendipity is the value at which this ratio converges during these repetitions. Consequently, this probability is determined solely by the uncertainty associated with the neutrino. This means that events with a high degree of uncertainty will be less significant, as there will be more sources of gamma rays present within that portion of the solid angle. \\

\begin{figure*}[ht!]
 \centering
 \includegraphics[width=1.\linewidth]{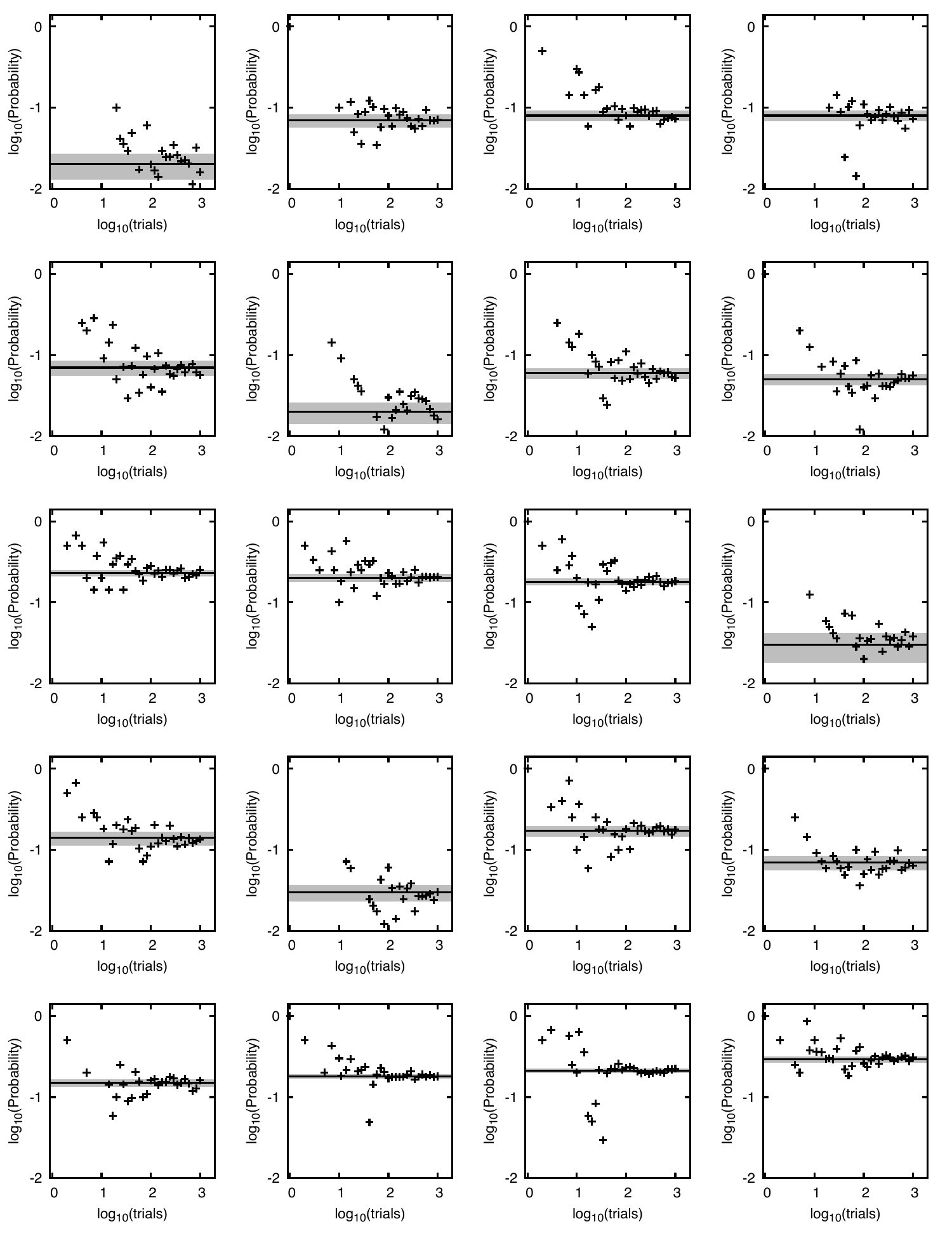}
 \caption{Serendipity probability for each spatial correlation found. As it can be appreciated, the match rate remains low ($\sim1\%$) for all the sources. The black line shows the probability obtained and the gray filled area denotes 1$\sigma$ of the mean value.}
 \label{fig:Serendipy}
\end{figure*}

\begin{table}[!t]\centering
 \setlength{\tabnotewidth}{0.2\columnwidth}
 \tablecols{3}
 \caption{Average value of the Monte Carlo processes done to compute each spatial association found. The first column list the IceCube event. The second column list the FRSQ associated to the gamma-ray source listed in the 4LAC-DR3 and finally, the third column show the average value with their standard deviation of the values obtained in the Monte Carlo Processes.} \label{tab:Serendipy}
 \begin{tabular}{clc}
 \toprule
 IceCube event & FSRQ Associated & \begin{tabular}[c]{@{}c@{}}Serendipity \\ probability\end{tabular}   \\
 \midrule
 IC110807A & PKS B2224+006 & $0.02 \pm 0.007$ \\
 IC110930A & PKS 1741-03 & $0.07 \pm 0.013$ \\
 IC120515A & OP 313 & $0.08 \pm 0.012$ \\
 IC120515A & RX J131058.8+323335 & $0.08 \pm 0.012$ \\
 IC120916A & MG1 J120448+0408 & $0.07 \pm 0.015$ \\
 IC130127A & PKS B2330-017 & $0.02 \pm 0.006$ \\
 IC130127A & PKS 2332-017 & $0.06 \pm 0.009$ \\
 IC140114A & PKS B2224+006 & $0.05 \pm 0.008$ \\
 IC141012A & PKS 0420+022 & $0.23 \pm 0.020$ \\
 IC141210A & PMN J2118+0013 & $0.20 \pm 0.023$ \\
 IC150104A & B2 1811+29 & $0.18 \pm 0.017$ \\
 IC150904A & B2 0849+28 & $0.03 \pm 0.012$ \\
 IC150919A & 4C +31.51 & $0.14 \pm 0.027$ \\
 IC170308A & TXS 1015+057 & $0.03 \pm 0.007$ \\
 IC181212A & PKS 2058-297 & $0.17 \pm 0.026$ \\
 IC201130A & PMN J0206-1150 & $0.07 \pm 0.014$\\
 IC211216A & OX 110 & $0.15 \pm 0.016$\\
 IC220509A & TXS 2210+065 & $0.18 \pm 0.010$ \\
 IC220928A & NVSS J134240+094752 & $0.21 \pm 0.013$\\
 IC230914A & B2 1100+30B & $0.29 \pm 0.026$ \\
 \bottomrule
 \end{tabular}
\end{table}

The values in Table \ref{tab:Serendipy} are obtained by averaging each of the 1,000 steps, the error indicating the standard deviation of these distributions. The convergence value of thousands of trials is visually determined, indicating that the number is large enough for a clear trend to become apparent. \cite{2011EPJC...71.1554C} considered rejection of the background hypothesis with a significance of at least 5$\sigma$ as a suitable threshold to establish a discovery, which is equivalent to p = $2.87\times10^{-7}$.  To exclude a signal hypothesis, a threshold p-value of 0.05 (equivalent to a confidence level of 95\%) is commonly used, corresponding to $1.64 \, \sigma$. \\

Finally, we investigate evidence of a flaring condition in the gamma-ray source coinciding with IceCube's discovery of the neutrino. The temporal interval designated for the search for an electromagnetic equivalent in this study spans one week before and one week after the discovery of a neutrino by IceCube.  For this investigation, we will utilize Fermi-LAT rate light curves within the energy range of 0.1 to 200 GeV of the LAT 4FGL-DR2 Catalog\footnote{\url{https://fermi.gsfc.nasa.gov/ssc/data/access/lat/10yr_catalog/ap_lcs.php}}. We investigate flares within a data-driven framework. According to the baseline methodology. \cite{2019ApJ...877...39M} presented the foundational technique and then compared it within the light curves derived in Section \ref{subsec:photons} within the energy range of 100 MeV to 1 TeV. The authors used a continuous flux level criterion to determine the flare's onset and end, depending on whether the flux exceeded or dropped below that threshold. This indicates that contiguous blocks constitute a peak if they exceed the baseline, and only those peaks that surpass this baseline are considered flares.  In this study, we utilize the implementation created by \citep{2022icrc.confE.868W}, who developed a code in Python\footnote{\url{https://github.com/swagner-astro/lightcurves/blob/main/README.md}} to identify flares utilizing Bayesian blocks and the Eisenstein-Hut HOP algorithm, which integrates a baseline state to detect flares exhibiting a rise and fall pattern superimposed on a quiescent state.\\


\subsection{IC110807A}\label{subsec:IC110807A}

The event IC110807A was detected on 7 August 2011 at 21:36:00 UTC (MJD 55779.90), with an arrival direction RA, Dec = (336.80\deg$\substack{+1.36 \\ -1.98}$, 1.53\deg$\substack{+0.93 \\ -0.78}$). Within this region, three gamma sources have been reported in the 4LAC. Sorted by the nearest angular distance, the gamma ray sources: 4FGL J2226.6+0210, 4FGL J2226.8+0051, and 4FGL J2223.3+0102, with an angular separation of 0.65\deg, 0.67\deg and 1.07, respectively. The first and the late one are associated with blazars, but 4FGL J2226.8 + 0051, located at RA, Dec = (336.8\deg, 1.53\deg) \citep{2020A&A...644A.159C} is a gamma-ray source associated with a quasar, the PKS B2224+006 \citep{2011MNRAS.410..860A, 2002A&A...386...97J}, with a redshift z = 2.25.\\

\begin{figure}[!ht]
 \includegraphics[width=\columnwidth]{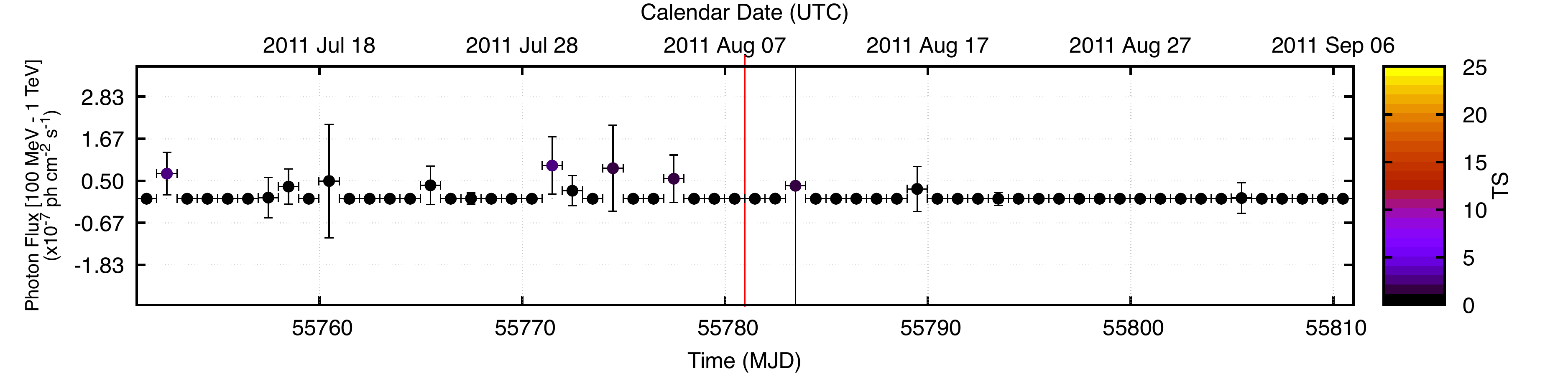}
 \caption{Photon rate detected by Fermi-LAT from 4FGL J2226.8+0051 over 15 years of data acquisition. The vertical dashed line denotes the time of detection of the neutrino by IceCube. In other hand, the vertical shaded region represents the area in which is detected a flare period accord on Hop criteria.}
 \label{fig:LC_IC110807A}
\end{figure}

\noindent The figure \ref{fig:LC_IC110807A} displays the ligth curve in of the 4FGL J2226.8+0051 over 15 years of data acquisition by Fermi-LAT. The Hop method finds a flare activity from MJD 59063 to MJD 59243, with a duration of 180 days. From this figure, we can see that at the time of neutrino detection, the light curve does not show signals of activity different from the baseline recorded up to the time of neutrino detection. In a time window containing one week after and before the detection done by IceCube, we found that the statistical significance at the neutrino position is $0.0\sigma$ in the energy band of 0.1-100 $\rm GeV$. Meanwhile, the 4FGL J2226.8+0051 source has a statistical significance of $1.17\sigma$. These values impede the claim of direct confirmation of 4FGL J2226.8+0051 as a neutrino progenitor. Also, for an event with an angular separation like this, we found that it is poorly associated with statistical arguments.\\

\subsection{IC110930A}\label{subsec:IC110930A}

On 30 September 2011 at 10:40:59 UTC (MJD 55833.44), IceCube detected the neutrino IC110930A, with an arrival position in RA, Dec = (267.01\deg$\substack{+1.19 \\ -1.14}$, -4.44\deg$\substack{+0.60 \\ -0.79}$). Within the error region is a gamma-ray source identified in the 4LAC, designated as 4FGL J1744.2-0353, located at R.A. 266.05\deg and Dec -3.88\deg. This source exhibits an angular separation of 1.10\deg from the neutrino position. At lower energies, the source 4FGL J1744.2-0353 is linked to the FSRQ PKS 1741-03 \citep{2009ApJS..180..283W}, which has a redshift of approximately z = 1.05 \citep{2017ApJS..233....3T}.\\

\begin{figure}[!ht]
 \includegraphics[width=\columnwidth]{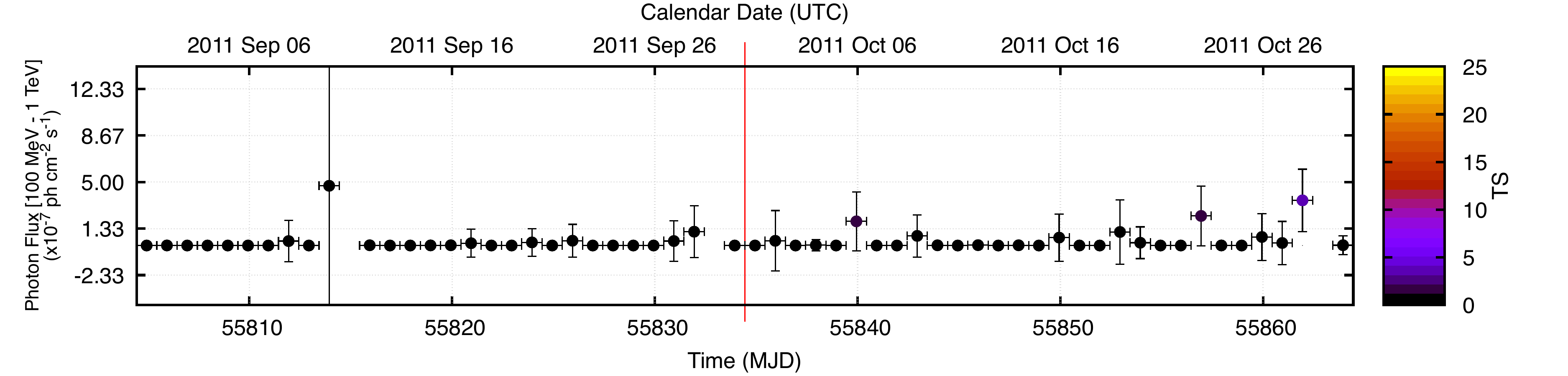}
 \caption{The same as Figure \ref{fig:LC_IC110807A} but for 4FGL J1744.2-0353 and the detection of the neutrino IC110930A.}
 \label{fig:LC_IC110930A}
\end{figure}

\noindent  Figure \ref{fig:LC_IC110930A} displays the ligth curve of the 4FGL J1744.2-0353. The vertical dashed line represents the neutrino time arrival at IceCube. The Hop algorithm applied to this light curve shows three flare episodes. The first from MJD 55403 to MJD 56543, covering 1140 days, the second from MJD 58763 to MJD 59243 with a duration of 480 days, and the last from MJD 59423 to MJD 60100, holding 677 days. We note that the neutrino event occurred on MJD 55834.44, when the first flare was found. Despite this, we can see from Figure \ref{fig:LC_IC110930A} that the light curve does not show a great increase in photon rate from this source in this flare period. The statistical significance recorded from this sky region in the gamma-ray band in energies from 0.1-100 $\rm GeV$, we found that in the spot from the neutrino arrival is $1.28 \sigma$. On the other hand, the statistical significance recorded for the 4FGL J1744.2-0353 is $ 0.0\sigma$. With these values, we find that there is no excess emission for an electromagnetic progenitor associated with this event. This is consistent with the probability of serendipity calculated with a low value of 0.07 reported in Table \ref{tab:Serendipy}. \\

\noindent It is worth mentioning that IceCube triggered a posterior alert, detecting the neutrino IC220205B, which coincides with this gamma-ray source on 5 February 2022 at 20:08:10.59 UTC. A golden alert with an arrival direction of R.A., Dec = ($266.80\pm0.51$ \deg, $-3.58\pm0.51$\deg) \citep{2022GCN.31554....1I}. Nevertheless, the Fermi-LAT collaboration reports that in this sky region, in the energy band of 0.1-300 ${\rm GeV}$, they did not find evidence of an excess on gamma-rays that could be considered a possible progenitor for this neutrino \citep{2022GCN.31558....1G}.\\

\subsection{IC120515A}\label{subsec:IC120515A}

On 15 May 2012 at 23:00:59.27 UTC (MJD 56061.95), the IceCube observatory detected the neutrino IC120515A, which had a reconstructed energy of approximately 194 TeV and an arrival direction of R.A., Dec = (198.94$\substack{+1.71 \\ -1.41}$ \deg, 32.00$\substack{+0.97 \\ -1.09}$ \deg). Within the error region of the arrival direction, three gamma-ray sources are identified in the 4LAC catalog. The sources include 4FGL J1310.5+3221, located at R.A. 197.6324\deg and Dec 32.35\deg, with an angular distance of 1.16\deg from the neutrino position. This source is related to FSRQ OP 313 \citep{2020arXiv200511208B}, with a redshift of z = 0.997 \citep{2010AJ....139.2360S}. The second object is 4FGL J1311.0+3233, positioned in R.A. 197.75\deg, Dec 32.55\deg, with an angular separation of approximately 1.14\deg from the neutrino location. The gamma-ray source is linked to RX J131058.8+323335, identified as an FRSQ with a redshift of z = 1.63 \citep{2015ApJS..219...12A}. The final source is 4FGL J1321.9+3219, a Blazar located in R.A. 200.48\deg, Dec 32.33\deg, exhibiting an angular separation of approximately 1.35\deg. \\

\begin{figure*}[!ht]
\centering{
\subfloat[c][\centering{OP 313}]{\includegraphics[width=\columnwidth]{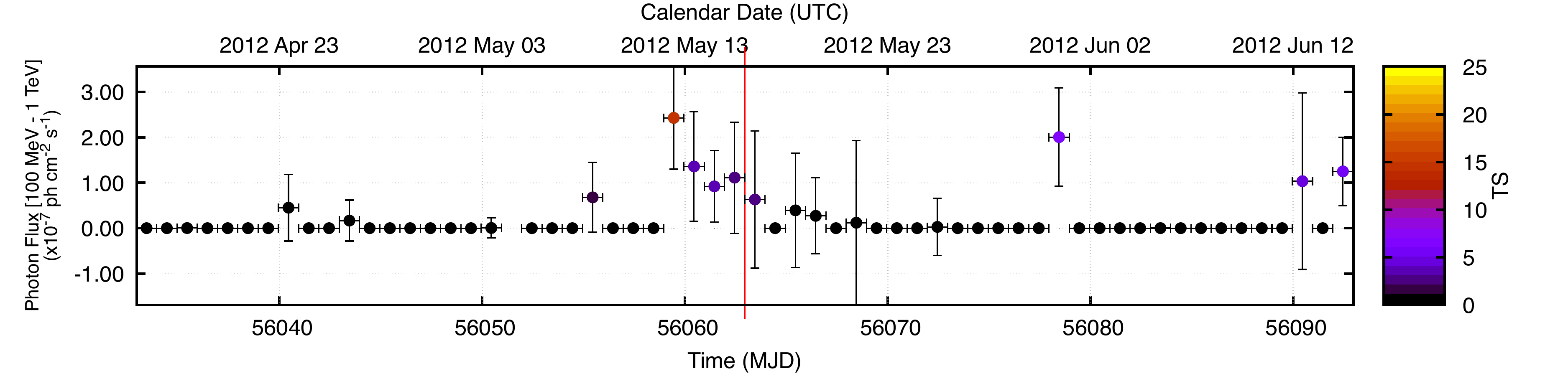} \label{fig:LC_OP313}}

\subfloat[c][\centering{RX J131058.8+323335}]{\includegraphics[width=\columnwidth]{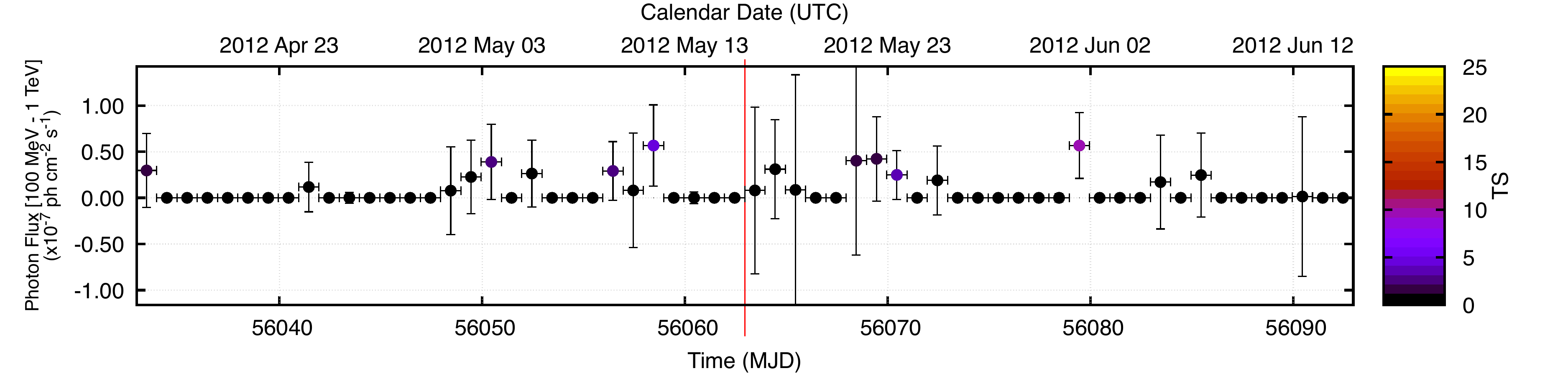}\label{fig:LC_RXJ13}}
}
\caption{The same as Figure \ref{fig:LC_IC110807A} but for 4FGL J1310.5+3221 at top. At bottom 4FGL J1311.0+3233. From the two cases with the detection of the neutrino IC120515A.}
\label{fig:IC120515A}
\end{figure*}

\noindent  Figure \ref{fig:IC120515A} displays the light curves detected from Fermi-LAT in the energy range of 0.1-100 ${\rm GeV}$ for the two FSRQs inside the error region. At the top of the figure, in the subpanel \ref{fig:LC_OP313}, we can see the 4FGL J1310.5+3221 count rate. Meanwhile, 4FGL J1311.0+3233 is displayed in the lower sub-panel \ref{fig:LC_RXJ13}. During the periods; from 55763 to 55793 with a duration of 30 days, from 56183 to 56303 with a duration of 120 days, from MJD 56723 to MJD 56963 with a duration of 240 days, from MJD 58613 to MJD 58943 with a duration of 330 days, from MJD 59243 to MJD 59303 with a duration of 60 days, and the last one from MJD 59453 to MJD 60116 with a duration of 663 days. At the time of the neutrino detection, the source was not in a flare period. On the other hand, we found that seven days before and after neutrino detection, the statistical significance of 4FGL J1310.5+3221 is $ 0.0\sigma$. Meanwhile, the statistical significance recorded at the neutrino position is $1.13\sigma$, which shows that there is no evidence that 4FGL J1310.5+3221 could be the progenitor of this neutrino. IceCube triggered multiple alerts from the position of this source in April (MJD 58949), May (MJD 58970, 58986), and August (MJD 59088) in 2020 and February (MJD 60365) and March (MJD 60399) in 2024 \citep{2022icrc.confE.960T, BRTHESIS}. Despite this, \citet{BRTHESIS} points out that the 2012 alert is the most post-trial relevant event concerning OP 313. We found that the serendipity probability of 0.08 is not statistically significant enough to claim a real association under these considerations.\\

\noindent On the other hand, the sub panel \ref{fig:LC_RXJ13} shows a similar behavior of RX J131058.8+323335, where the Hop algorithm find six flare periods; from MJD 55763 to MJD 55793 with a duration of 30 days, from MJD 56183 to MJD 56303 with a duration of 120 days, from MJD 56723 to MJD 56873 with a duration of 150 days, from MJD 58613 to MJD 58943 with a duration of 330 days, from MJD 59243 to MJD 59303 with a duration of 60 days and finally from MJD 59453 to MJD 60116 with a duration of 663 days. In the case of 4FGL J1310.5+3221, neutrino detection was outside of any flare activity on the electromagnetic counterpart. In a time window centered at the neutrino detection time, one week before and after, we found that the point of the sky where the neutrino was detected have a statistical significance of $0.0\sigma$, meanwhile, the 4FGL J1311.0+3233 have a statistical significance of $4.85\sigma$ which is a weak evidence of gamma-ray activity on the energy range form 0.1-100 ${\rm GeV}$ from this source. Finally, we found that the serendipity probability associated with this event is 0.08, which is also inconclusive.

\subsection{IC120916A}\label{subsec:IC120916A}

On 16 September 2012 at 07:19:38 UTC (MJD 56185.30), IceCube observed the neutrino event IC120916A, with an arrival direction of R.A., Dec = (182.24\deg$\substack{+1.36 \\ -1.71}$, 3.88\deg$\substack{+0.67 \\ -0.82}$) and an energy of 174 TeV. Within the error zone of the reconstructed position, a gamma-ray source identified in the 4LAC, designated as 4FGL J1204.8+0407, is situated at R.A. 181.20\deg, Dec. 4.12\deg, with an angular separation of 1.06\deg. This source is related to MG1 J120448+0408, classified as an FSRQ, with a redshift of z = 1.94.\\

\begin{figure}[!ht]
 \includegraphics[width=\columnwidth]{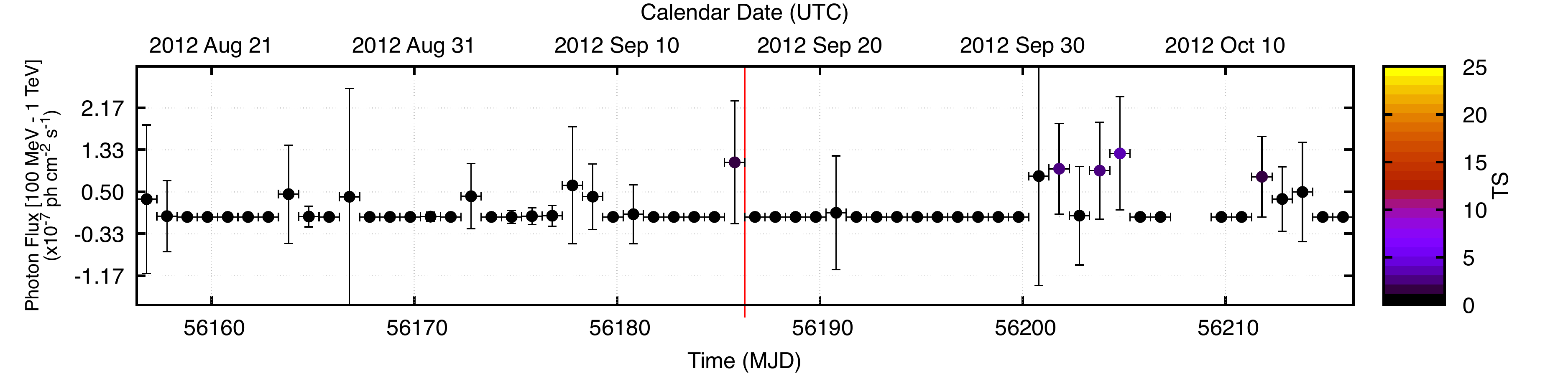}
 \caption{The same as Figure \ref{fig:LC_IC110807A} but for 4FGL J1204.8+0407 and the detection of the neutrino IC120916A.}
 \label{fig:IC120916A}
\end{figure}

\noindent The Figure \ref{fig:IC120916A} displays the light curve from this source, the Hop algorithm does not show any presence of a flare state. In the previous time window centered on neutrino detection, we found that in the neutrino position the statistical significance is $0.0\sigma$, while in the Fermi-LAT source the statistical significance is $0.06\sigma$. This scenario is not favorable for considering MG1 J120448+0408 as a neutrino progenitor if we consider the serendipity probability of 0.07 obtained.\\

\subsection{IC130127A}\label{subsec:IC130127A}

On 27 January 2013 at 06:43:11 UTC (MJD 56318.27), the IceCube observatory detected the neutrino IC130127A. The reconstructed arrival position is R.A., Dec = (337.59\deg $\substack{+0.57 \\ -0.92}$, 0.71\deg $\substack{+0.97 \\ -0.86}$), with an energy of 235 TeV. Within the error region of the reconstructed neutrino position, two sources are identified in the 4LAC. The first object, 4FGL J2333.4-0133, is situated at R.A. 353.36\deg, Dec -1.55\deg with an angular separation of approximately 0.58\deg. The second object, 4FGL J2335.4-0128, is located at R.A. 353.86\deg, Dec -1.47\deg, exhibiting an angular separation of 1.03\deg. Both are linked to a Flat Spectrum Radio Quasar (FSRQ). The initial object is linked to PKS B2330-017, a quasar at a distance of z= 1.05. Another FSRQ is PKS 2332-017 \citep{2010A&A...518A..10V}, located at a distance of z = 1.19 \citep{2020ApJS..250....8L}. \\ 

\begin{figure*}[!ht]
\centering{
\subfloat[c][\centering{PKS B2330-017}]{\includegraphics[width=\columnwidth]{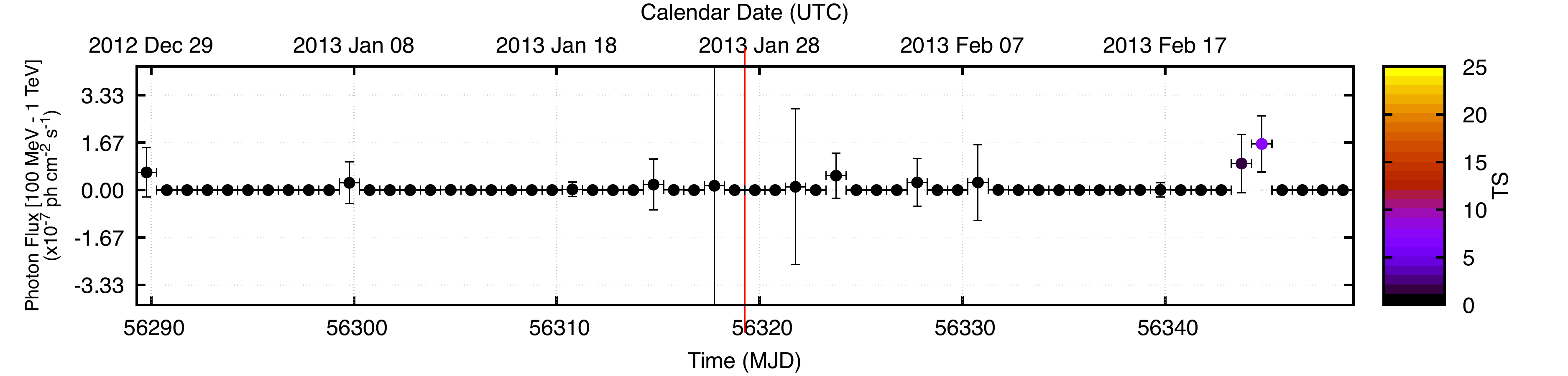} \label{fig:LC_B2330-017}}

\subfloat[c][\centering{PKS 2332-017}]{\includegraphics[width=\columnwidth]{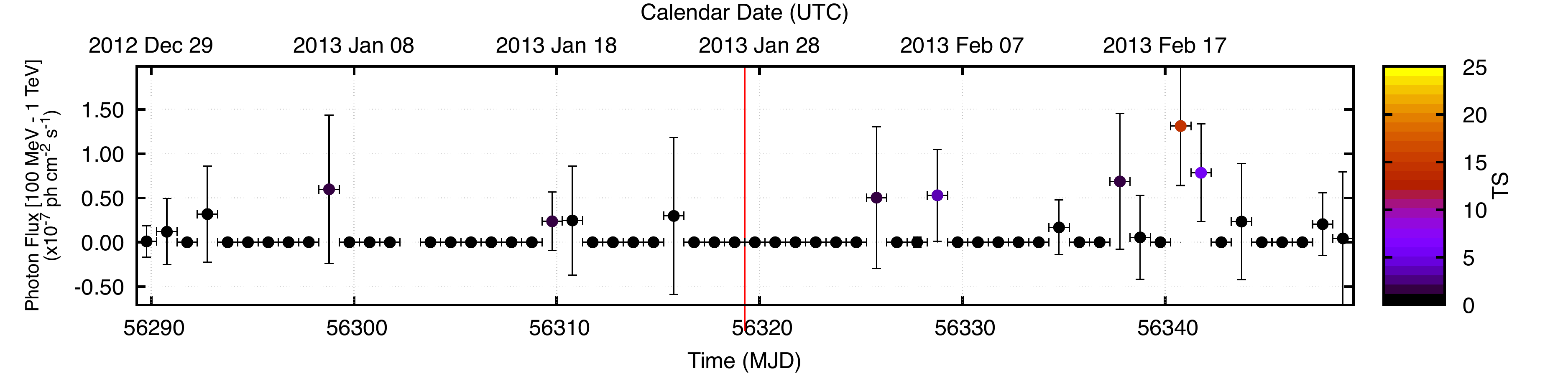}\label{fig:LC_2332-017}}
}
\caption{The same as Figure \ref{fig:LC_IC110807A} but for 4FGL J2333.4-0133 at top. At bottom 4FGL J2335.4-0128. From the two cases with the detection of the neutrino IC130127A.}
\label{fig:IC130127A}
\end{figure*}

\noindent The Figure \ref{fig:IC130127A} displays the rate light curve from these sources. At the top sub-panel \ref{fig:LC_B2330-017} we can appreciate the case of PKS B2330-017 in the gamma-rays band of 0.1-100 $\rm GeV$. The Hop algorithm finds three flare states: from MJD 55223 to MJD 57173 with a duration of 1950 days, from MJD 58313 to MJD 58853 with a duration of 540 days, and finally from MJD 59423 to MJD 60055 with a duration of 632 days. The neutrino time detection lies in the first period. In a time window from one week before and after the detection, we found that the LAT source had a statistical significance of $0.71\sigma$. Meanwhile, the position of the best fit reconstruction had a $0.0\sigma$. Moreover, the serendipity probability that we found is 0.02, which is also a weak argument to tell if this source is a neutrino progenitor.\\

\noindent In the other hand, the sub panel \ref{fig:LC_2332-017} shows the rate light curve of PKS 2332-017 in the energy band of 0.1-100 $\rm GeV$. With the Hop algorithm, we found four flare periods; from MJD 55253 to MJD 56003 with a duration of 750 days, from MJD 56333 to MJD 56873 with a duration of 540 days, from MJD 58673 to MJD 58733 with a duration of 60 days, and finally from MJD 59544 to MJD 60085 with a duration of 541 days. The neutrino detection lies near the start of the second period. In this case, the statistical significance of the Fermi-LAT source is $1.11\sigma$, in the time interval of one week before and after from the neutrino detection, meanwhile the neutrino arrival point has a statistical significance of $0.0\sigma$.

\subsection{IC140114A}\label{subsec:IC140114A}

The IceCube observatory recorded the neutrino event IC140114A on 14 January 2014 at 21:04:09 UTC (MJD 56669.87). The optimal fit reconstructed for this neutrino is R.A., Dec = (337.59\deg $\substack{+0.57 \\ -0.92}$, 0.71\deg $\substack{+0.97 \\ -0.86}$) with an energy of 54 TeV. Within the error region of this neutrino, two gamma-ray sources are identified by LAT and documented in the 4LAC. The first object is 4FGL J2227.9+0036, located at R.A. 336.98\deg, Dec 0.61\deg, associated with PMN J2227+0037, a Blazar \citep{2010A&A...518A..10V}, exhibiting an angular distance of 0.61\deg from the neutrino position. The second object is 4FGL J2226.8+0051, located at an angular distance of 0.89\deg from the neutrino position. This source was also a candidate for the IC110807A neutrino.\\

\begin{figure}[!ht]
 \includegraphics[width=\columnwidth]{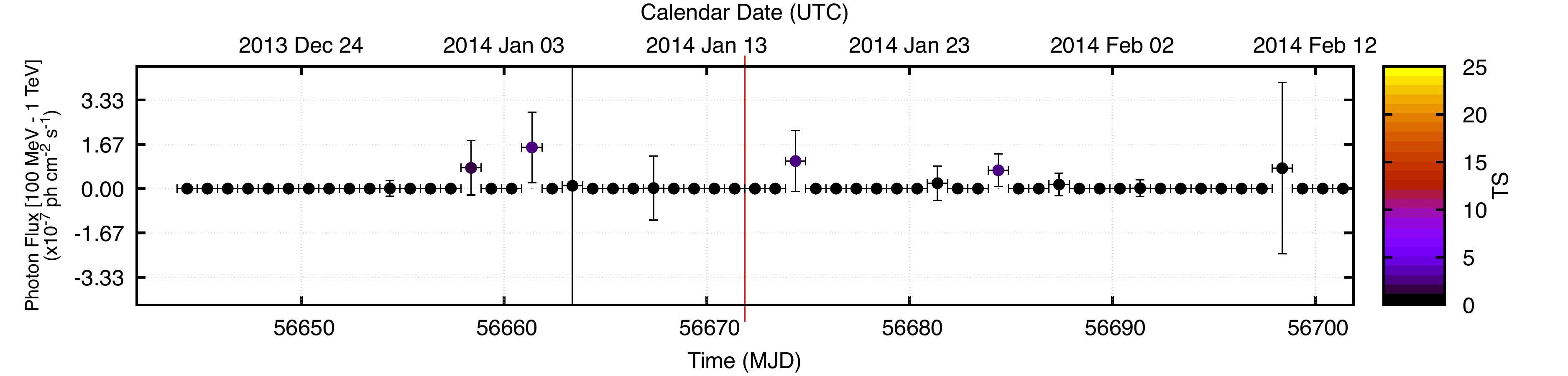}
 \caption{The same as Figure \ref{fig:LC_IC110807A} but for 4FGL J2226.8+0051 and the detection of the neutrino IC140114A.}
 \label{fig:IC140114A}
\end{figure}

\noindent Figure \ref{fig:IC140114A} displays the rate light curve from this source. The Hop algorithm shows a flare period from MJD 59063 to MJD 59243 with a duration of 180 days. The time detection from the neutrino by IceCube is far away from this period. Searching in a time window from one week before and after, centered on the time of neutrino detection, we found that the point of arrival direction of the neutrino has a statistical significance of $0.0\sigma$. Meanwhile, 4FGL J2226.8+0051 has a statistical significance of $1.62\sigma$. Taking into consideration this and the low probability of serendipity of 0.05, we found that this source has weak criteria to be associated with this neutrino.

\subsection{IC141012A}\label{subsec:IC141012A}

On 12 October 2014 at 18:01:14 UTC (MJD 56940.75), the neutrino IC141012A was detected by the IceCube observatory. Its energy was 173 $\rm TeV$, and its arrival direction was R.A., Dec = (63.85\deg $\substack{+2.24 \\ -1.36}$, 3.21\deg$\substack{+0.90 \\ -1.08}$). There are two known gamma-ray sources in the 4LAC inside the error region of this neutrino. The Blazar candidate PKS 0409+025 \citep{2021MNRAS.505.5853G} is linked to 4FGL J0412.3+0239, which is located in R.A., Dec = 63.09\deg, 2.65\deg, with an angular separation of 0.94\deg. The second is the 4FGL J0422.8+0225 in R.A., Dec = 65.70\deg, 2.42\deg, which has a greater angular separation of 2.01\deg. This source is related to the farthest source in our sample, PKS 0420+022, a Quasar \citep{2010ApJ...715..429A} located at z = 2.27 \citep{2017ApJS..233....3T}.\\

\begin{figure}[!ht]
 \includegraphics[width=\columnwidth]{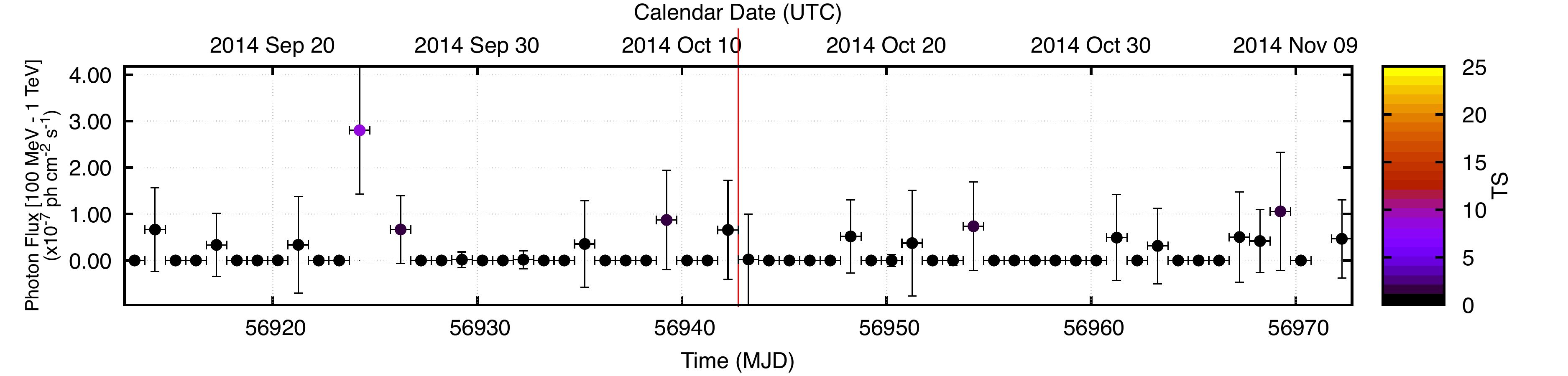}
 \caption{The same as Figure \ref{fig:LC_IC110807A} but for 4FGL J0422.8+0225 and the detection of the neutrino IC141012A.}
 \label{fig:IC141012A}
\end{figure}

\noindent Figure \ref{fig:IC141012A} shows the rate light curve of PKS 0420+022 detected by Fermi-LAT. We note that the Hop algorithm did not find any possible flare periods. Also, in the time window centered on the neutrino position, one week before and after, we found that the neutrino position has a statistical significance of $0.0\sigma$, the same as the 4FLG point source. This event has a chance of being associated with 0.05. The event is not associated with them.

\subsection{IC141210A}\label{subsec:IC141210A}

With an energy of 154 $\rm TeV$ and an arrival position from R.A., Dec = (318.12\deg$\substack{+2.33 \\ -1.93}$, 1.57\deg$\substack{+1.57 \\ -1.72}$), IceCube detects the neutrino IC141210A on 10 December 2014 at 20:20:47 UTC (MJD 57000.84). The 4FGL J2118.0+0019, a 4LAC source, is located inside the error region of the arrival position at (R.A., Dec) = (319.50\deg, 0.32\deg), with an angular separation of 1.86\deg from the neutrino position. This item is located at z = 0.46 \citep{2017ApJS..233...25A} and is associated as an FRSQ at lower energies with PMN J2118+0013 \citep{2007ApJS..171...61H}. \\

\begin{figure}[!ht]
 \includegraphics[width=\columnwidth]{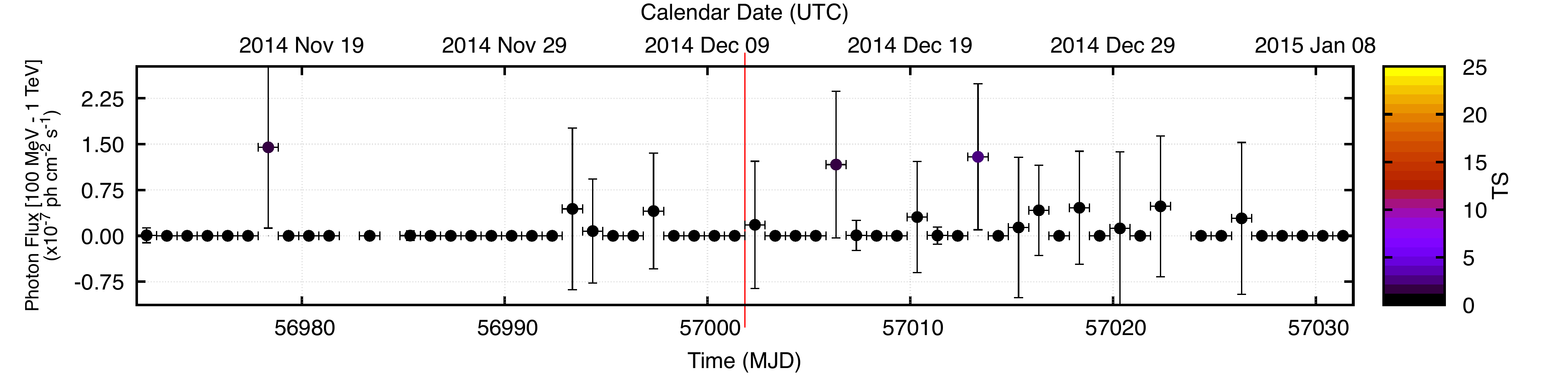}
 \caption{The same as Figure \ref{fig:LC_IC110807A} but for 4FGL J2118.0+0019 and the detection of the neutrino IC141210A.}
 \label{fig:IC141210A}
\end{figure}

\noindent Figure \ref{fig:IC141210A} displays the count rate light curve from 4FGL J2118.0+0019 in the energy range of 0.1-100 $\rm GeV$. As can be seen in it, the Hop algorithm suggests a flare period from MJD 54713 to MJD 55043 with a duration of 330 days. The neutrino detection is far away from this suggested period. Also, the statistical significance, computed in a time window of one week before and after the neutrino detection at the neutrino position, is $0.0\sigma$, while the statistical significance of 4FGL J2118.0+0019 in the same time window is $0.14\sigma$. The probability of serendipity from this source is 0.20. 

\subsection{IC150104A}\label{subsec:IC150104A}

On 4 January 2015 at 09:34:31 UTC (57025.39), IceCube detected the neutrino IC150104A, which possessed an energy of 133 TeV and an arrival position of R.A., Dec = (272.11\deg$\substack{+1.71 \\ -1.54}$, 28.76\deg$\substack{+2.41 \\ -1.86}$). Within the error region of the arrival position, there exist three 4LAC sources, specifically 4FGL J1803.5+2756, 4FGL J1809.7+2910, and 4FGL J1814.4+2953. Starting with 4FGL J1809.7+2910, which is placed at R.A., Dec = 272.44\deg, 29.17\deg, holding an angular separation of 0.50\deg from the neutrino position. This source is classified as BL Lac at lower energies with MG2 J180948+2910 \citep{2009A&A...495..691M}. On the other hand, 4FGL J1814.4+2953 is a gamma-ray source located at R.A., Dec = 273.61\deg, 29.89\deg, which produces an angular separation of 1.73\deg. This source is associated with an FSRQ in lower energies to B2 1811+29 \citep{2005ApJ...626...95S}, with a distance of z = 1.36 \citep{2005ApJ...626...95S}. Finally, the 4FGL J1803.5+2756 located at R.A., Dec = 270.87\deg, 27.94\deg with an angular separation from the neutrino of approximately 1.36\deg, associated with BL LAC, NVSS J180341+275404 \citep{2019ApJS..242....4D}.\\

\begin{figure}[!ht]
 \includegraphics[width=\columnwidth]{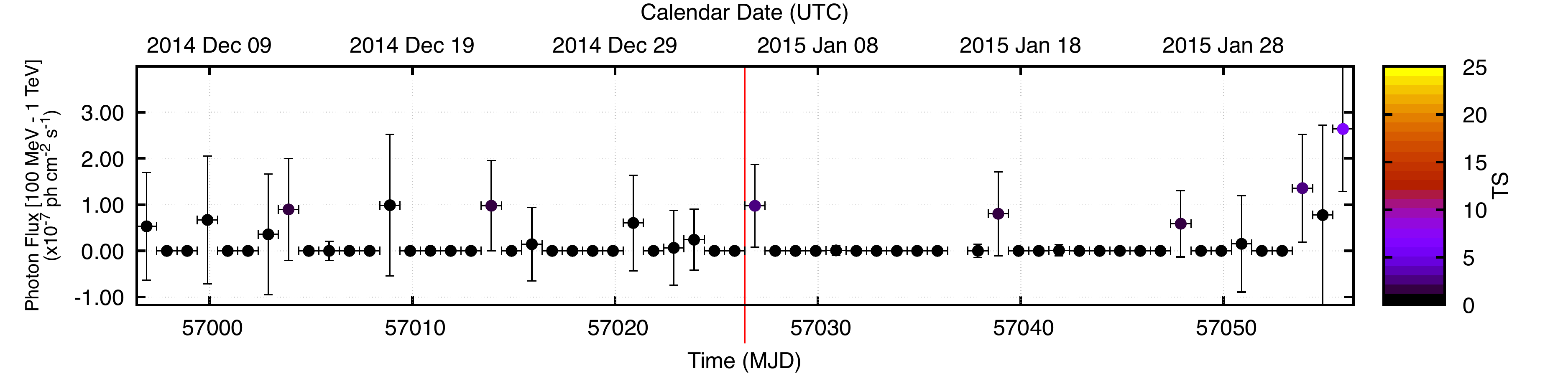}
 \caption{The same as Figure \ref{fig:LC_IC110807A} but for 4FGL J1814.4+2953 and the detection of the neutrino IC150104A.}
 \label{fig:IC150104A}
\end{figure}

\noindent Figure \ref{fig:IC150104A} shows the rate light curve from the gamma-ray source 4FGL J1814.4+2953 in the energy range of 0.1-100 $\rm GeV$. The Hop algorithm did show a flare period on the light curve, from MJD 57442 to MJD 57505 during 63 days, but the neutrino arrival time has a difference of ~440 days. In the gamma-ray band, we find that the 4FGL J1814.4+2953 has a statistical significance of $0.0\sigma$. Meanwhile, the neutrino arrival point has a statistical significance of $1.3\sigma$. Furthermore, the serendipity probability for this source is 0.18. These arguments are not strong enough to claim an association between these two phenomena.

\subsection{IC150904A}\label{subsec:IC150904A}

With an energy of 228 $\rm TeV$ and an arrival position from R.A., Dec = (133.77\deg$\substack{+0.53 \\ -0.88}$, 28.08\deg$\substack{+0.51 \\ -0.55}$), IceCube detects the neutrino IC150904A on 4 September 2015 at 18:13:54 UTC (MJD 57268.75). A 4LAC source, the 4FGL J0852.2+2834, located at R.A., Dec = 133.06\deg, 28.57\deg, with an angular separation of 0.79\deg from the neutrino position, is located inside the arrival position error region. A counterpart to 4FGL J0852.2+2834 in lower energies is located at z = 1.28 \citep{2015ApJS..219...12A}, with the FRSQ B2 0849+28 \citep{2011MNRAS.410..860A}. Another source also lies in the position error region, the 4FGL J0854.0+2753 placed at R.A., Dec = 133.51\deg, 27.88\deg, which is associated at lower energies to the source SDSS J085410.16+275421.7, which is classified as BL Lac \citep{2008AJ....135.2453P}.\\

\begin{figure}[!ht]
 \includegraphics[width=\columnwidth]{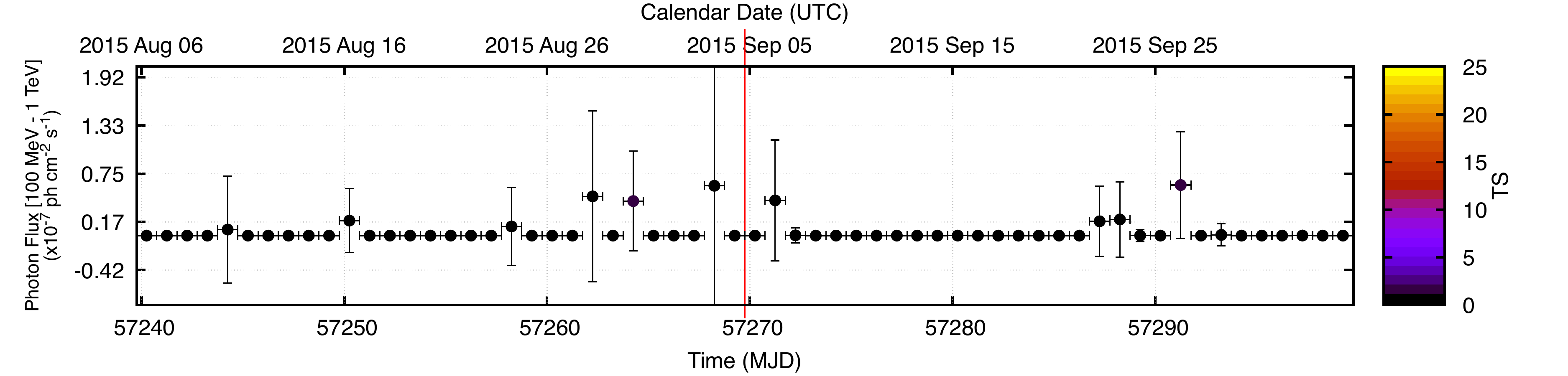}
 \caption{The same as Figure \ref{fig:LC_IC110807A} but for 4FGL J0852.2+2834 and the detection of the neutrino IC150904A.}
 \label{fig:IC150904A}
\end{figure}

\noindent We can appreciate the rate light curve in Figure \ref{fig:IC150904A} in the range of 0.1-100 $\rm GeV$. We note that the Hop algorithm does not show any flare period on the gamma-ray light curve from 4FGL J0852.2+2834. Also, in the week before and after the neutrino detection time. We find that 4FGL J0852.2+2834 has a statistical significance of $0.0\sigma$. Meanwhile, the neutrino position has a statistical significance of $1.48\sigma$. This event has a serendipity probability of 0.03. This evidence is insufficient to support the hypothesis that these sources are associated.

\subsection{IC150919A}\label{subsec:IC150919A}

IceCube identified the neutrino IC150919A on 19 September 2015 at 04:56:09 UTC (MJD 57283.20), with an arrival position at R.A., Dec = (267.01\deg$\substack{+1.19 \\ -1.14}$, -4.44\deg$\substack{+0.60 \\ -0.79}$). Within the error region, there exist four gamma-ray sources identified in the 4LAC. There are; 4FGL J1836.4+3137, 4FGL J1841.3+2909, 4FGL J1834.2+3136, 4FGL J1841.8+3218. We pinpoint that J1836.4+3137, J1841.3+2909, and J1841.8+3218 are sources that are associated with the BL Lac Blazar. J1836.4+3137 is associated with RX J1836.2+3136 \citep{2013ApJS..206...12D}, with an angle separation about 1.32\deg from the neutrino position. J1841.3+2909 is associated with MG3 J184126+2910 \citep{2007ApJS..171...61H}, with a distance of 1.39\deg from the best fit neutrino position. In addition, J1841.8+3218, which is associated with RX J1841.7+3218, a BL Lac Blazar \citep{2014ApJS..215...14D} with an angular separation of 2.10\deg. Finally, the 4FGL J1834.2+3136, located at coordinates R.A. 278.56\deg and Dec 31.60\deg. This source exhibits an angular separation of 1.51\deg from the neutrino position. At lower energies, the source 4FGL J1834.2+3136 is linked to FSRQ 4C +31.51 \citep{1991ApJS...75....1B}, which has a redshift of z = 0.59 \citep{2022ApJ...940...20G}. \\

\begin{figure}[!ht]
 \includegraphics[width=\columnwidth]{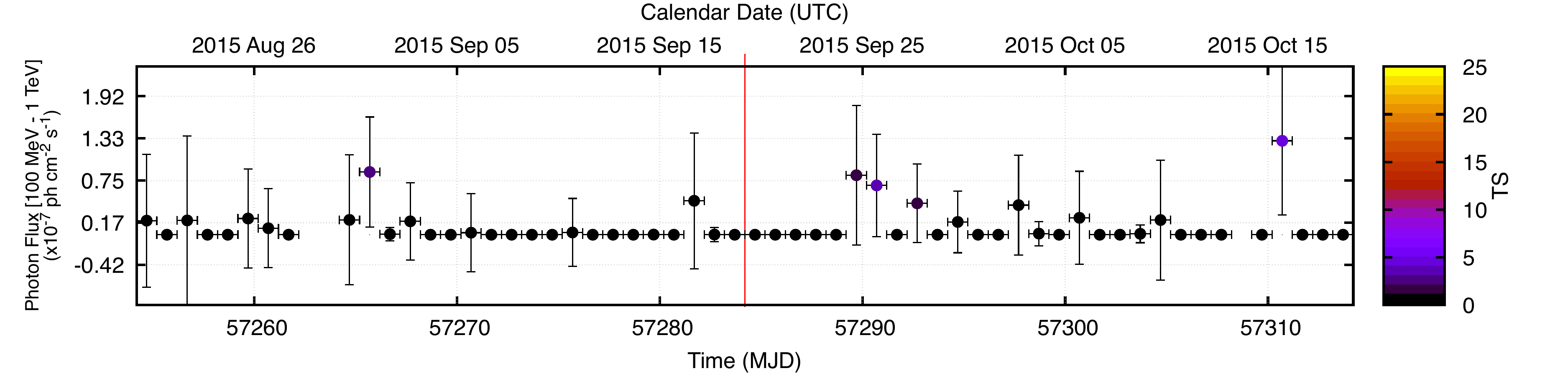}
 \caption{The same as Figure \ref{fig:LC_IC110807A} but for 4FGL J1834.2+3136 and the detection of the neutrino IC150919A.}
 \label{fig:IC150919A}
\end{figure}

\noindent Figure \ref{fig:IC150919A} displays the rate light curve from 4FGL J1834.2+3136. We can see that the Hop algorithm does not suggest any flare activity. Additionally, within the one-week window preceding and following the neutrino observation, we find that the neutrino arrival position and also the 4FGL J1834.2+3136 have a statistical significance of $0.0\sigma$.\\

\subsection{IC170308A}\label{subsec:IC170308A}

On 8 March 2017 at 22:11:49 UTC (MJD 57820.92), IceCube detected the neutrino IC170308A, located at R.A., Dec = (155.35\deg $\substack{+2.02 \\ -1.19}$, 5.53\deg $\substack{+0.98 \\ -0.90}$) . Three gamma-ray sources have been identified within the error region in the 4LAC. The identified sources are: 4FGL J1019.7+0511, 4FGL J1026.9+0608, and 4FGL J1018.4+0528. The J1019.7+0511 and J1026.9+0608 are identified as sources associated with the Blazar category. 4FGL J1019.7+0511 is associated with NVSS J101948+051327 a candidate Blazar \citep{2023arXiv230712546B} with an angular separation of 0.53\deg, from the neutrino position. In the case of 4FGL J1026.9+0608, a gamma-ray source associated with NVSS J102703+060934, a BL Lac \citep{2016ApJS..224...26M} with an angular separation of 1.51\deg, from the neutrino position. Finally, the gamma-ray source 4FGL J1018.4+0528 is associated with an FRSQ in the low-energy band with TXS 1015+057 \citep{2005ApJ...626...95S}. This source is located at R.A., Dec = 154.61\deg, 5.47\deg, separating about 0.73\deg, from the best-fit neutrino position. The distance from this source is about z = 1.94 \citep{2015ApJS..219...12A}.\\

\begin{figure}[!ht]
 \includegraphics[width=\columnwidth]{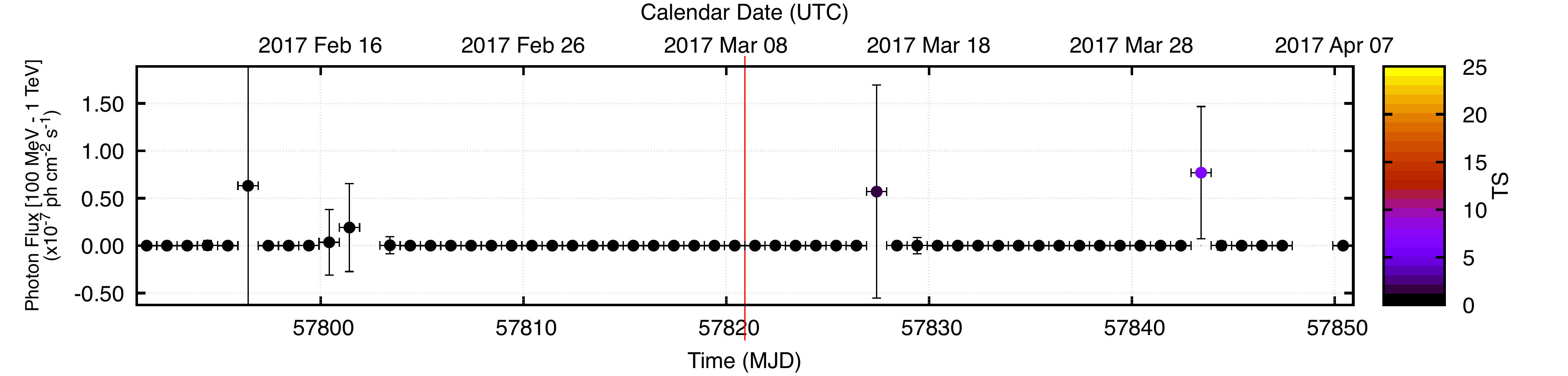}
 \caption{The same as Figure \ref{fig:LC_IC110807A} but for 4FGL J1018.4+0528 and the detection of the neutrino IC170308A.}
 \label{fig:IC170308A}
\end{figure}

\noindent Figure \ref{fig:IC170308A} displays the rate light curve from 4FGL J1018.4+0528 in the energy range of 0.1-100 $\rm GeV$, where the Hop algorithm does not give clues of flaring activity. Additionally, within one week between and after neutrino detection, the statistical significance of 4FGL J1018.4+0528 is $0.0\sigma$. Meanwhile, the best-fit position from this neutrino has a statistical significance of $0.85\sigma$. The serendipity probability of this association is extremely weak, reaching a value of 0.03.

\subsection{IC181212A}\label{subsec:IC181212A}

On 12 December 2018 at 02:02:43 UTC (MJD 58464.08), IceCube detected the neutrino IC181212A, which possessed an energy of 162 $\rm TeV$ and was located at R.A., Dec = (316.41\deg$\substack{+1.85 \\ -2.02}$, -31.0\deg$\substack{+1.68 \\ -1.58}$). The 4LAC source, 4FGL J2101.4-2935, is located at R.A. 315.36\deg, Dec -29.59\deg, with an angular separation of 1.67\deg, from the neutrino position, and is encompassed within the arrival position error region. This object is located at z = 1.50 \citep{2017ApJS..233....3T} and is identified as an FRSQ at lower energies in association with PKS 2058-297 \citep{2010A&A...518A..10V}.\\

\begin{figure}[!ht]
 \includegraphics[width=\columnwidth]{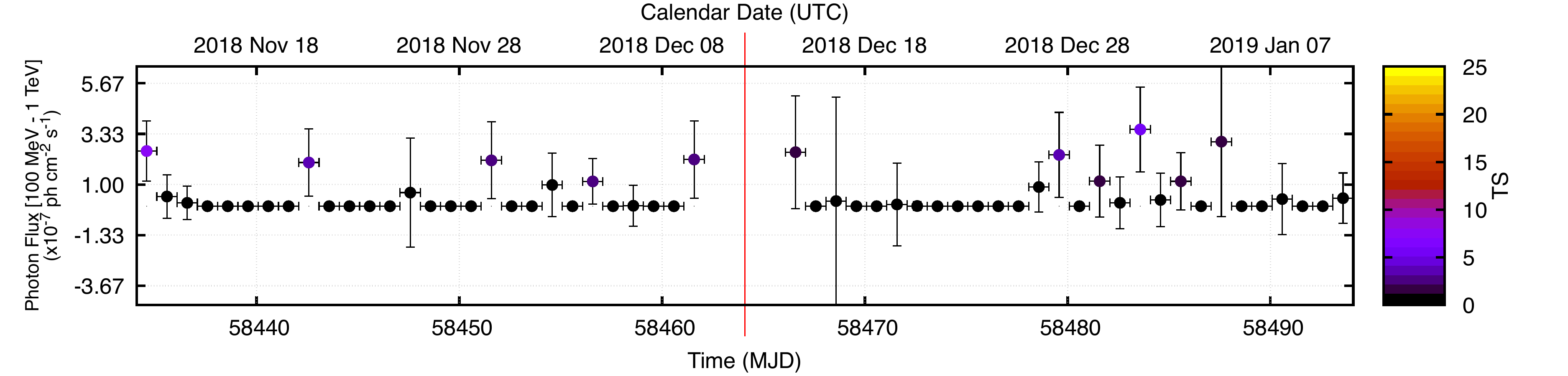}
 \caption{The same as Figure \ref{fig:LC_IC110807A} but for 4FGL J2101.4-2935 and the detection of the neutrino IC181212A.}
 \label{fig:IC181212A}
\end{figure}

\noindent We analyze the rate light curve in Figure \ref{fig:IC181212A} in the energy range of 0.1-100 $\rm GeV$. The Hop method indicates that there are no flare periods in the gamma-ray light curve of 4FGL J2101.4-2935. In the period of one-week preceding and before neutrino detection, 4FGL J2101.4-2935 exhibits a statistical significance of $0.0\sigma$, but the position of the neutrino demonstrates a statistical significance of $2.82\sigma$. This occurrence has a serendipity probability of 0.03. This data undermines the premise that these sources are correlated.

\subsection{IC201130A}\label{subsec:IC201130A}

On 30 November 2020 at 20:21:46 UTC (MJD 59183.84), IceCube identified the neutrino IC201130A, which had an energy of 203 $\rm TeV$ and was positioned at R.A., Dec = (30.54\deg$\substack{+1.10 \\ -1.27}$, -12.10\deg$\substack{+1.14 \\ -1.11}$). The 4LAC source, 4FGL J0206.4-1151, is located at R.A. 31.60\deg, Dec -11.85\deg, with an angular separation of 1.07\deg, from the neutrino position, and lies inside the arrival position error zone. This object is located at z = 1.66 \citep{2018ApJS..237...32A} and is classified as an FRSQ at lower energies in connection with PMN J0206-1150 \citep{2007ApJS..171...61H}. \\

\begin{figure}[!ht]
 \includegraphics[width=\columnwidth]{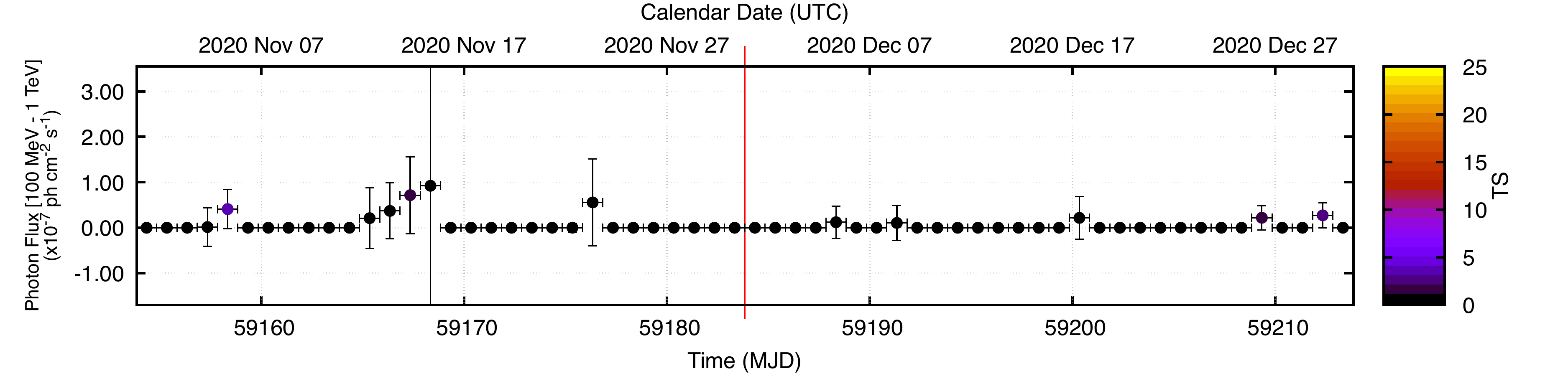}
 \caption{The same as Figure \ref{fig:LC_IC110807A} but for 4FGL J0206.4-1151 and the detection of the neutrino IC201130A.}
 \label{fig:IC201130A}
\end{figure}

\noindent We can examine the rate light curve in Figure \ref{fig:IC201130A} within the range of 0.1-100 $\rm GeV$. The Hop approach suggests three flare phases in the gamma-ray light curve of J0206.4- 1151, from MJD 54983 to MJD 55133 with a duration of 150 days, from MJD 55373 to MJD 55733 with a duration of 360 days, and finally from MJD 57473 to MJD 58523 with a duration of 1050 days. However, the time detection of the neutrino was out of these flare episodes. During the week prior to and following the neutrino detection, J0206.4-1151 shows a statistical significance of $0.0\sigma$, whereas the neutrino location indicates a statistical significance of $1.28\sigma$. The likelihood of this occurrence is 0.07. 

\subsection{IC211216A}\label{subsec:IC211216A}

On 16 December 2021 at 07:07:38 UTC (MJD 59564.29), IceCube detected the neutrino IC211216A, which exhibited an energy of 113 $\rm TeV$ and was located at R.A., Dec = (316.05\deg$\substack{+2.55 \\ -1.93}$, 15.79\deg$\substack{+1.62 \\ -1.24}$). The 4LAC source, 4FGL J2108.5+1434, is situated at R.A. 317.14\deg and Dec 14.58\deg. It has an angular separation of 1.61\deg from the neutrino position and is located within the arrival position error zone. The object is located at z = 2.02 \citep{2018A&A...618A..80G} and is categorized as an FRSQ at lower energies in relation to OX 110 \citep{2007ApJS..171...61H}.\\

\begin{figure}[!ht]
 \includegraphics[width=\columnwidth]{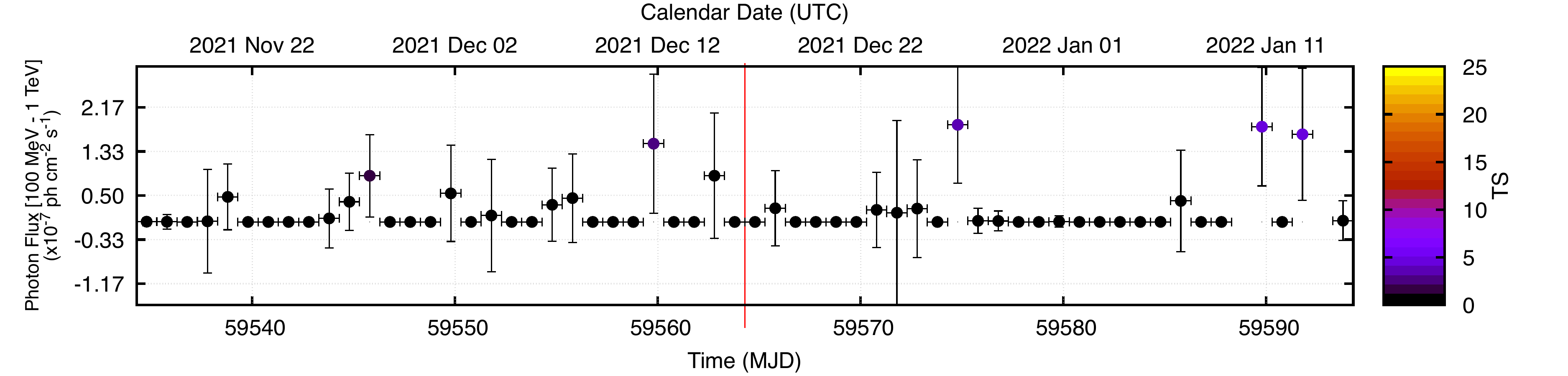}
 \caption{The same as Figure \ref{fig:LC_IC110807A} but for 4FGL J2108.5+1434 and the detection of the neutrino IC211216A.}
 \label{fig:IC211216A}
\end{figure}

\noindent The light curve of the rate  shown in Figure \ref{fig:IC211216A} can be analyzed in the range of 0.1-100 $\rm GeV$. The Hop approach indicates the presence of a moderate flare phase in the gamma-ray light curve of 4FGL J2108.5+1434, from MJD 55595 to MJD 56033, with a duration of 438 days, but the neutrino detection was far away from this activity. In the week preceding and following neutrino detection, 4FGL J2108.5+1434 exhibits a statistical significance of $0.96\sigma$, while the location of the neutrino indicates a statistical significance of $0.18\sigma$. The probability of this event is 0.15. This data contradicts the claim that multiple sources are correlated.

\subsection{IC220509A}\label{subsec:IC220509A}

On 9 May 2022 at 18:19:04 UTC (MJD 59709.76), IceCube identified the neutrino IC220509A, positioned at R.A., Dec = 334.25\deg$\substack{+1.93 \\ -1.41}$, 5.38\deg$\substack{+1.65 \\ -1.58}$). Three gamma-ray sources have been detected within the error zone of the neutrino. The recognized sources are: 4FGL J2215.4+0544, 4FGL J2212.8+0647, and 4FGL J2224.5+0353. J2224.5+0353 and J2215.4+0544 are classified as sources under the Blazar category. 4FGL J2224.5+0353 is linked to 1RXS J222426.5+035445, a candidate Blazar \citep{2009ApJS..184..138H}, exhibiting an angular separation of 2.39\deg from the neutrino location. For 4FGL J2215.4+0544, a gamma-ray source linked to NVSS J221513+054454, a BL Lac \citep{2018AJ....155..189D}, there exists an angular separation of 0.53\deg from the neutrino position. The gamma-ray source 4FGL J2212.8+0647 is linked to an FRSQ in the low-energy band with TXS 2210+065 \citep{2007ApJS..171...61H}. This source is situated at R.A., Dec = 333.21\deg, 9.73\deg, separating 1.89\deg from the optimal neutrino position. The distance from this source is z = 1.12 \citep{2021AJ....161..196P}.\\

\begin{figure}[!ht]
 \includegraphics[width=\columnwidth]{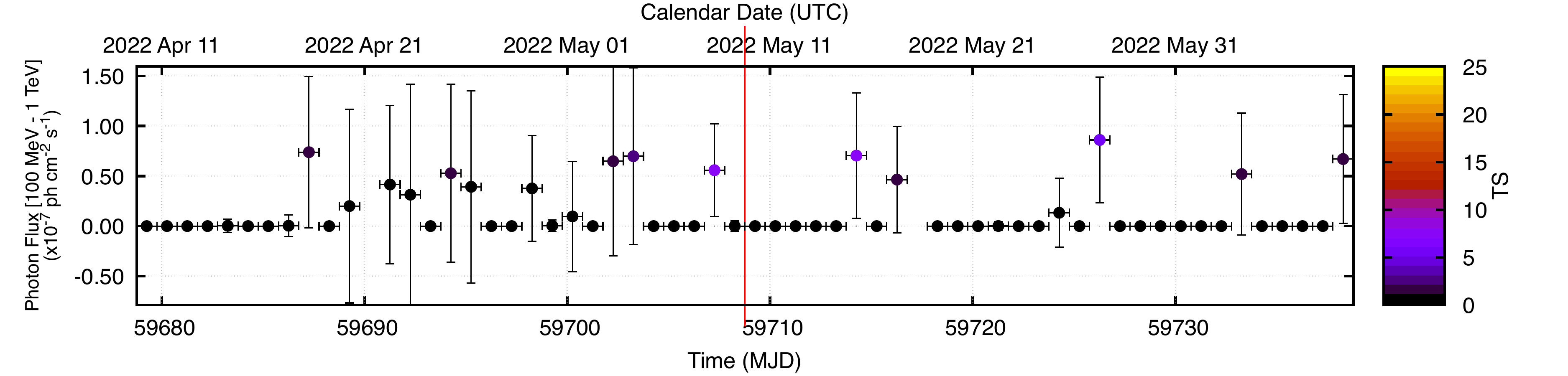}
 \caption{The same as Figure \ref{fig:LC_IC110807A} but for 4FGL J2212.8+0647 and the detection of the neutrino IC220509A.}
 \label{fig:IC220509A}
\end{figure}

\noindent Figure \ref{fig:IC220509A} illustrates the light curve of 4FGL J2212.8+0647 in the energy range of 0.1-100 $\rm GeV$, indicating that the Hop algorithm suggests a discrete increase of activity, from MJD 56873 to 57803 with a duration of 930 days. Furthermore, throughout the one-week interval surrounding the neutrino detection, the statistical significance of 4FGL J2212.8+0647 is $3.64\sigma$, whereas the position of the best-fit from this neutrino exhibits a statistical significance of $0.0\sigma$. The likelihood of serendipity associated with this correlation is relatively low, attaining a value of 0.18.

\subsection{IC220928A}\label{subsec:IC220928A}

 The IceCube neutrino observatory detected the neutrino IC220928A on 28 September 2022 at 12:32:38 UTC (MJD 59851.52), with an energy of 143 TeV and an arrival direction of R.A., Dec = (207.42\deg$\substack{+1.41 \\ -2.46}$, 10.43\deg$\substack{+0.91 \\ -0.91}$). Within the error region of this neutrino, two gamma-ray sources are reported in the 4LAC. The object 4FGL J1351.3+1115, positioned at R.A. 207.84\deg and Dec 11.25\deg, is linked to the BL LAC: RX J1351.3+1115 \citep{2008AJ....135.2453P}, exhibiting an angular separation of 0.92\deg. The second object, exhibiting a larger angular separation of 1.86\deg, is identified as 4FGL J1342.6+0944, positioned at R.A., Dec = 205.67\deg, 9.73\deg. This source is linked to NVSS J134240+094752, a FSRQ \citep{2019ApJS..240....6Y} with a redshift of z = 0.28. \\

\begin{figure}[!ht]
 \includegraphics[width=\columnwidth]{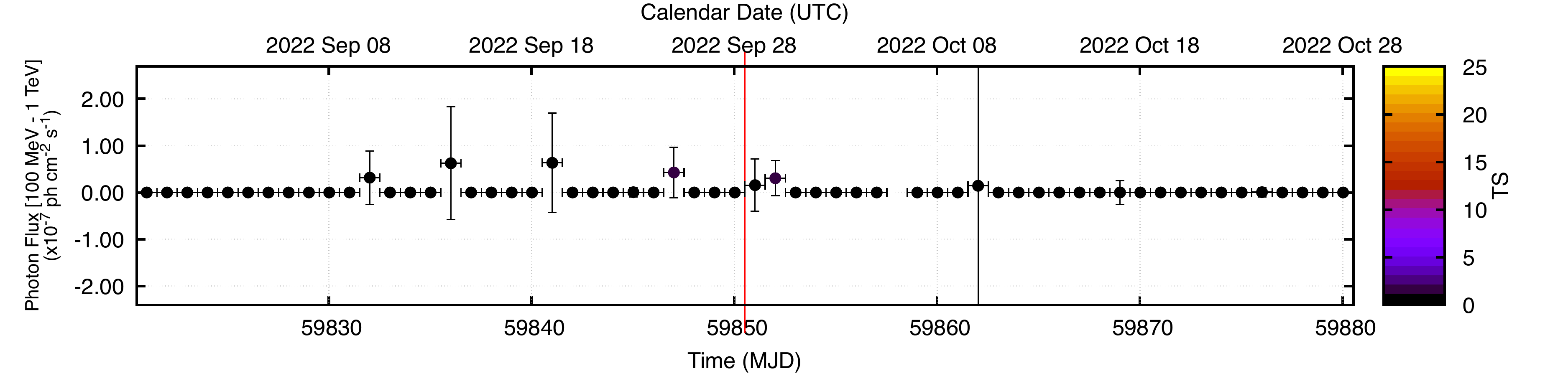}
 \caption{The same as Figure \ref{fig:LC_IC110807A} but for 4FGL J1342.6+0944 and the detection of the neutrino IC220928A.}
 \label{fig:IC220928A}
\end{figure}

 \noindent Figure \ref{fig:IC220928A} illustrates the light curve rate of NVSS J134240+094752 as detected by Fermi-LAT. The Hop algorithm did not identify any potential flare periods. Additionally, our analysis revealed that within the time window centered on the neutrino position, specifically one week before and after, the statistical significance of the neutrino position was $0.98\sigma$. Also, the 4FLG point source has a statistical significance of $0.47\sigma$. The probability of this event being associated is 0.05. This event does not appear to be associated with them.

\subsection{IC230914A}\label{subsec:IC230914A}

The event IC230914A was detected on 14 September 2023 at 05:21:03 UTC (MJD 60202.22), with an arrival direction of RA, Dec = (163.83\deg$\substack{+2.55 \\ -2.02}$, 31.83\deg$\substack{+2.08 \\ -1.77}$). One gamma ray source has been reported in this region in the 4LAC. The gamma-ray sources are: 4FGL J1102.9+3014, with angular separations of 2.28\deg. These sources are linked to FSRQ. 4FGL J1102.9+3014, positioned at R.A.=165.74\deg and Dec=30.24\deg, is a gamma-ray source associated with the B2 1100+30B \citep{2005ApJ...626...95S}, exhibiting a redshift of z = 0.38 \citep{2012ApJS..203...21A}.\\

\begin{figure}[!ht]
 \includegraphics[width=\columnwidth]{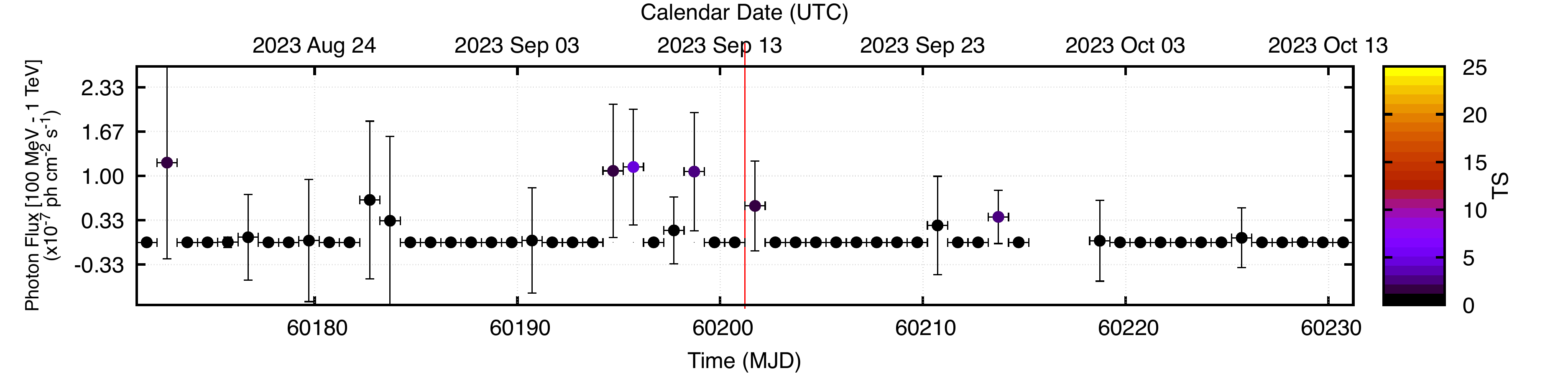}
 \caption{The same as Figure \ref{fig:LC_IC110807A} but for 4FGL J1102.9+3014 and the detection of the neutrino IC230914A.}
 \label{fig:IC230914A}
\end{figure}

\noindent The figure \ref{fig:IC230914A} presents the light curve of 4FGL J1102.9+3014. The Hop method does not identify flare activity that occurs from this source. The figure indicates that at the time of neutrino detection, the light curve does not exhibit any activity signals beyond the baseline recorded before the detection. Within one week surrounding the detection by IceCube, the statistical significance at the neutrino position was found to be $1.20\sigma$ in the energy range of 0.1-100 $\rm GeV$. Additionally, the 4FGL J1102.9+3014 source exhibits a statistical significance of $1.49\sigma$. The values prevent a direct confirmation of 4FGL J1102.9+3014 as a neutrino progenitor. The probability of an event with this angular separation is found to be weak, reaching a value of 0.29, which does not provide a strong argument to believe that these two objects are correlated with statistical arguments.

\section{Model}\label{sec:model}

In a pure leptonic scenario, in which the non thermal emission of the SED of the AGNs is described by a relativistic population of electrons in which the first peak of the spectra is described by Synchrotron radiation of the electrons. Meanwhile, the second peak would be attributed to Synchrotron Self-Compton radiation as a result of the interactions of the electrons and Synchrotron photons. In this scenario we cannot expect neutrino emission, thus a hadronic component is injected. We will assume that a relativistic population of protons are co-accelerated within the electron population. In this paradigm, the protons can interact with low energy photons and produce an hadronic flux of $\gamma$-rays with a neutrino flux, product of neutral and charged pion decay, respectively \citep{2020MNRAS.497.5318F}. \\

For this, we propose that the extended Spectral Energy Distribution of the FRSQ that we propose as a neutrino progenitor can be described as a one zone emission region with a lepto-hadronic contribution. A relativistic population of electrons and protons is injected in the emission region, where the protons interact with low energy photons via photo-pion ($p\gamma$) interactions. Assume a standard spherical blob of radius $R_{b}^{\prime}$ moving with a Lorentz factor $\Gamma = (1-\beta^{2})^{-1/2}$, with $\beta$ the velocity normalized to the speed of light. Thus, the comoving volume of the blob is $V_{b}^{\prime} = \frac{4}{3}\pi R_{b}^{\prime 3}$. Since the blazars are astrophysical objects that show a high variability which can range from time-scales of the order of seconds up to years, this implies that the emitting volume is restricted by the light travel time given by $t_{\rm var}^{\prime} \gtrsim R_{b}^{\prime}/c$. There is an angle $\theta$ between the velocity of the jet of the FSRQ and the line of sight of the observer defined as $\theta = \rm arccos(\mu)$, and the doopler factor is defined as $\delta_{D} = [\Gamma(1-\beta \mu)]^{-1}$. \\

The model assumes that the electrons cool due to synchrontron radiation and synchrontron self-Compton (SSC), and this radiation could explain the extended spectral energy distribution. The protons cool down via photo-hadronic processes, producing a non-thermal of hadronic $\gamma$-rays contribution with a signature of high energy neutrinos. The protons need a low energy radiation field to produce p$\gamma$ interactions, we propose that these radiation should come from the broadline region. \\

Inside the blob, both electrons and protons are accelerated up to ultra relativistic regime, which must produce the observing radiation. 
Electrons are assumed to have a homogeneous and isotropic distribution given by a broken power-law function \citep[e.g., see][]{2017APh....89...14F, 2020MNRAS.497.5318F}
\begin{equation}\label{eq_eDistritbution}
N_e'(\gamma_e') = K_e'
\begin{cases}
{\gamma'_e}^{-\alpha_{e,1}}, 
\qquad \qquad \qquad \quad \;\; {\gamma'_{e, \rm min}} \leq {\gamma'_e} \leq {\gamma'_{e, \rm br}} 
\\
{\gamma'_{e, \rm br}}^{\alpha_{e,2}-\alpha_{e,1}} {\gamma'_e}^{-\alpha_{e,2}}, 
\qquad {\gamma'_{e, \rm br}} \leq {\gamma'_e} \leq {\gamma'_{e, \rm max}}\,,
\end{cases}
\end{equation}
where $\gamma'_{\rm e, min}$, $\gamma'_{\rm e, br}$ , $\gamma'_{\rm e, max}$ are the minimum, break and maximum Lorentz factor of ultrarelativisc electrons, respectively, and $K_e'$ is the normalization constant. On the other hand, we assume protons are distributed homogeneously and isotropically during the equilibrium stage, given by 
 
\begin{equation}\label{eq_pDistribution}
{N_p'}({\varepsilon_p'}) = {K_p'} \left( \frac{\varepsilon'_p}{m_p c^2} \right)^{-\alpha_{p}} \, 
\qquad \varepsilon'_{p, \rm min} \leq \varepsilon_p' \leq \varepsilon'_{p, \rm max}\,,
\end{equation}

where $K_p'$ is the normalization constant, $\alpha_p$ is the proton spectral index and ${\varepsilon}'_{p, \rm min}$ and ${\varepsilon}'_{p, \rm max}$ correspond to the minimum and maximum energy in the comoving frame, respectively.

\subsection{Synchrotron radiation}
In the presence of a magnetic field ($B$), relativistic charged particles move along them while losing their energy via synchrotron radiation. The power radiated by an electron distribution is \citep{RevModPhys.42.237, 2008ApJ...686..181F, 2004ApJ...616..136S, 2023Galax..11..117A}

\begin{equation}
 J_{\rm s}' (\varepsilon'_s) = \frac{\sqrt{3} e^3 B'}{ 2 \pi \hbar m_e c^2} 
 \int_{ \gamma'_{e,{\rm min}} }^ { \gamma'_{e,{\rm max}} } d\gamma_e' N'_e (\gamma_e') \, R_{\rm syn}(x) \, ,
\end{equation}\label{eq_syn_emission}

where $\hbar$ is the reduced Planck constant $e$ is the electron charge, $x = \varepsilon'_s/\varepsilon'_{\rm ch}$, the characteristic energy is given by $\varepsilon'_{\rm ch}= \frac{3 e B' \hbar }{2 m_e c} {\gamma'_e}^ 2 \sin{\theta}$ and $\sin{\theta}$ is the pitch angle, the function $R_{\rm syn}$ is given by \citet{2008ApJ...686..181F}.

\subsection{Inverse-Compton scattering}

An electron moving inside a radiation field produces Compton scattering. As well, synchrotron electron loss energy for each collision produces high-energy photons. The total emissivity coefficient produced by an isotropic electron population is \citep {RevModPhys.42.237}

\begin{equation}
J'_{ic} (\varepsilon'_c) = \frac{3}{4} c \sigma_{\rm T} \varepsilon'_c \int_{\frac{\varepsilon'_c}{m_e c^2}} d\gamma'_e \frac{N'_e(\gamma''_e)}{{\gamma'_e}^2} \int d\varepsilon' \frac{n'_{\rm ph}(\varepsilon')}{\varepsilon'} F_c(q,\Gamma_e) \, .
\end{equation}

Finally, the inverse Compton radiation is calculated using the above equation but a different photon spectrum for SSC and EIC emission. The total seed photon distribution is 

\begin{equation}
 n'_{\rm ph}(\varepsilon') \simeq \frac{J'_{s}(\varepsilon')}{4\pi {R'}^{2} c \, \varepsilon'} + \delta_D \left( n_{\rm BLR}(\varepsilon') + n_{\rm DT}(\varepsilon') \right)
\end{equation}

\subsection{Photopion process}

The photopion process results in the production of nonstable secondary products, i.e., $\pi^0$ and $\pi^\pm$ mesons \citep{2008PhRvD..78c4013K}:

\begin{equation}
 p+\gamma \rightarrow n_0 \pi^0 + n_+ \pi^+ + n_- \pi^- + ...
\end{equation}

where $n_0, n_-, n_+$ are the multiplicities of neutral, negative, and positive charged pions, respectively. Pions decay into final stable particles as

\begin{align}
 \pi^0 &\rightarrow \gamma\gamma
 \\
 \pi^+ &\rightarrow e^+ + \nu_{e} + \bar{\nu}_{\mu}+\nu_{\mu}
 \\
 \pi^- &\rightarrow e^- + \bar{\nu}_{e} +\nu_{\mu} + \bar{\nu}_{\mu}\,.
\end{align}

The proton energy threshold to photopion production is given by the condition to produce a single rest mass pion; by considering the seed's photons in the blob frame we obtain

\begin{equation}\label{eq_pion_th_gamma}
{\varepsilon'}_{p,\rm th}^{p\pi} = \frac{ m_\pi^2 + 2 m_\pi m_p }{4 \varepsilon'} \approx 70 \, {\rm PeV} \, \left( \frac{ {\varepsilon'} }{\rm eV} \right)^{-1} \, .
\end{equation}

The production rate of stable particles is given by \citet{2008PhRvD..78c4013K}

\begin{align}\label{eq_photopion_flux}
{Q'}^{p\pi}_{i}(\varepsilon'_{i}) 
&= \int \frac{d\varepsilon'_p}{\varepsilon'_p} N'_p(\varepsilon'_p) \,\int d\varepsilon' \, n'_{\rm ph}(\varepsilon') \, \Phi_{i} (\eta,x) \, ,
\end{align}

where the label ${\rm i}$ represents the particles $\gamma,\nu, \Bar{\nu},e^-,e^+$, the function $\Phi_{i}$ is parametrized by the authors, and they define the parameters $\eta=\frac{4\varepsilon'_p\varepsilon'}{m_p^2 c^4} > 0.303 $ and $x=\frac{\varepsilon'_i}{\varepsilon'_p}$.\\

\noindent Finally, the luminosity of the jet can be obtained as the contribution of the electron, proton, and magnetic field luminosities \citep{2008MNRAS.385..283C, 2023JHEAp..38....1A, 2023EPJC...83..338A};

\begin{equation}
 L_{j} = \sum_{i=e,p,B} L_{i} ,
\end{equation}

where $L_{i} \simeq \pi r_{d}^{2} \Gamma^{2} U_{i}$ where, $r_{d}$ is the emission region size defined as $r_{d} = \delta_{D}\tau_{\rm{v, min}}/(1+z)$ with $t_{\rm{var, min}}$ as the minimum time of variability. As we are dealing with blazars, is assume that $\delta_{D} \approx \Gamma$. So, $U_{e} = m_{e} N_{e} \langle\gamma_{e}\rangle \ = m_{e} \int_{\gamma_{\rm{min}}}^{\gamma_{\rm{max}}} \frac{dn_{e}}{d\gamma_e} d\gamma_e$, $U_{p} = N_{p}m_{p}$ and $U_{B} = B'^{2}/(8\pi)$ are the electron, proton and magnetic field densities, respectively.

\subsection{Modeling the Spectral Energy Distributions}\label{modeling}
The fitting method chosen is the Monte Carlo Markov Chain (MCMC) technique implemented in the \texttt{emcee} package \citep{2013PASP..125..306F} for Python. In order to obtain a suitable set of priori parameters for each source listed in Table \ref{table:Correlations} a previous fit was done with the \texttt{LMFIT} \citep{newville_matthew_2014_11813} python package. Once the prior parameters were obtained, an MCMC sampler was built using 512 walkers with a length of 12,000 steps. The parameters involved in the MCMC process were the parameters for the leptonic model, such as the magnetic field (B), the Doppler factor ($\delta_{D}$), the minimum, break, and maximum Lorentz factor from the electron population ($\rm{\gamma^{\prime}_{min}}$, $\rm{\gamma^{\prime}_{break}}$ and $\rm{\gamma^{\prime}_{max}}$ respectively) with the shape of a broken power law population with slope $p$ and normalization constant $\rm{K'_{e}}$. For the hadronic component, the power law index of the proton distribution was fixed to 2, and the target photons taken into account came from the Narrow and Broad emission lines on the seed photon distribution, allowing the constant normalization to be free $\rm{K_{p}}$.\\

\section{Results and Conclusion}\label{sec:discussion}

Our study analyzed data spanning 12 years from IceCube, along with 14 years of gamma-ray emissions detected by Fermi-LAT, as documented in 4FGL-DR4 \citep{2020ApJS..247...33A, 2020arXiv200511208B}. We focused on $\gamma$-ray sources with angular separations of less than two degrees ($2^{\circ}$) from the best-fit neutrino positions. This analysis revealed 20 spatial coincidences between the FSRQ $\gamma$-ray sources listed in 4LAC-DR3 and the neutrinos observed by IceCube. Additionally, motivated by the neutrino flare associated with the IceCube-170922A event, which was observed from TXS 0506+056 \citep{2018Sci...361..147I, 2018Sci...361.1378I}, we examined the $\gamma$-ray observations for each quasar within a time frame of one week ($\pm$ one week) following the trigger event.

\begin{figure*}[!ht]
\centering{
\subfloat[c][\centering{PKS B2224+006}]{\includegraphics[scale=0.26]{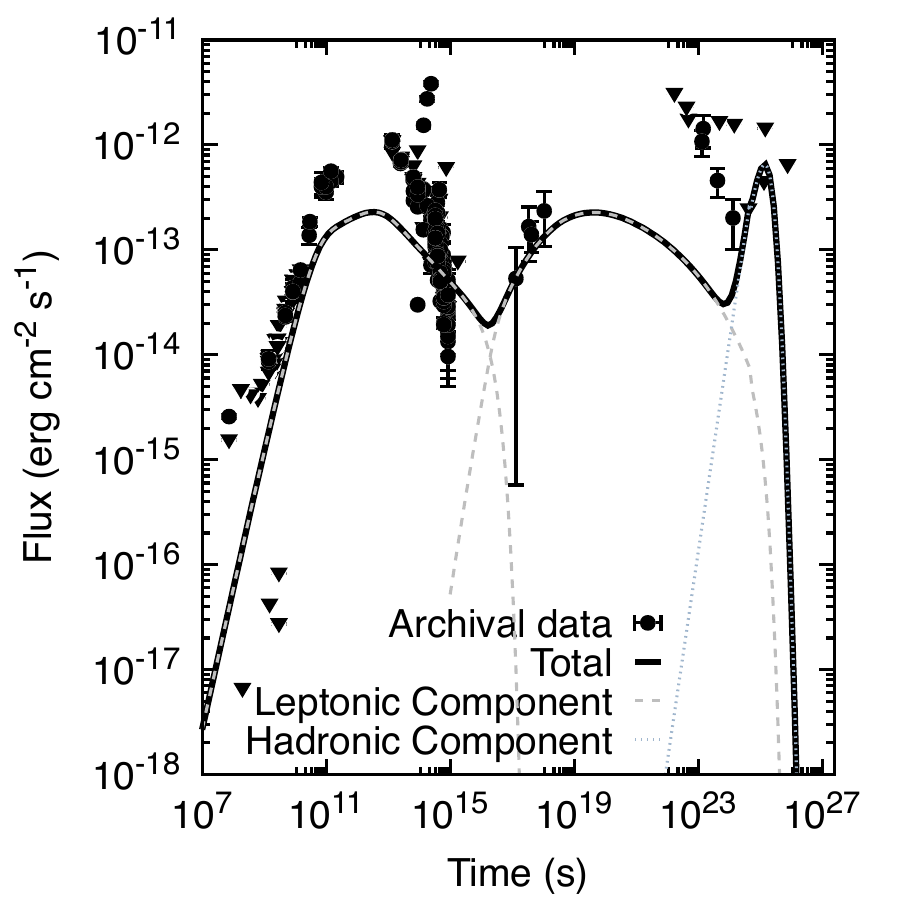} \label{fig:SED_PKSB2224+006}}
\subfloat[c][\centering{PKS 1741-03}]{\includegraphics[scale=0.26]{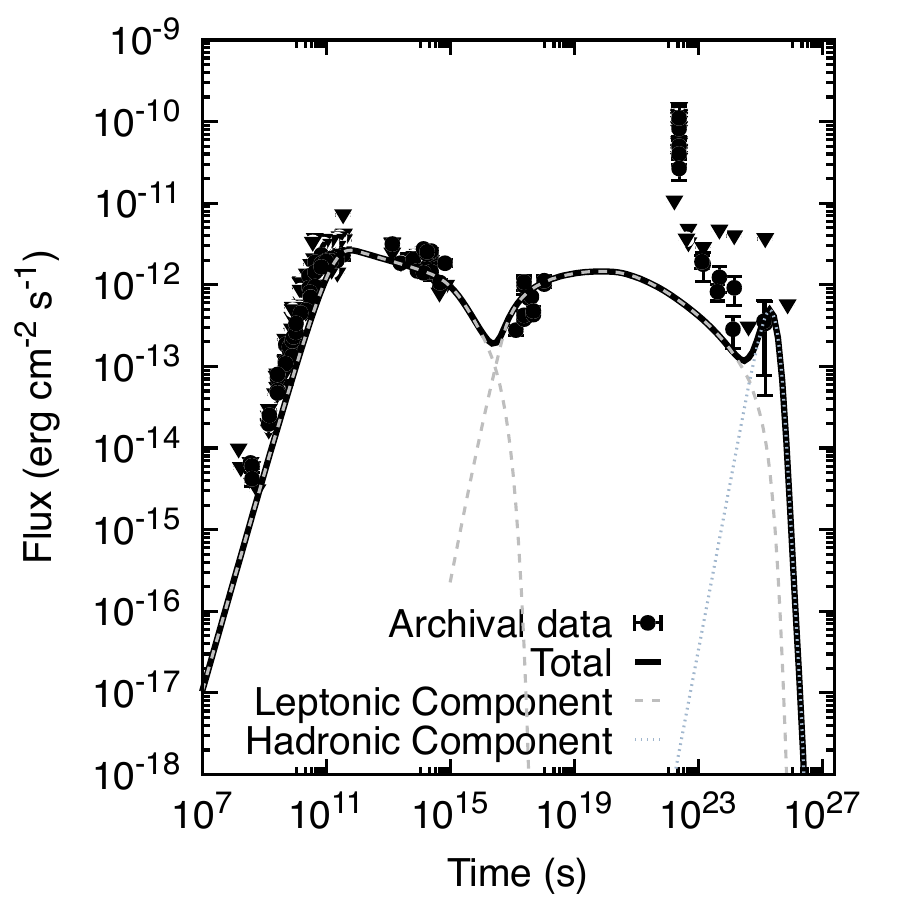}\label{fig:SED_PKS1741-03}}
\subfloat[c][\centering{OP 313}]{\includegraphics[scale=0.26]{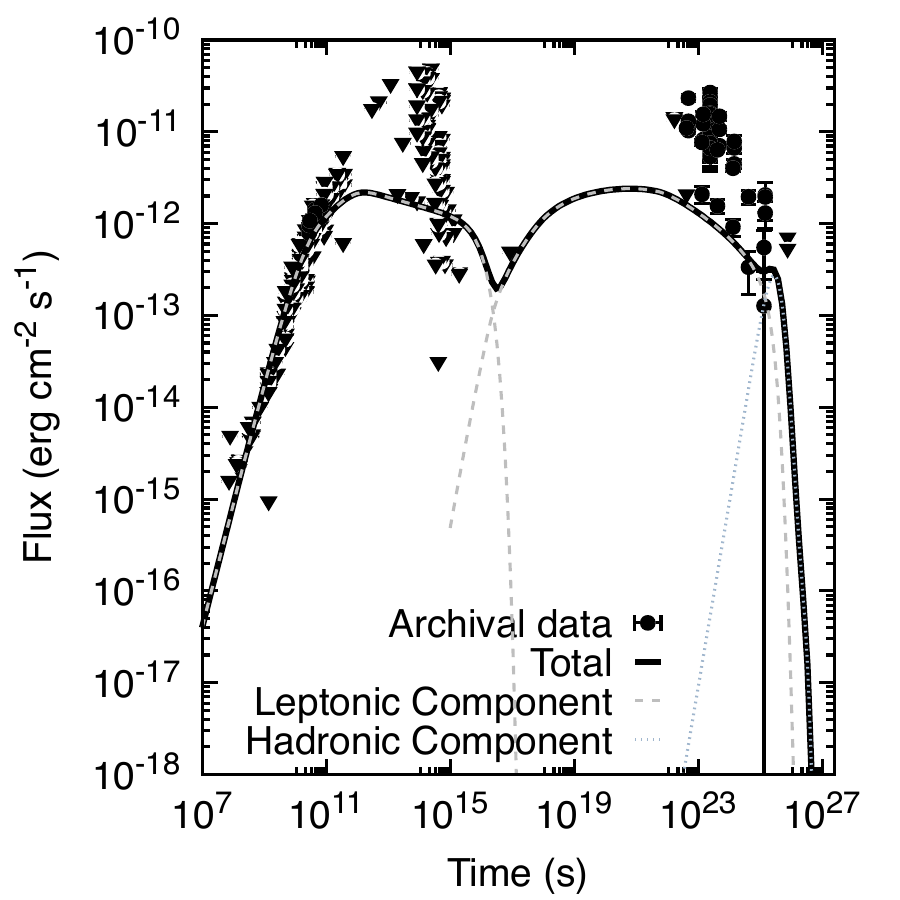} \label{fig:SED_OP 313}}
\subfloat[c][\centering{RX J131058.8+323335}]{\includegraphics[scale=0.26]{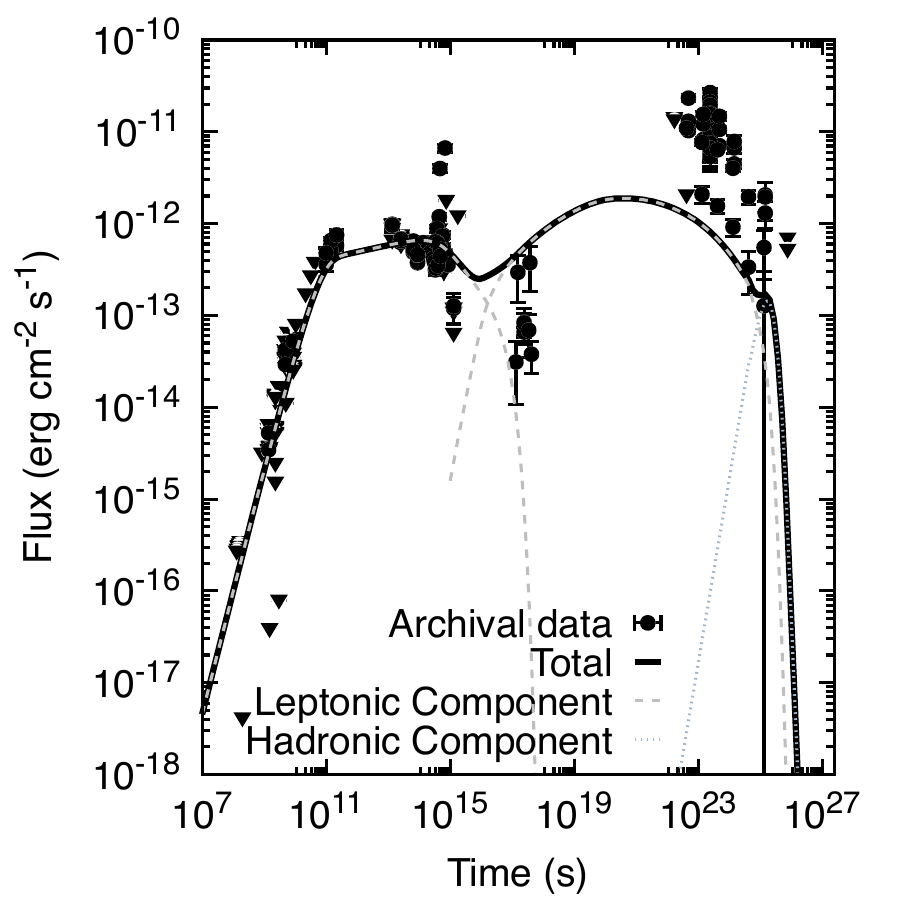} \label{fig:SED_RXJ1310588323335}}

\subfloat[a][\centering{MG1 J120448+0408}]{\includegraphics[scale=0.26]{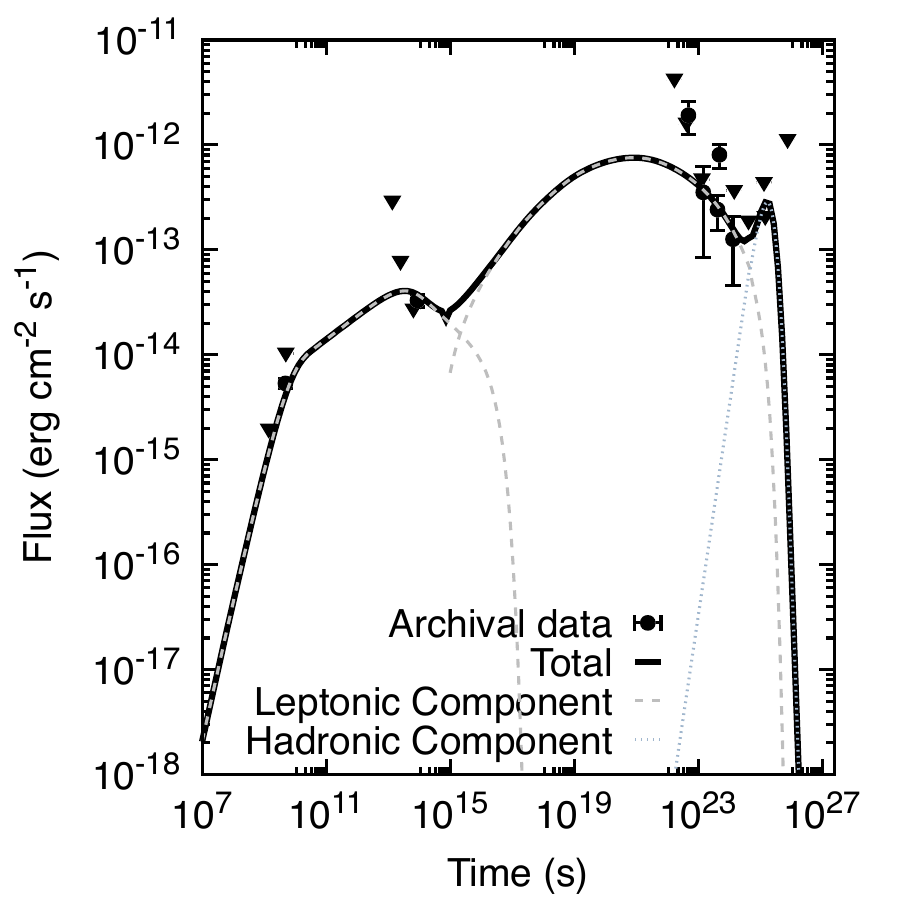} \label{fig:SED_MG1 J1204480408}}
\subfloat[a][\centering{PKS B2330-017}]{\includegraphics[scale=0.26]{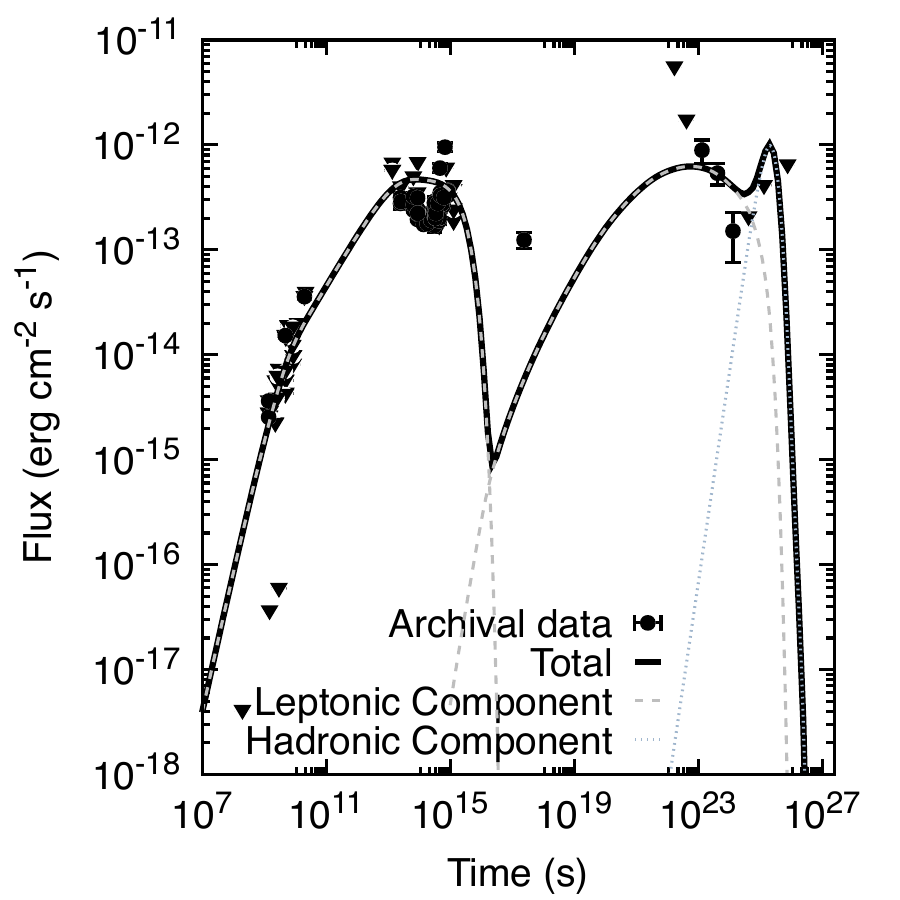} \label{fig:SED_PKSB2330017}}
\subfloat[a][\centering{PKS 2332-017}]{\includegraphics[scale=0.26]{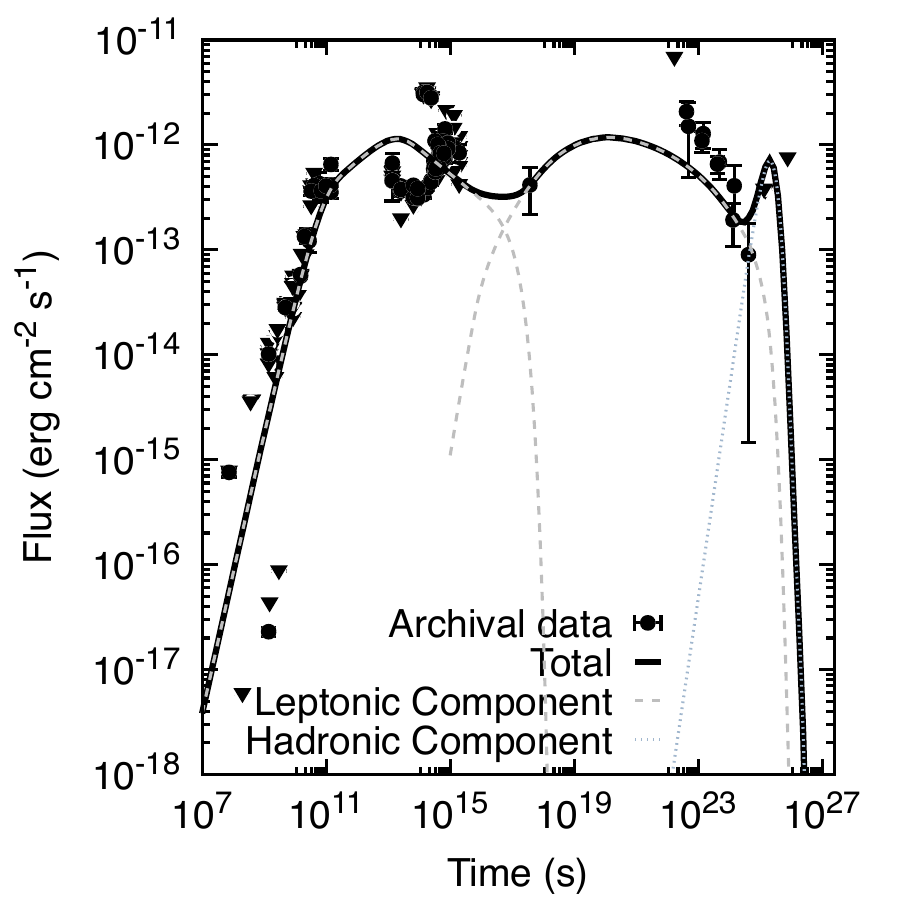} \label{fig:SED_PKS2330017}}
\subfloat[c][\centering{PKS B2224+006}]{\includegraphics[scale=0.26]{SEDV2/PKSB2224006_ModelSED_NLR.pdf} \label{fig:SED_PKSB2224+006b}}

\subfloat[b][\centering{PKS 0420+022}]{\includegraphics[scale=0.26]{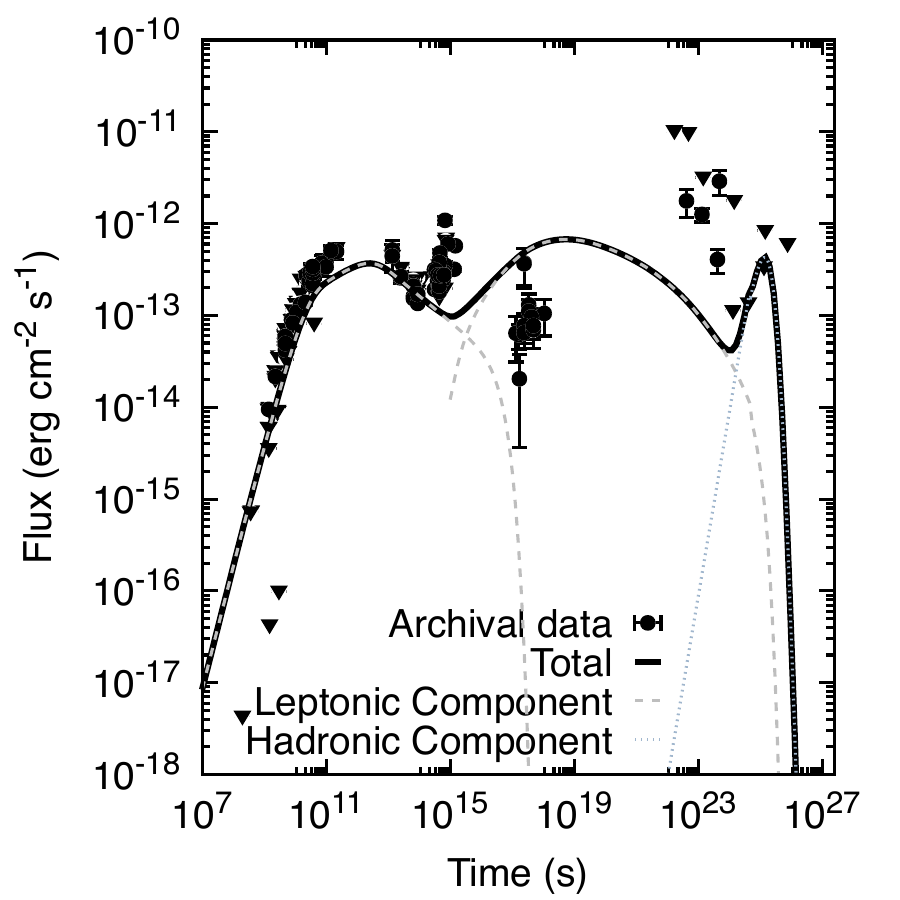}\label{fig:SED_PKS_0420022}}
\subfloat[c][\centering{PMN J2118+0013}]{\includegraphics[scale=0.26]{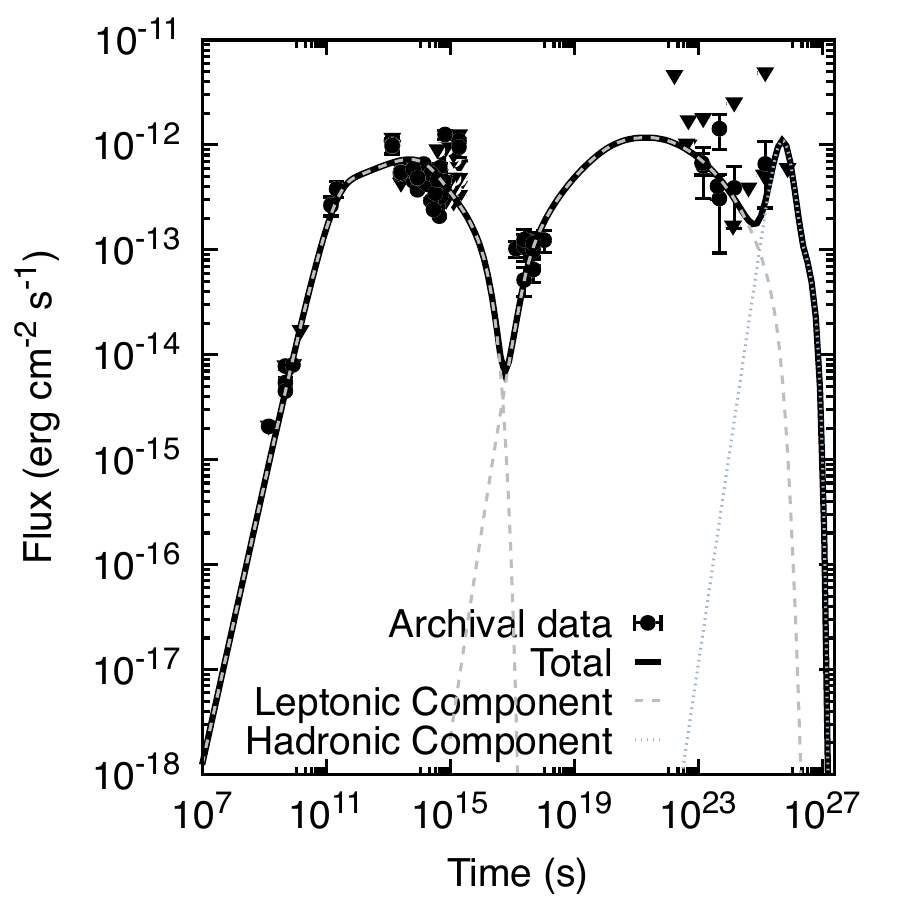} \label{fig:SED_PMNJ21180013}}
\subfloat[c][\centering{B2 1811+29}]{\includegraphics[scale=0.26]{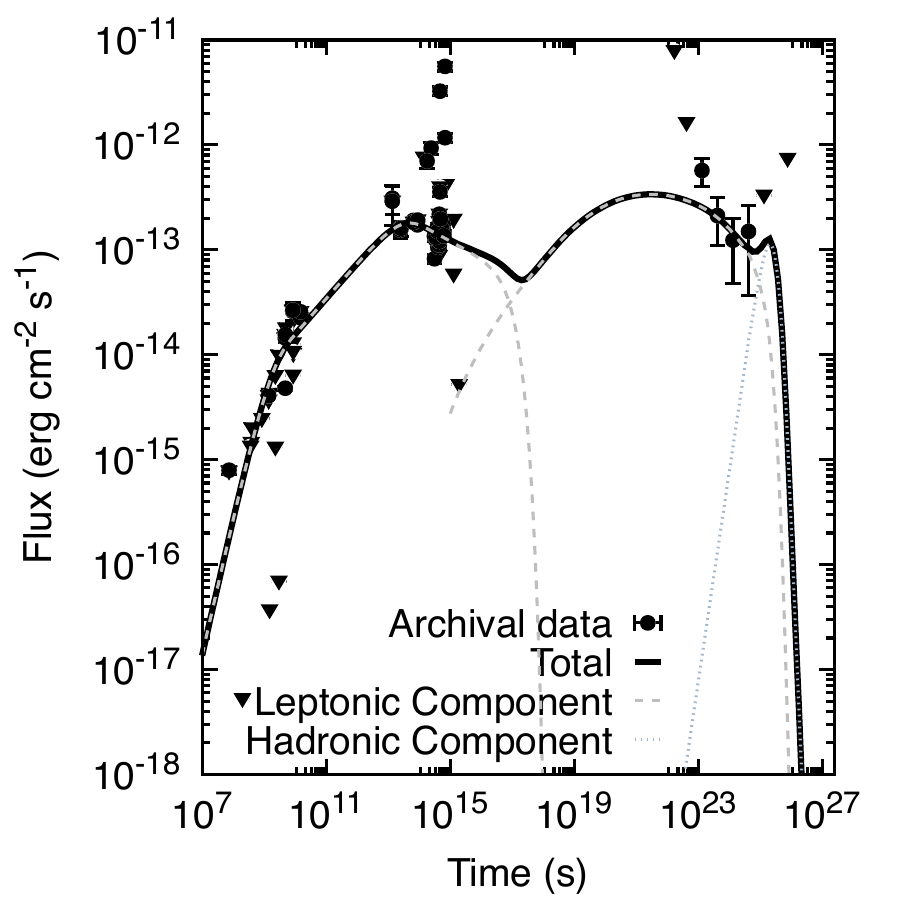} \label{fig:SED_B2181129}}
\subfloat[][\centering{B2 0849+28}]{\includegraphics[scale=0.26]{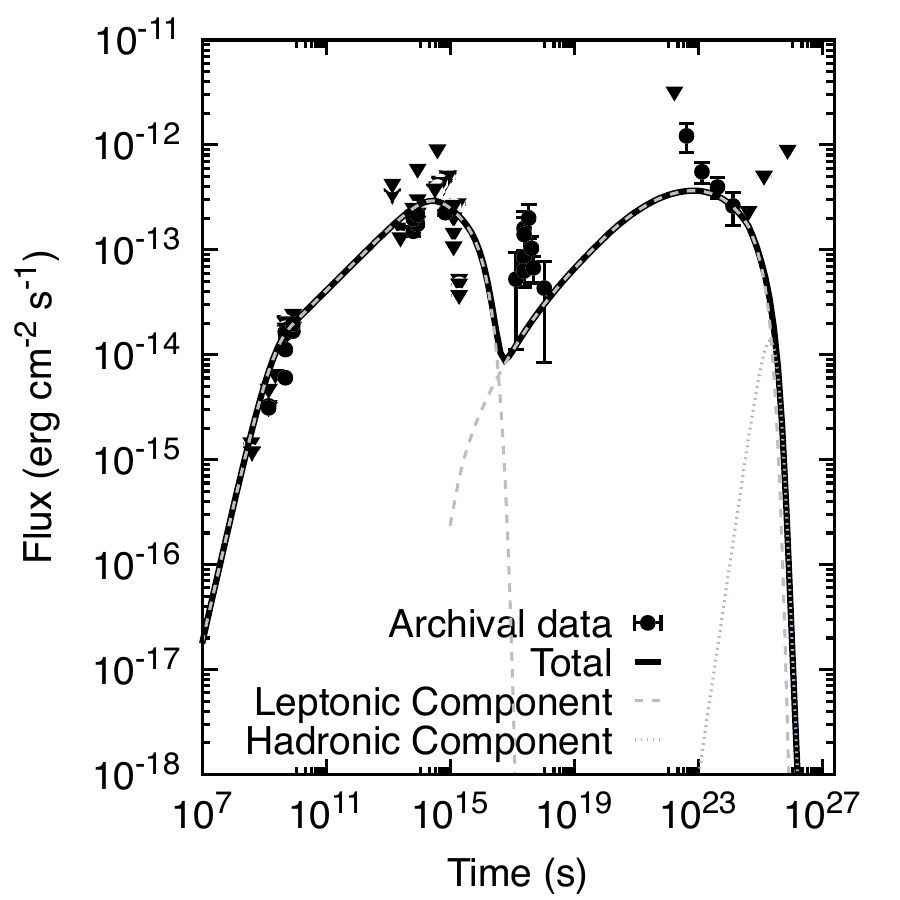} \label{fig:SED_B2084928}}

\subfloat[][\centering{4C +31.51}]{\includegraphics[scale=0.26]{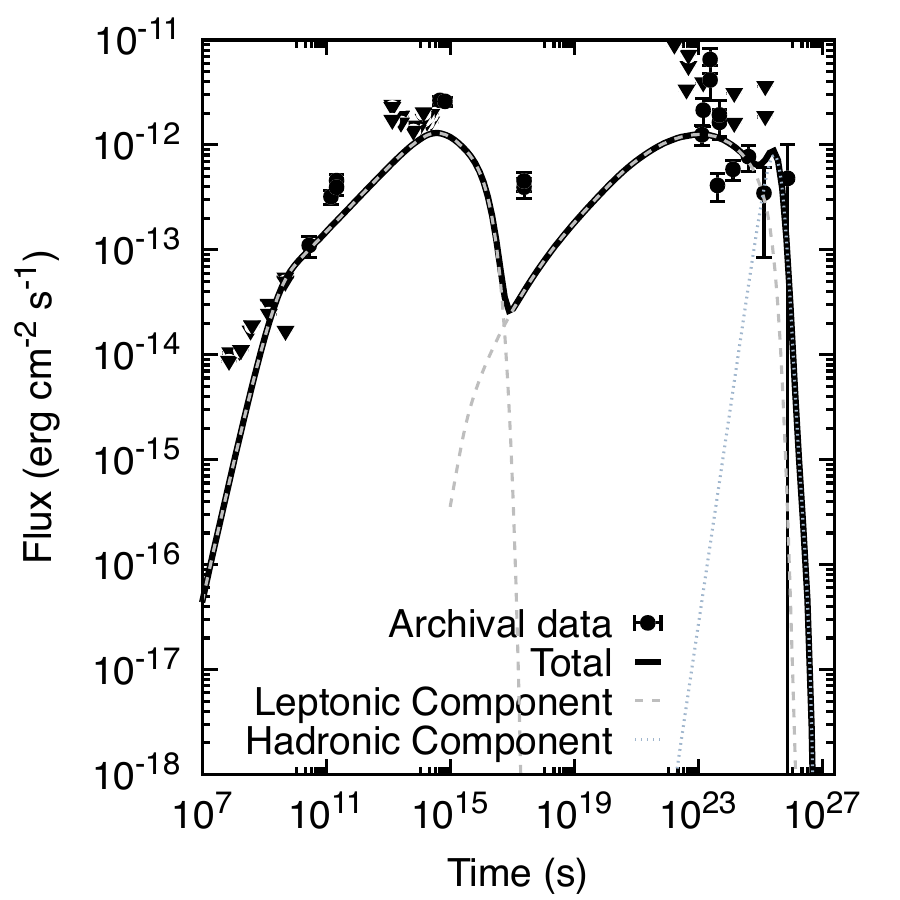} \label{fig:SED_4C31_5}}
\subfloat[a][\centering{TXS 1015+057}]{\includegraphics[scale=0.26]{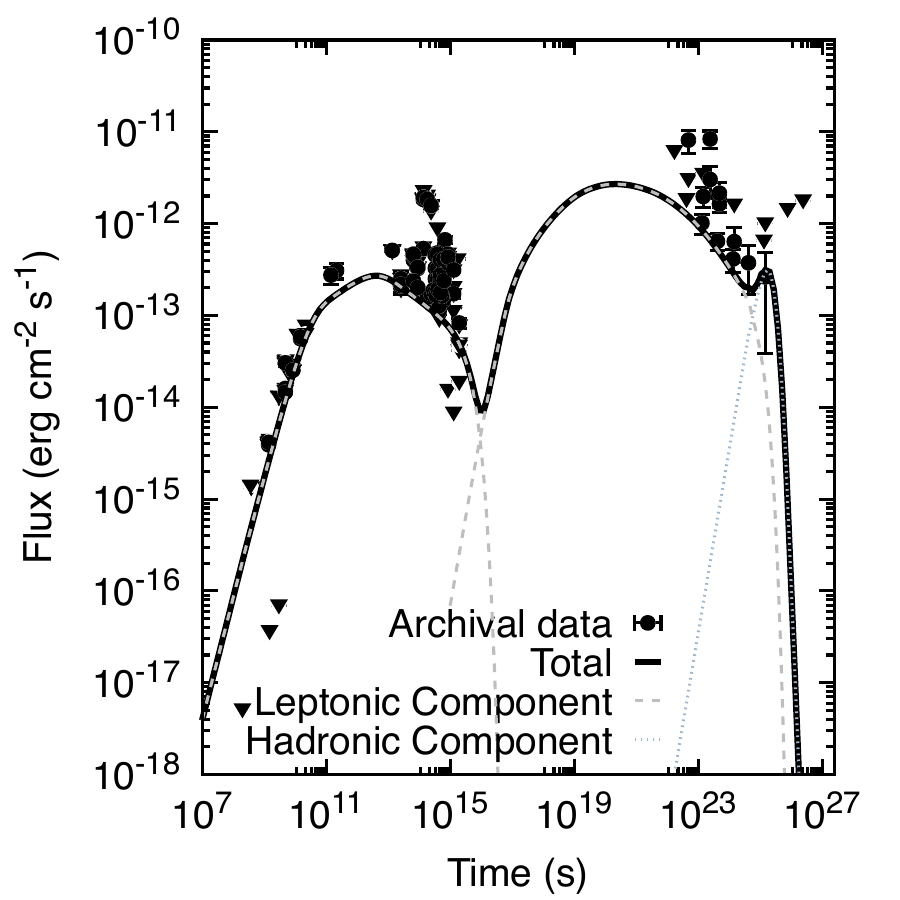} \label{fig:SED_TXS1015057}}
\subfloat[a][\centering{PKS 2058-297}]{\includegraphics[scale=0.26]{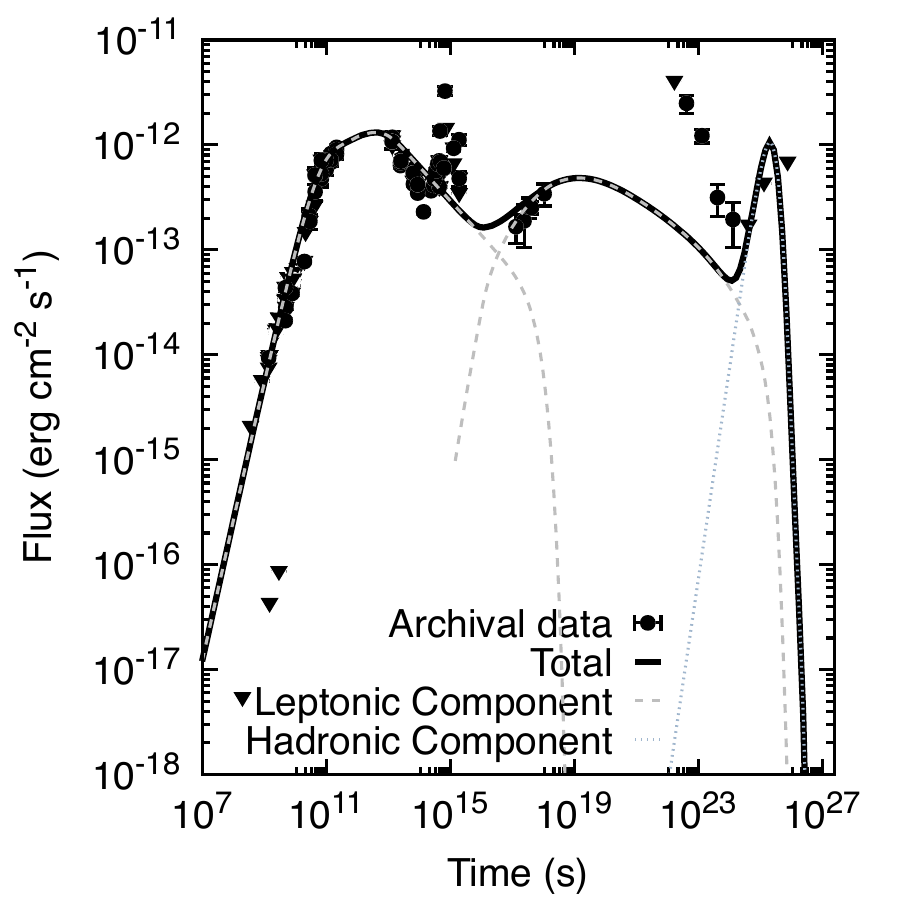} \label{fig:SED_PKS_2058297}}
\subfloat[b][\centering{PMN J0206-1150}]{\includegraphics[scale=0.26]{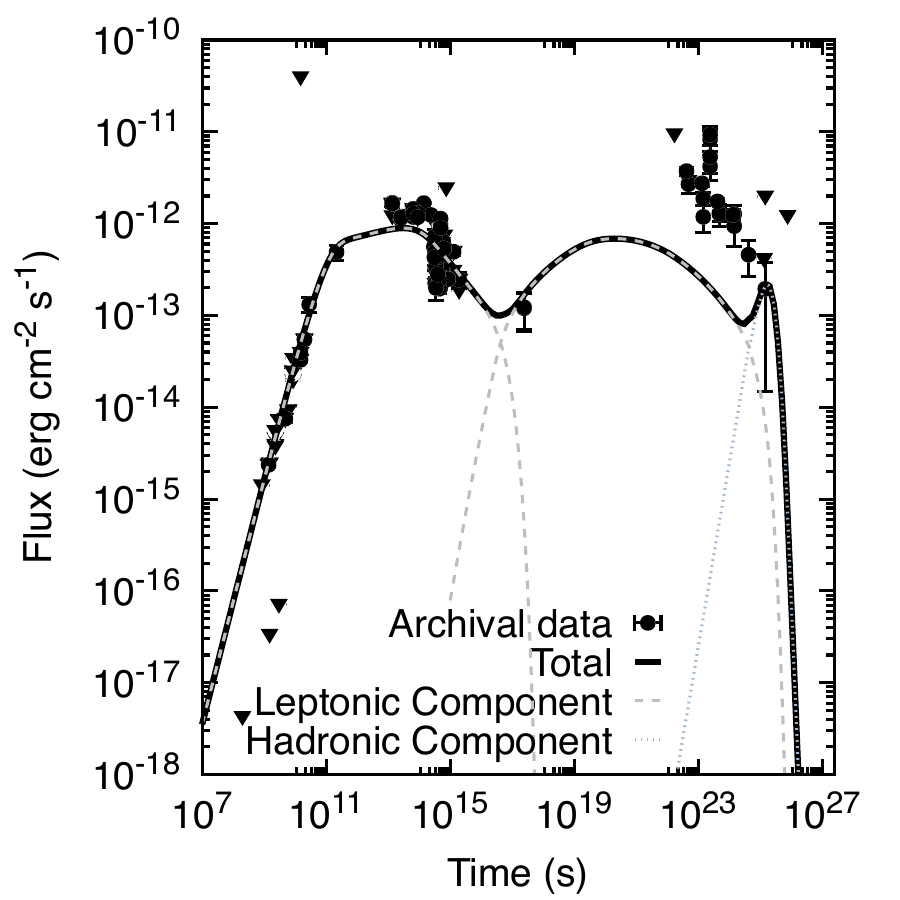} \label{fig:SED_PMN J02061150}}

\subfloat[a][\centering{OX 110}]{\includegraphics[scale=0.26]{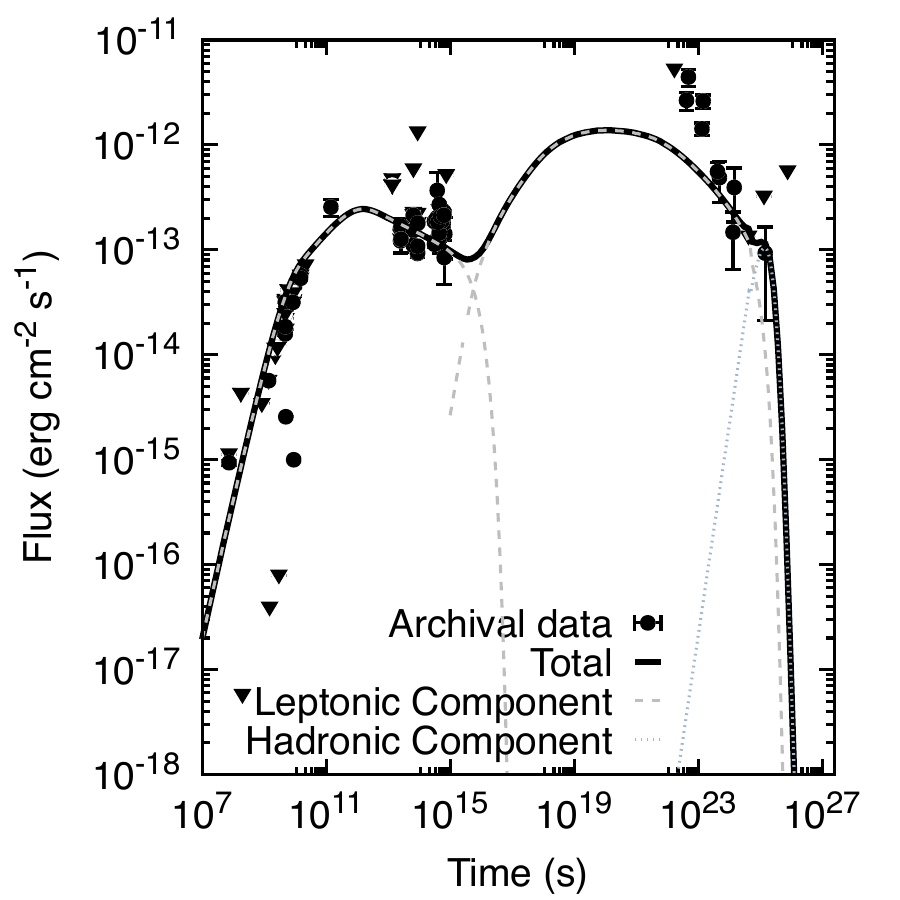}\label{fig:SED_OX110}}
\subfloat[b][\centering{TXS 2210+065}]{\includegraphics[scale=0.26]{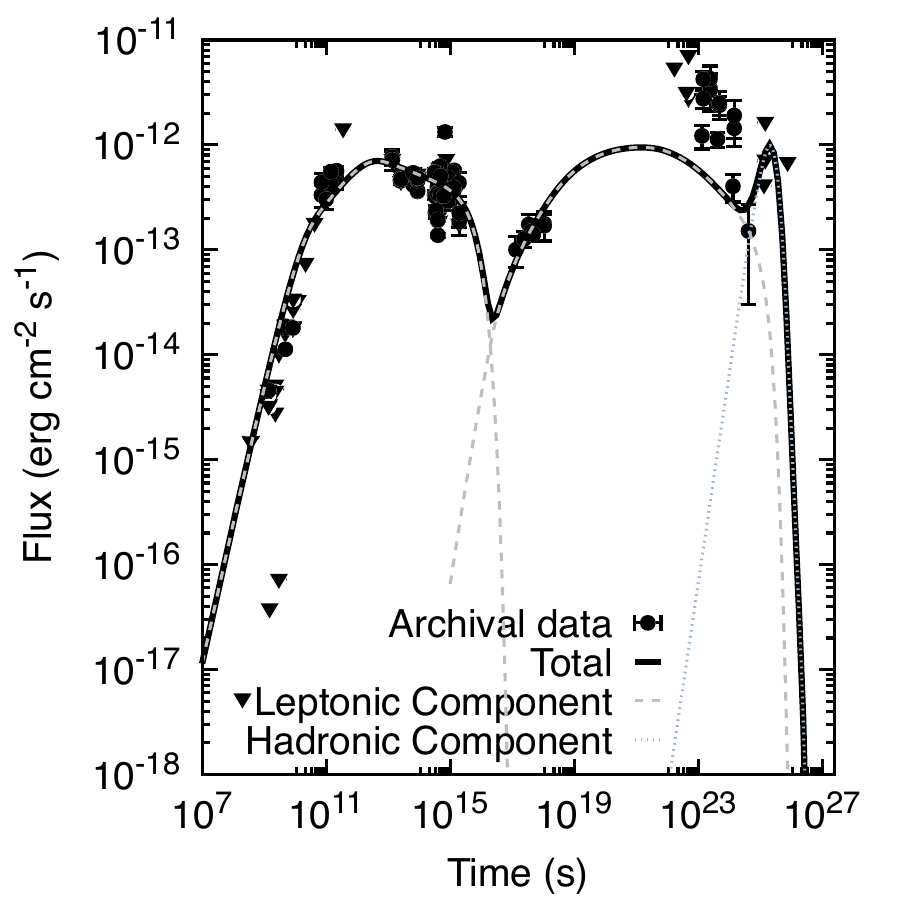} \label{fig:SED_TXS2210065}}
\subfloat[b][\centering{NVSS J134240+094752}]{\includegraphics[scale=0.26]{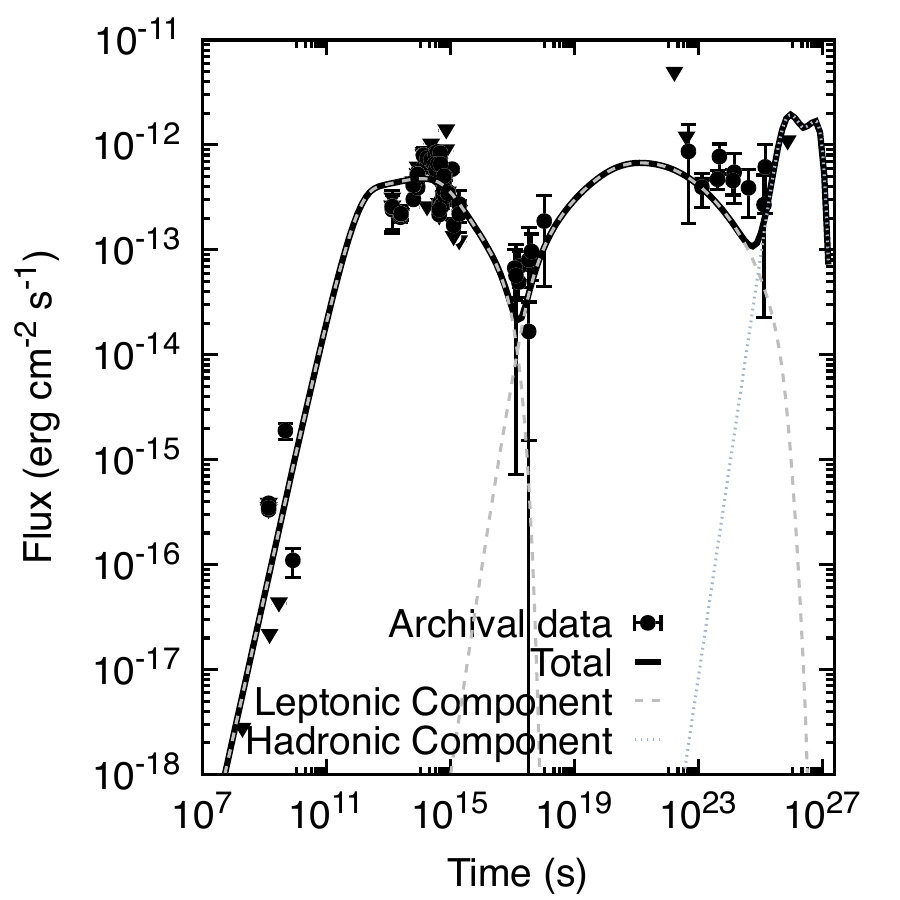} \label{fig:SED_NVSS_J134240094752}}
\subfloat[][\centering{B2 1100+30B}]{\includegraphics[scale=0.26]{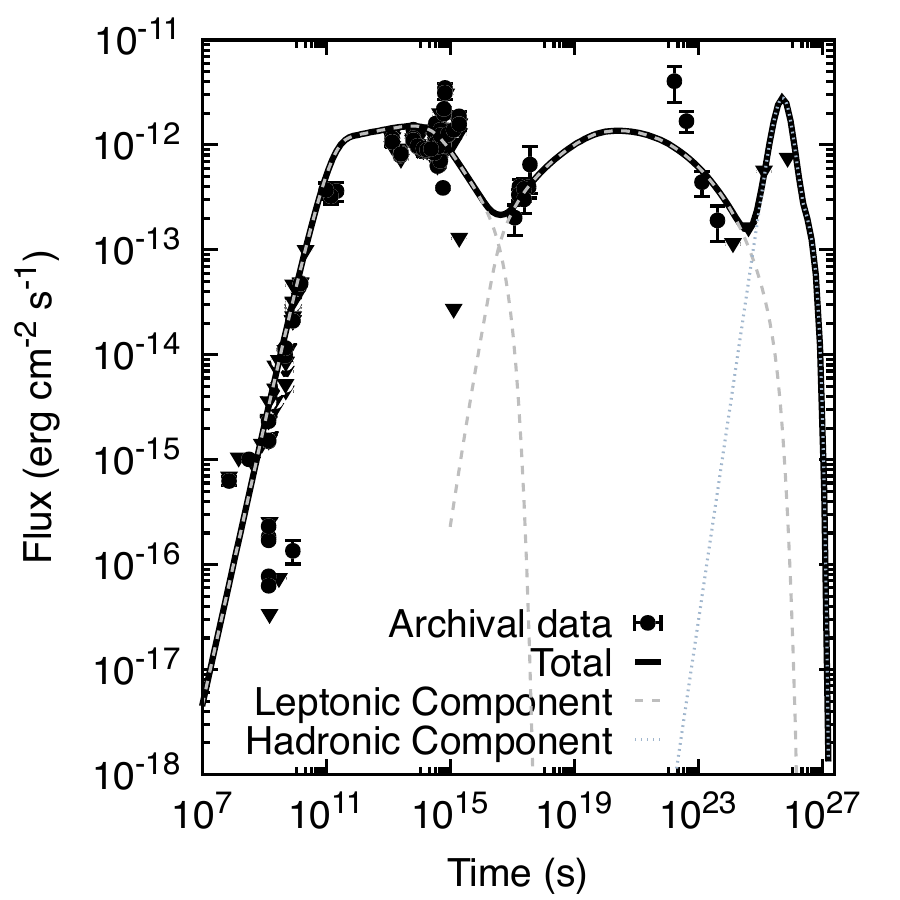} \label{fig:SED_B2110030B}}
}
\caption{Spectral Energy Distribution of the $\gamma$-ray sources detected by Fermi-LAT associated with Quasars within the sky region enclosed within the High Energy neutrino arrival direction. In this particular case, we have assumed that the target photons to produce hadronic reactions comes from the narrow line region.}\label{fig:SED1}
\end{figure*}
\begin{figure*}[!ht]
\centering{
\subfloat[c][\centering{PKS B2224+006}]{\includegraphics[scale=0.26]{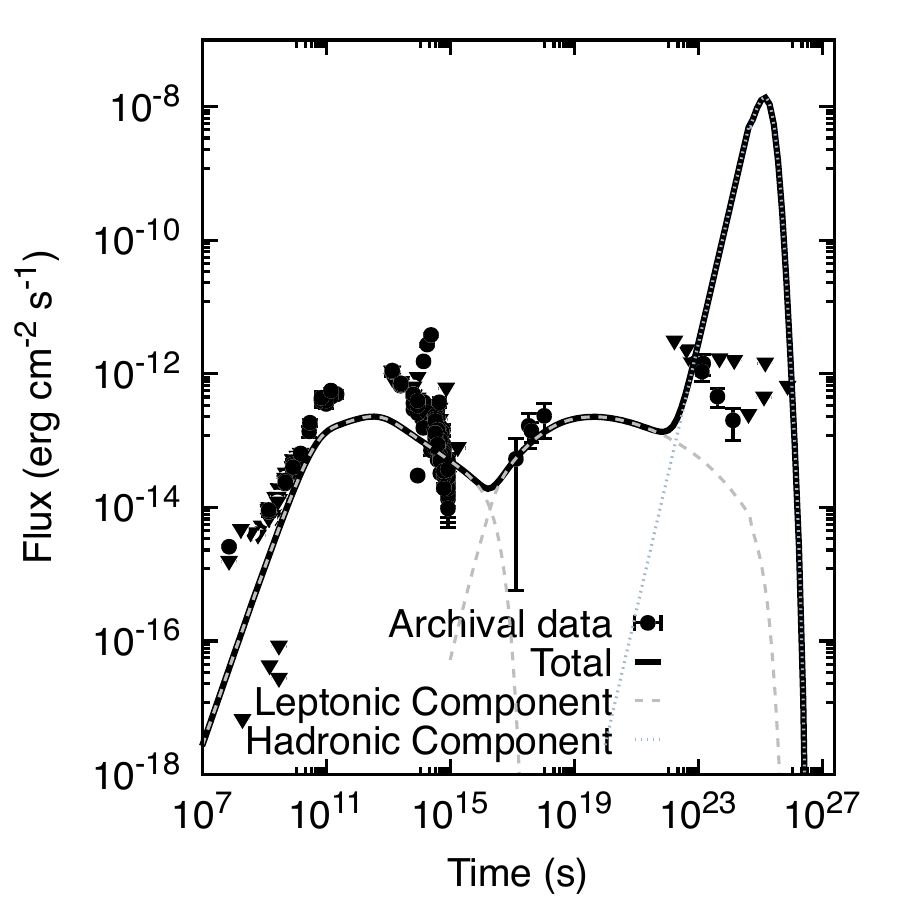} \label{fig:SED_PKSB2224+006_BLR}}
\subfloat[c][\centering{PKS 1741-03}]{\includegraphics[scale=0.26]{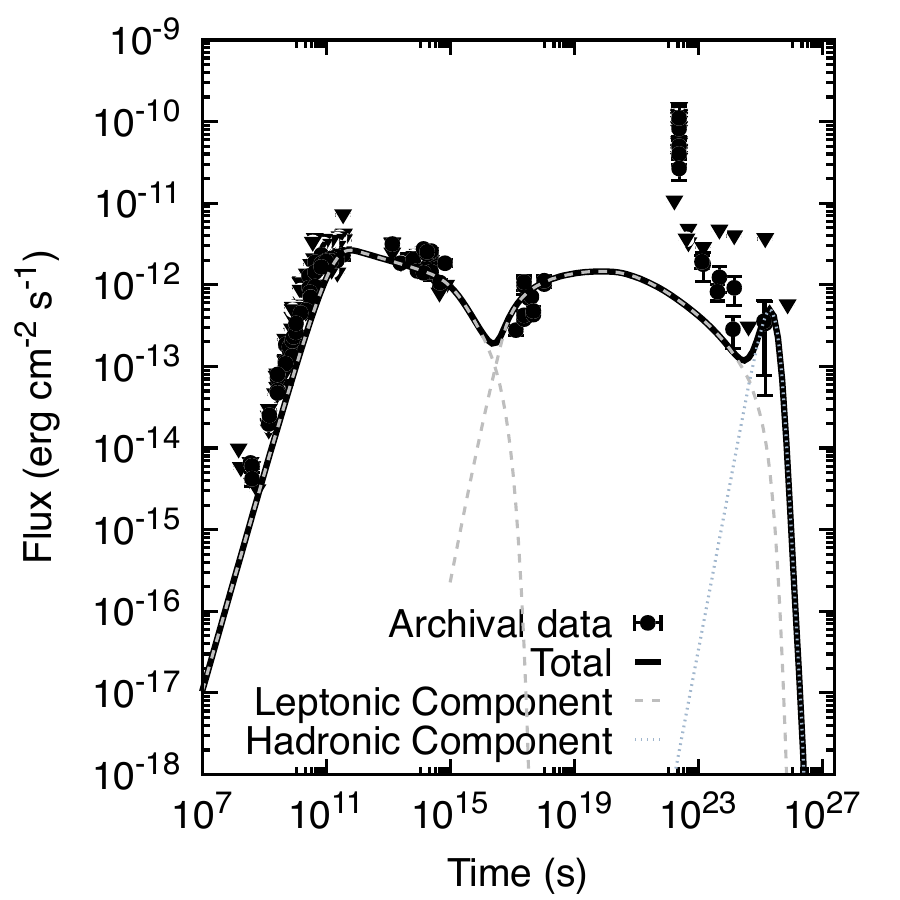}\label{fig:SED_PKS1741-03_BLR}}
\subfloat[c][\centering{OP 313}]{\includegraphics[scale=0.26]{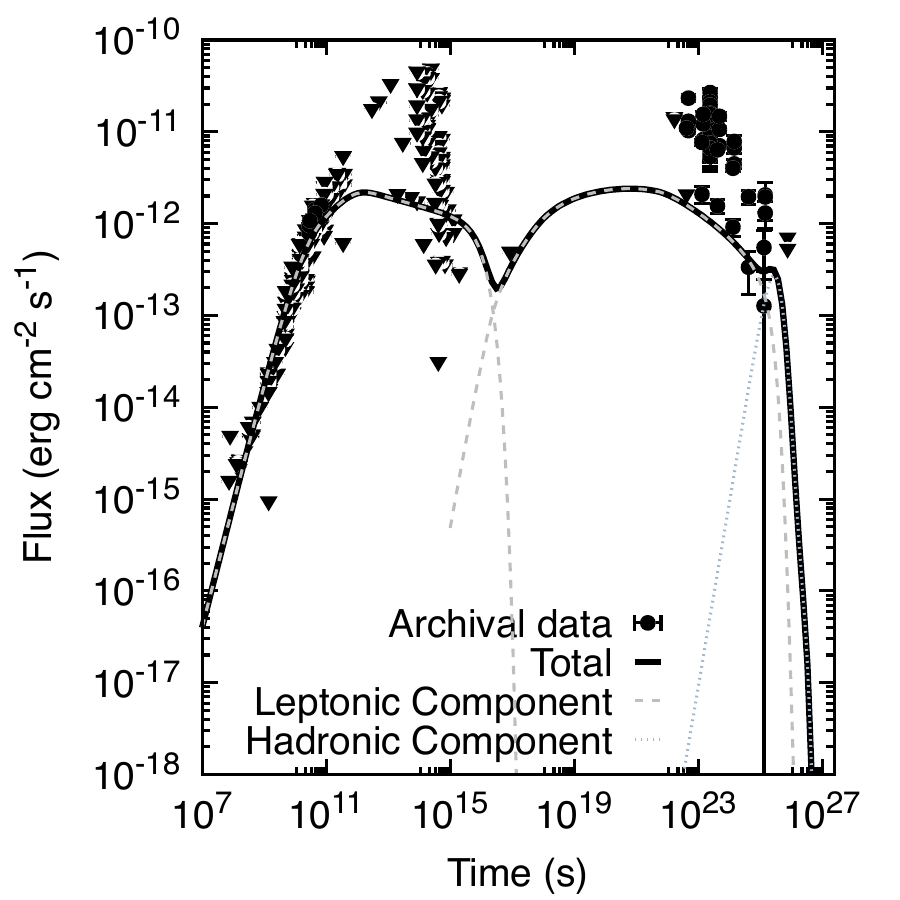} \label{fig:SED_OP 313_BLR}}
\subfloat[c][\centering{RX J131058.8+323335}]{\includegraphics[scale=0.26]{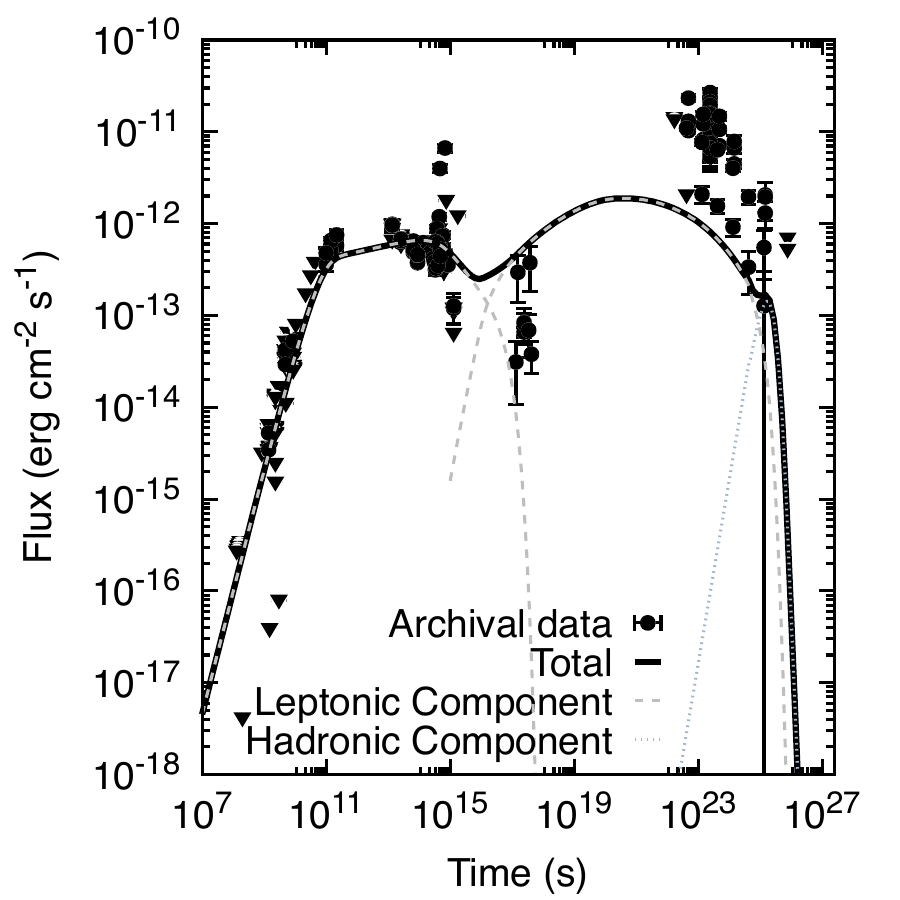} \label{fig:SED_RXJ1310588323335_BLR}}

\subfloat[a][\centering{MG1 J120448+0408}]{\includegraphics[scale=0.26]{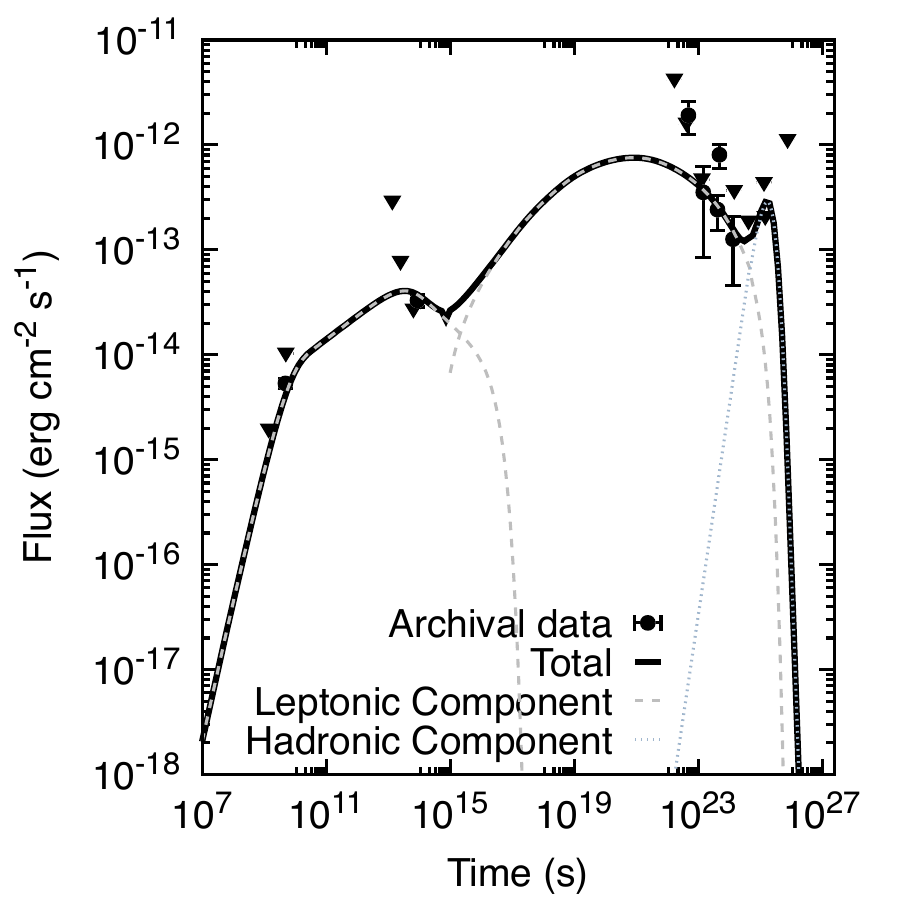} \label{fig:SED_MG1 J1204480408_BLR}}
\subfloat[a][\centering{PKS B2330-017}]{\includegraphics[scale=0.26]{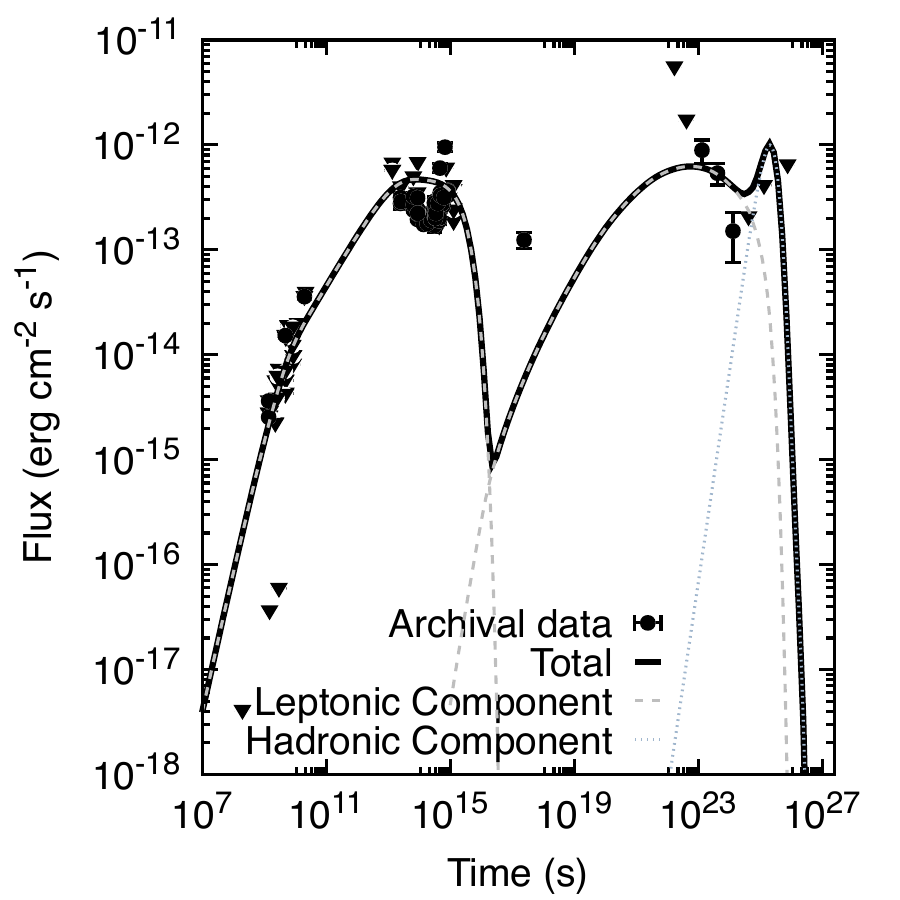} \label{fig:SED_PKSB2330017_BLR}}
\subfloat[a][\centering{PKS 2332-017}]{\includegraphics[scale=0.26]{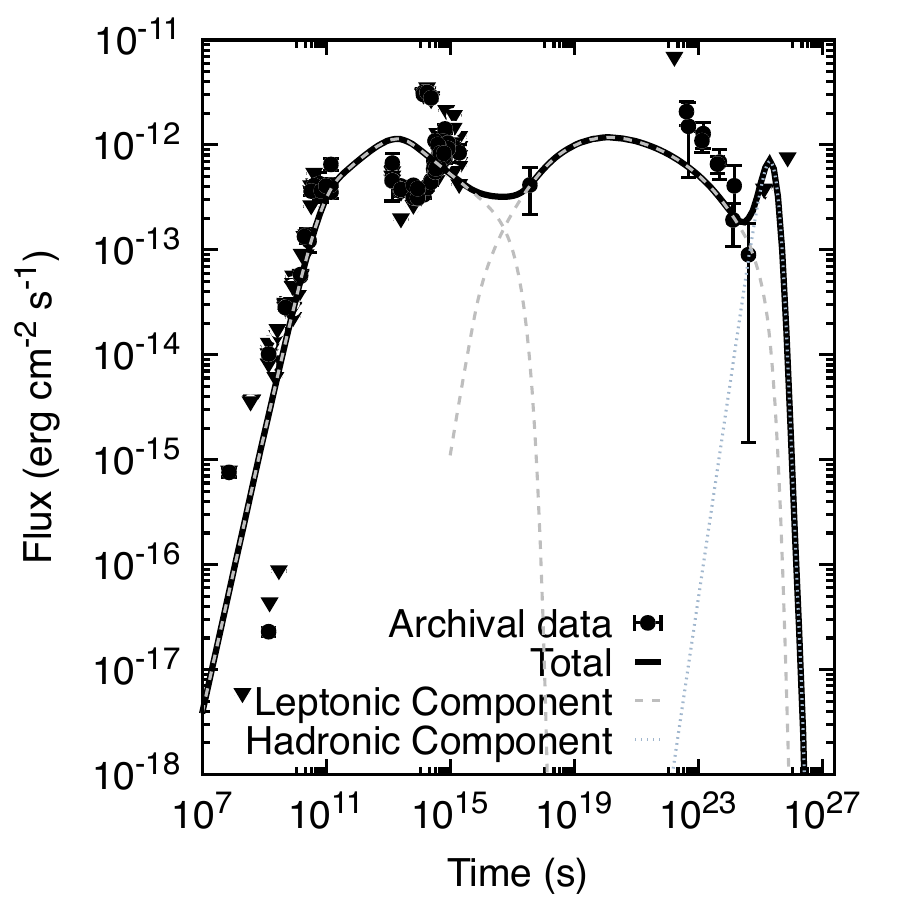} \label{fig:SED_PKS2330017_BLR}}
\subfloat[c][\centering{PKS B2224+006}]{\includegraphics[scale=0.26]{SEDV2/PKSB2224006_ModelSED_BLR.pdf} \label{fig:SED_PKSB2224+006b_BLR}}

\subfloat[b][\centering{PKS 0420+022}]{\includegraphics[scale=0.26]{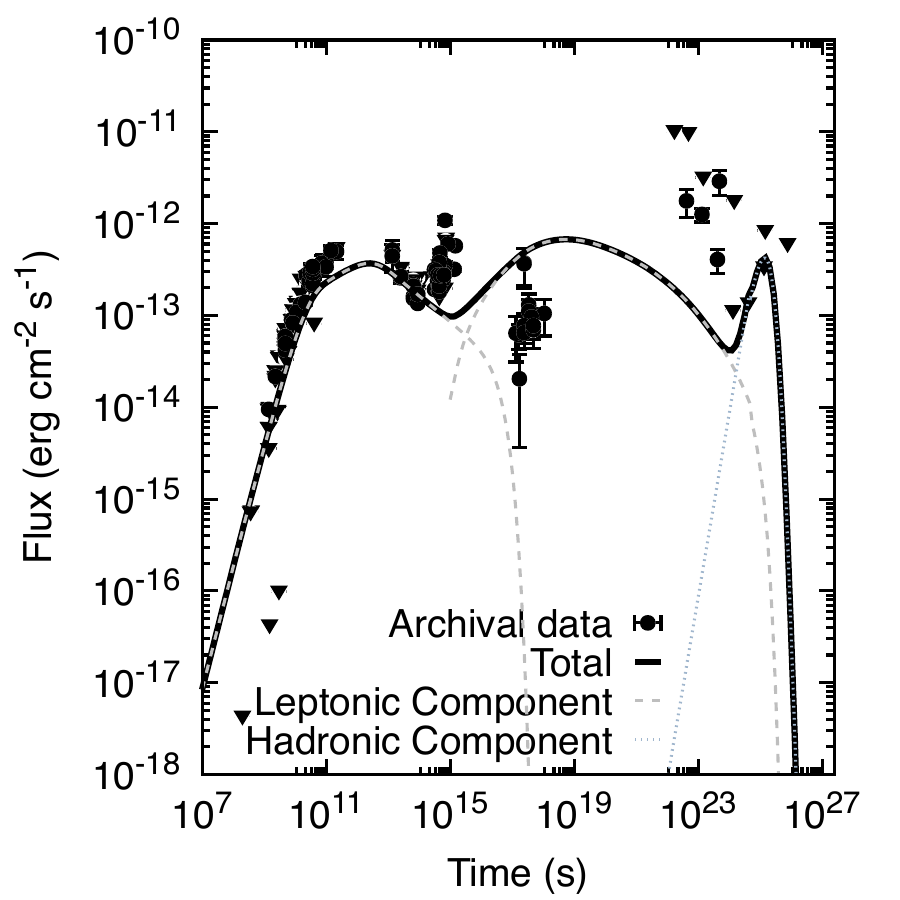}\label{fig:SED_PKS_0420022_BLR}}
\subfloat[c][\centering{PMN J2118+0013}]{\includegraphics[scale=0.26]{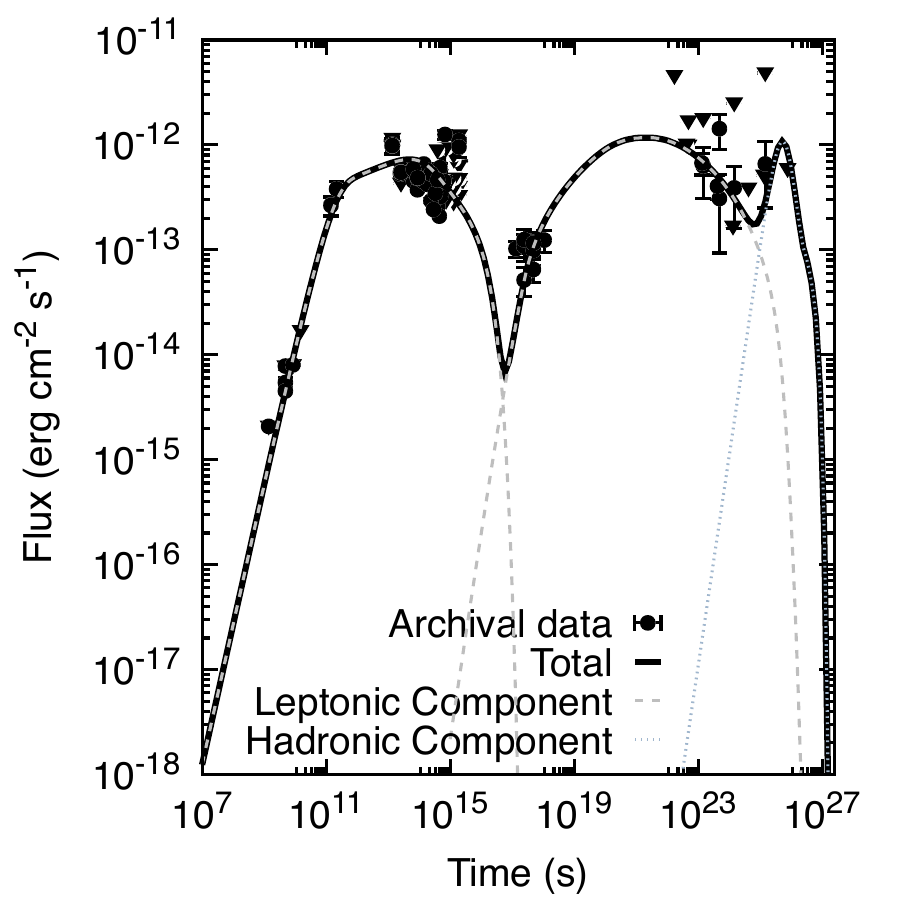} \label{fig:SED_PMNJ21180013_BLR}}
\subfloat[c][\centering{B2 1811+29}]{\includegraphics[scale=0.26]{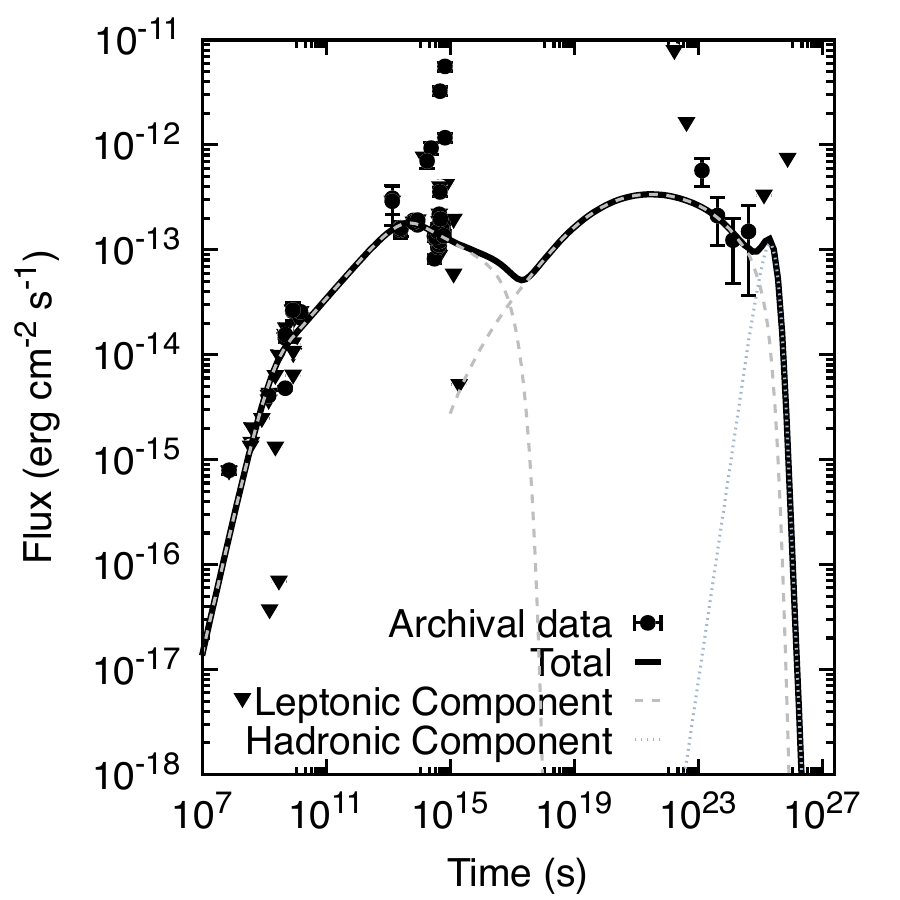} \label{fig:SED_B2181129_BLR}}
\subfloat[][\centering{B2 0849+28}]{\includegraphics[scale=0.26]{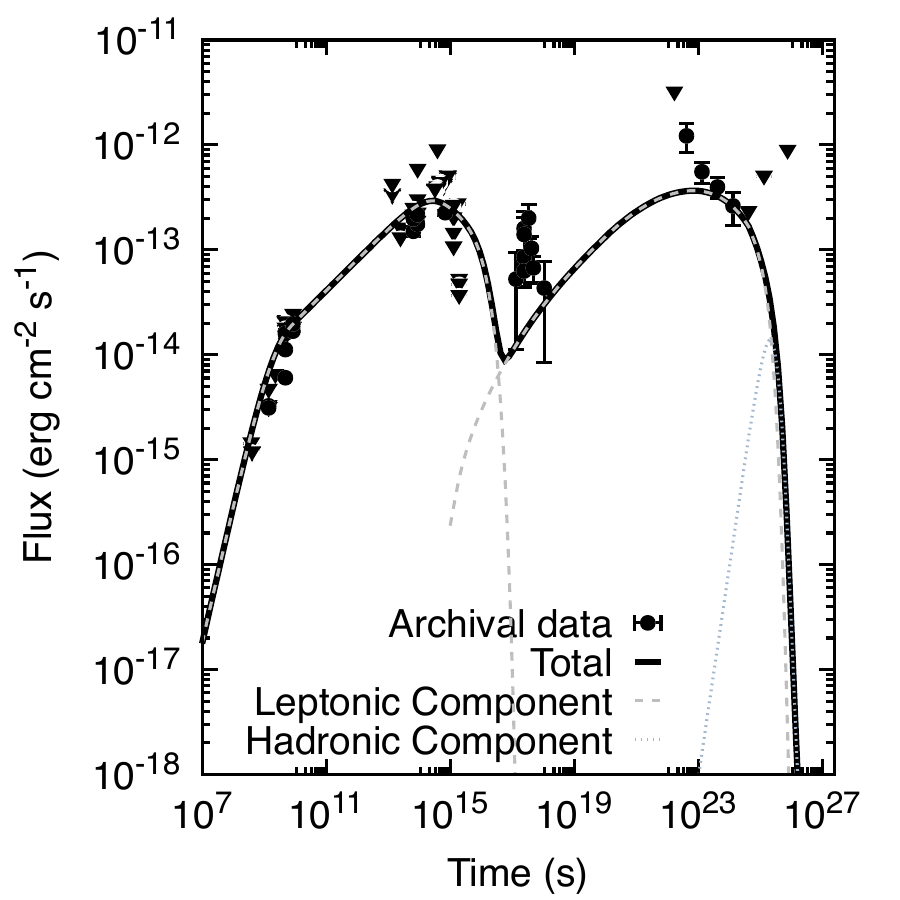} \label{fig:SED_B2084928_BLR}}

\subfloat[][\centering{4C +31.51}]{\includegraphics[scale=0.26]{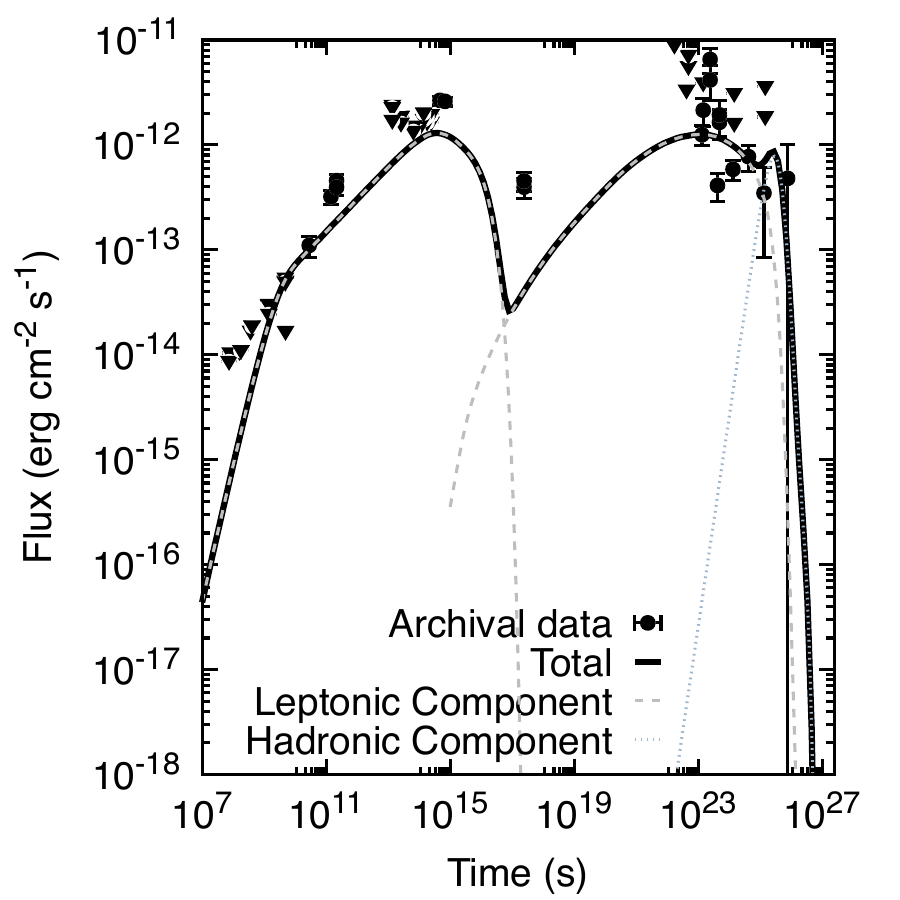} \label{fig:SED_4C31_5_BLR}}
\subfloat[a][\centering{TXS 1015+057}]{\includegraphics[scale=0.26]{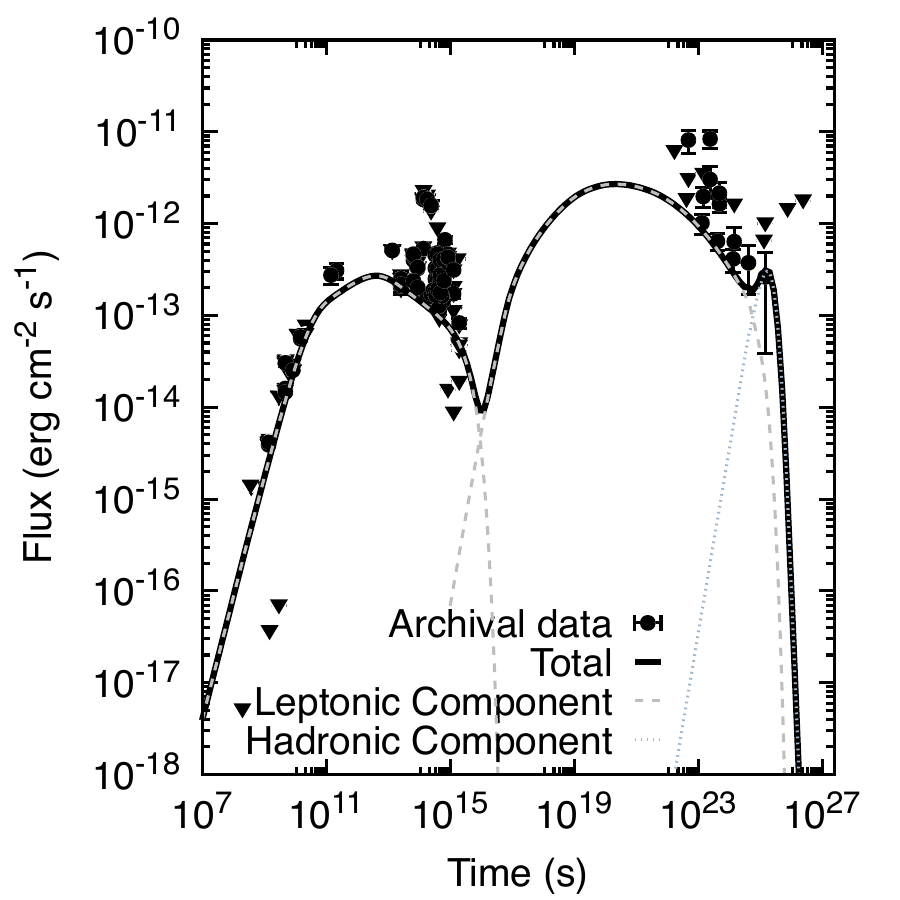} \label{fig:SED_TXS1015057_BLR}}
\subfloat[a][\centering{PKS 2058-297}]{\includegraphics[scale=0.26]{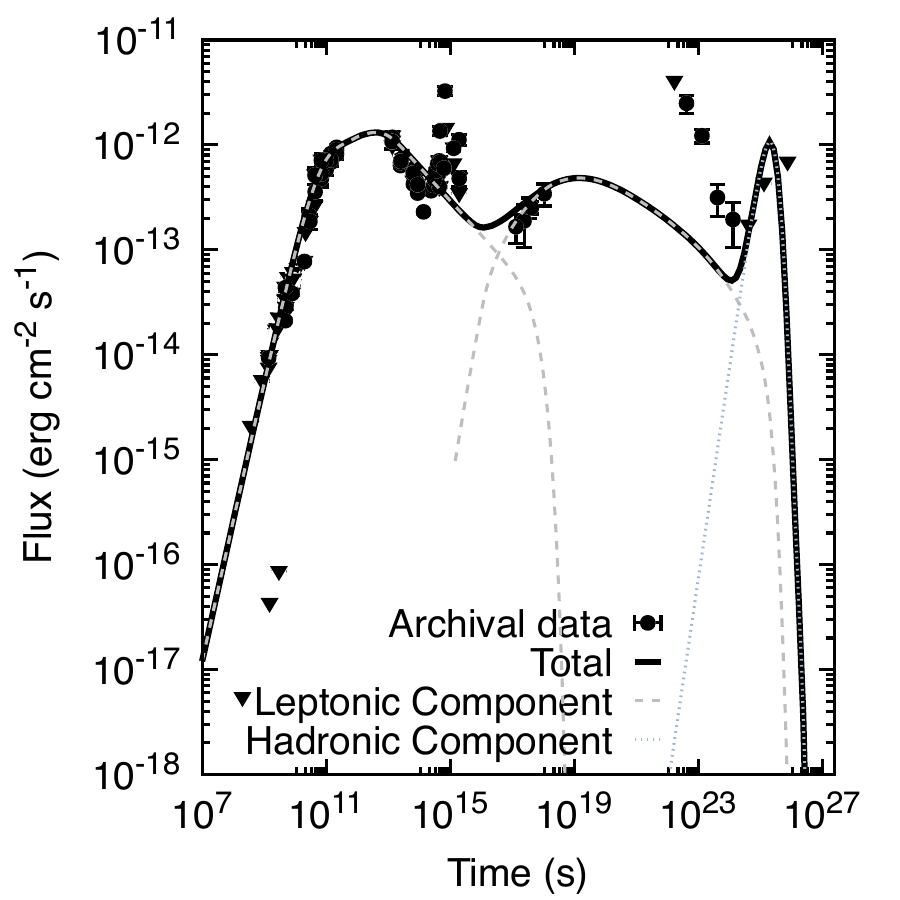} \label{fig:SED_PKS_2058297_BLR}}
\subfloat[b][\centering{PMN J0206-1150}]{\includegraphics[scale=0.26]{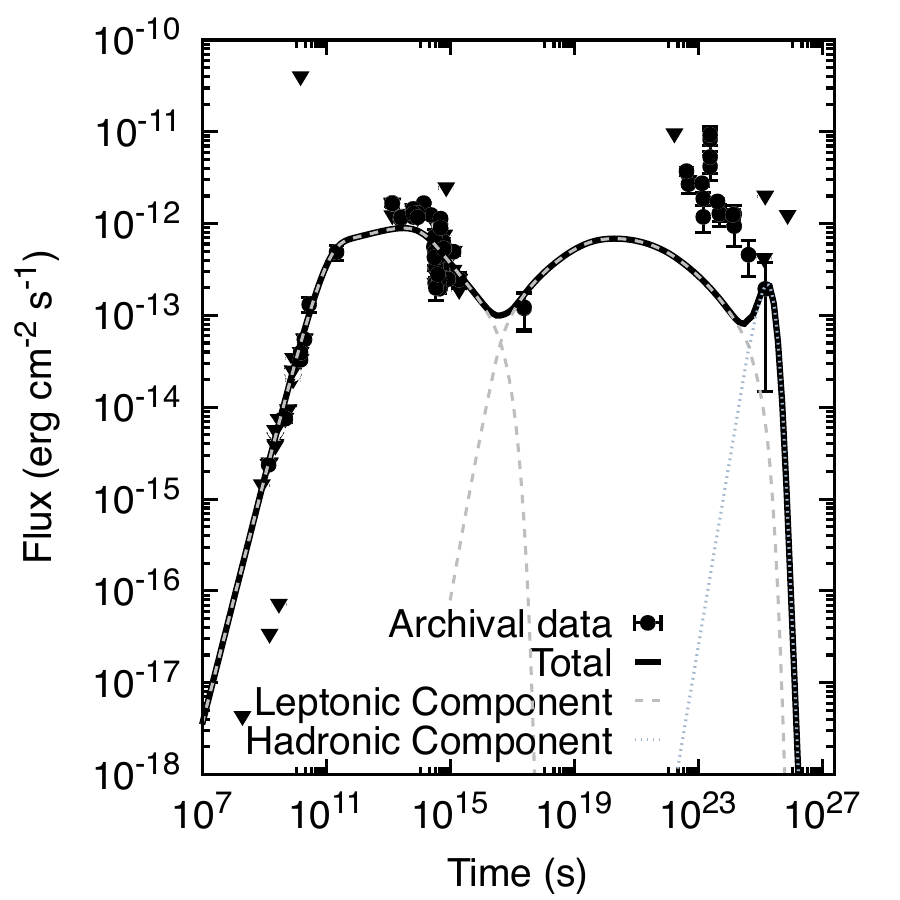} \label{fig:SED_PMN J02061150_BLR}}

\subfloat[a][\centering{OX 110}]{\includegraphics[scale=0.26]{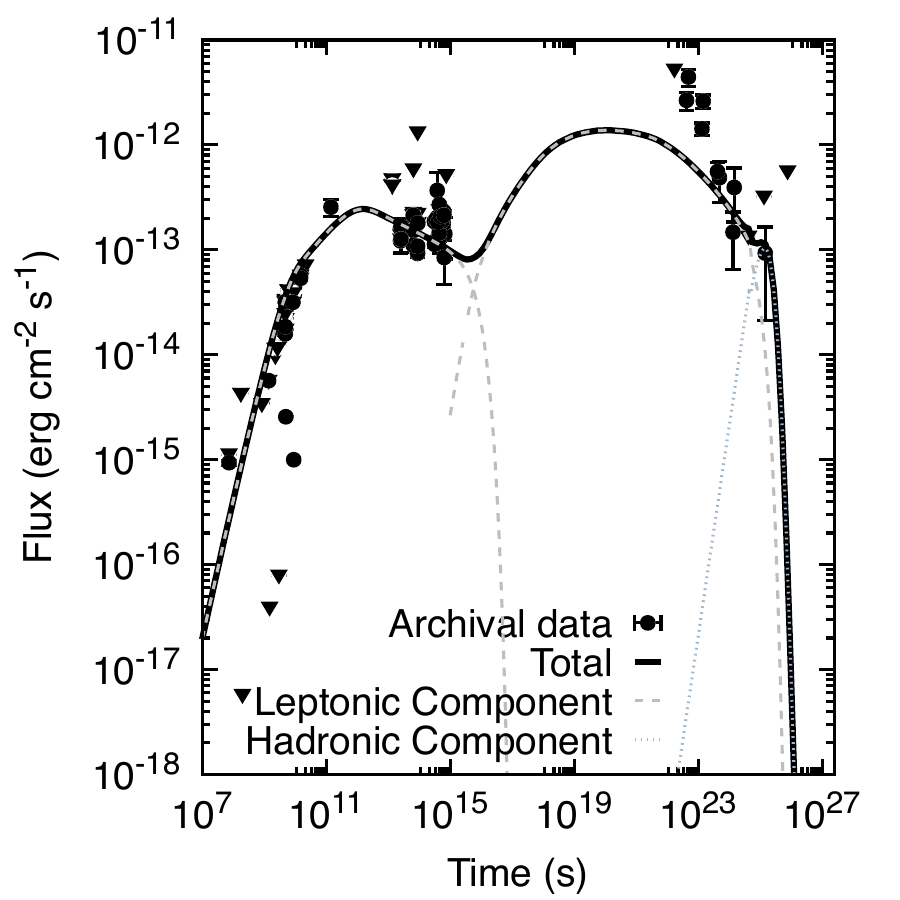}\label{fig:SED_OX110_BLR}}
\subfloat[b][\centering{TXS 2210+065}]{\includegraphics[scale=0.26]{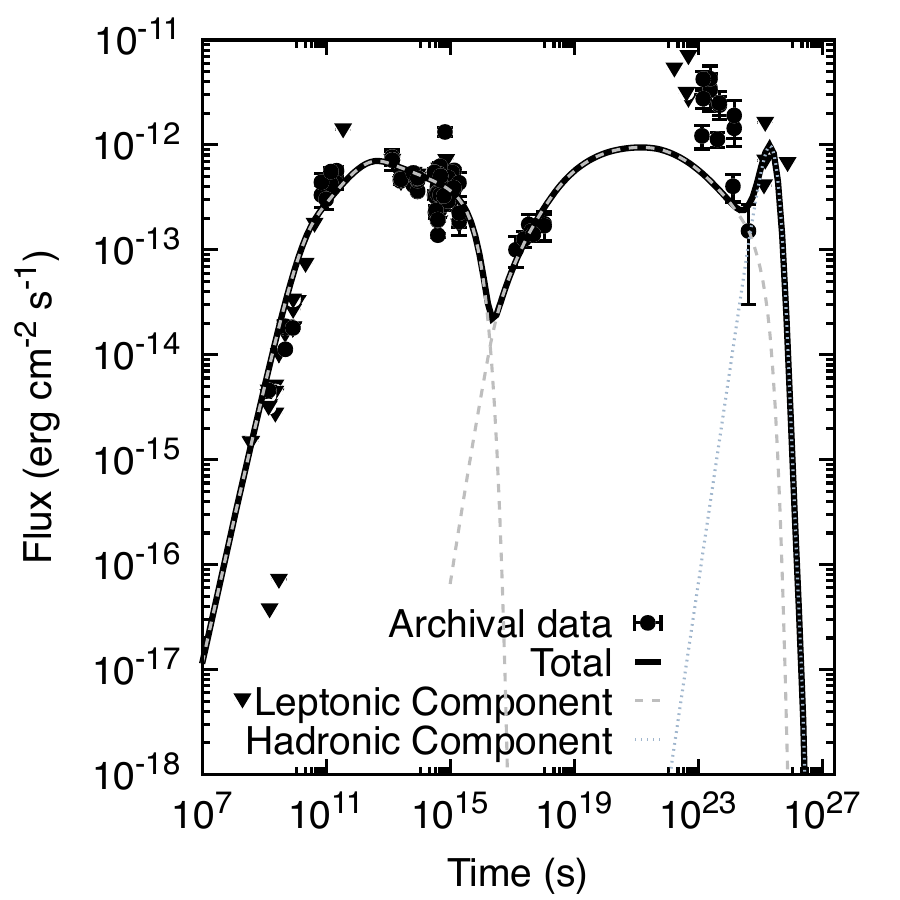} \label{fig:SED_TXS2210065_BLR}}
\subfloat[b][\centering{NVSS J134240+094752}]{\includegraphics[scale=0.26]{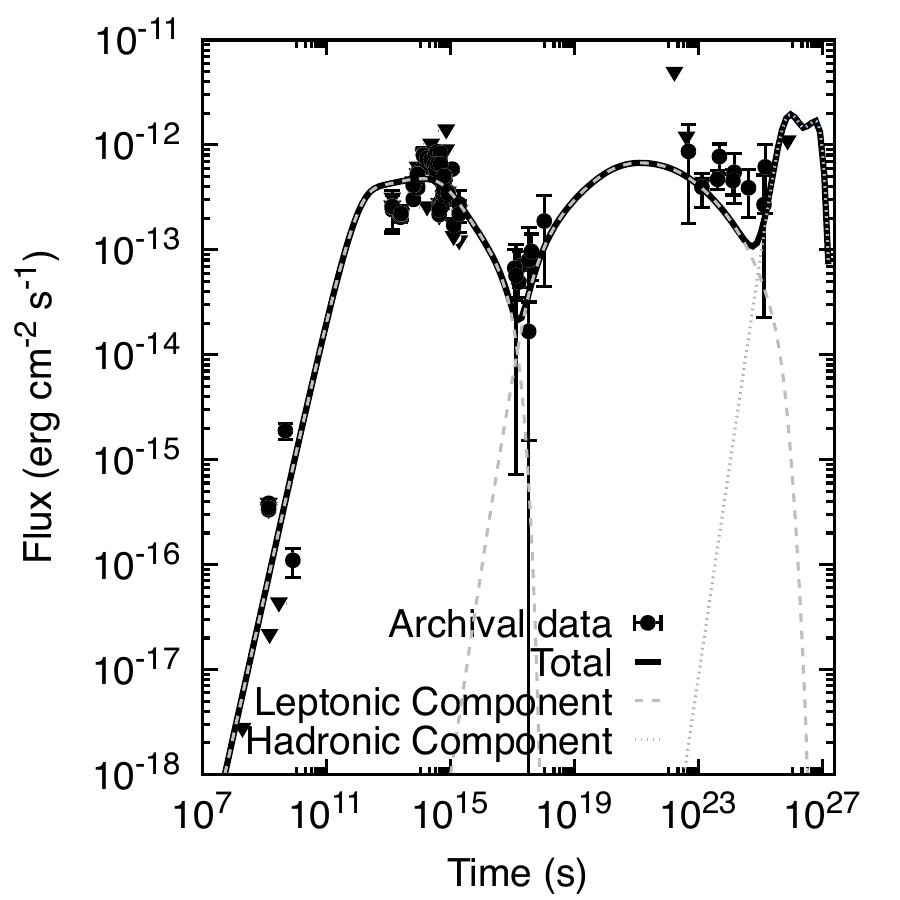} \label{fig:SED_NVSS_J134240094752_BLR}}
\subfloat[][\centering{B2 1100+30B}]{\includegraphics[scale=0.26]{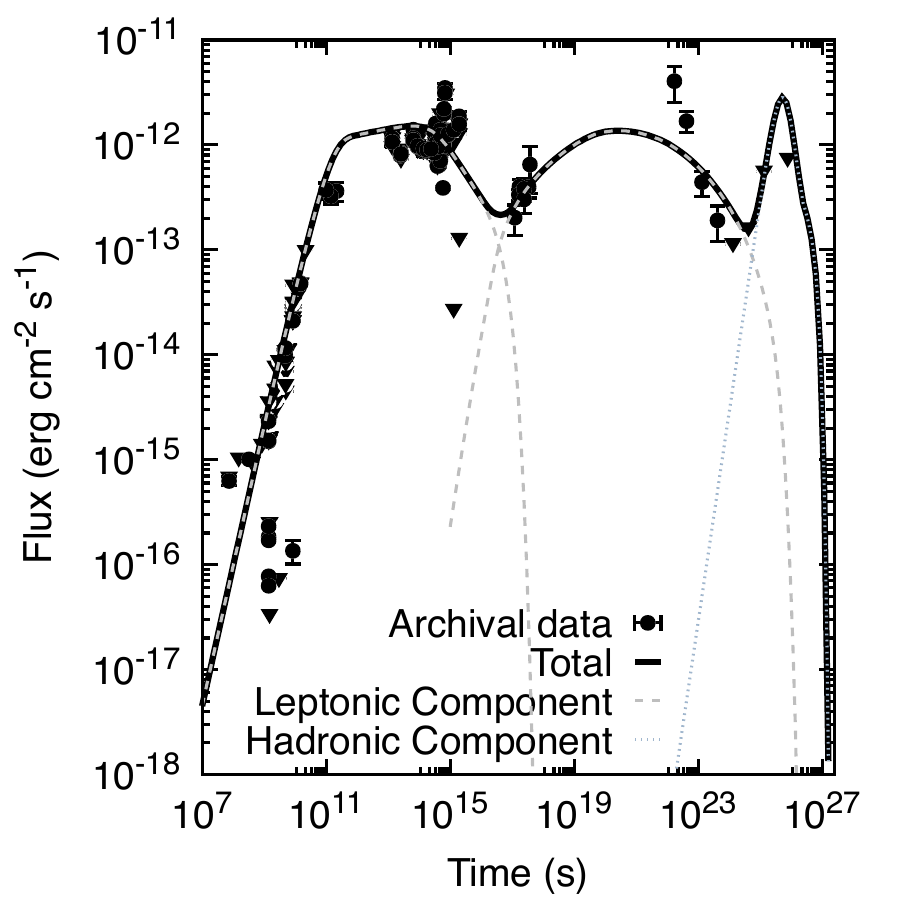} \label{fig:SED_B2110030B_BLR}}
}
\caption{Same as table \ref{fig:SED1} but considering the target photons of the $p\gamma$ interaction comes from the broadline region.}\label{fig:SED2}
\end{figure*}

From Figure \ref{fig:SED1} and Figure \ref{fig:SED2}, we can see the SED of each object found with a spatial coincidence with a neutrino. The SED was built with historical data. The values of the best-fit parameters are listed in Table \ref{Table:results}. Each row of these figures corresponds to the SED of the sources listed in Table \ref{table:Correlations}. For each SED archival data obtained from NED\footnote{\url{https://ned.ipac.caltech.edu/}} and from the CDS portal\footnote{\url{http://cdsportal.u-strasbg.fr/}} \citep{2020ApJS..247...33A}, were fitted assuming a lepto-hadronic model to describe the SED. Figure \ref{fig:SED1} shows the results of photo-pion decay products, considering the $p\gamma$ interactions with a low photon field  from the Narrow Line Region (NLR) and Figure \ref{fig:SED2} with the Broad Line Region (BLR). A dotted gray line in both figures indicates this contribution.  The leptonic component is shown in long-dashed lines, the first peak ($\sim 10^{12-14}\, {\rm Hz}$) is due to the synchrotron of the relativistic electrons, meanwhile the second peak ($\sim 10^{22-24}\, {\rm Hz}$) is described by a lepton component produced by Inverse Compton scattering together with a hadronic component of photons ($\sim 10^{25-26}\, {\rm Hz}$)  that decay from neutral pions created by interactions of protons with seed photons from the NLR and the BRL respectively.   In the case of the high energy component, the flux was attenuated due to Extragalactic Background Ligth (EBL) by a factor $\rm{\exp{(-\tau(E,z))}}$ with the factor $\tau$ provided by \citet{2008A&A...487..837F}, and parametrized by polynomial functions in this work.\\

\begin{table*}[htbp]

  \centering\small
  \caption{Best-fit values for the leptonic components in the SED of each source associated to a 4FGL in the arrival region of the neutrino detected by IceCube. \label{Table:results}}

    \begin{tabular}{l|cccccccccc}
    \toprule
    Name & $\log(B/{\rm G})$ & $\delta_D$ & $\log(K_e)$ & $p$ & $\log(\gamma_{\min})$ & $\log(\gamma_{\rm br})$ & $\log(\gamma_{\max})$ & $\log(t_{\rm var}/{\rm s})$ & $\log(K^{\rm BLR}_p)$ & $\log(K^{\rm NLR}_p)$ \\
    \midrule
    PKS B2224+006      & 1.46 & 1.56 & 49.00 & 3.34 & 2.78 & 3.33 & 5.68 & 5.93 & 30.89 & 17.52 \\
    PKS 1741-03         & 1.50 & 1.52 & 45.20 & 3.22 & 2.41 & 4.28 & 5.05 & 5.47 & 28.72 & 19.67 \\
    OP 313              & 1.04 & 1.60 & 49.91 & 2.18 & 2.15 & 2.82 & 5.00 & 5.52 & 20.35 & 11.30 \\
    RX J131058.8+323335 & 1.81 & 1.65 & 45.29 & 2.85 & 2.06 & 3.96 & 4.28 & 4.55 & 25.72 & 16.68 \\
    MG1 J120448+0408    & 1.77 & 1.32 & 46.54 & 2.55 & 2.15 & 3.68 & 5.04 & 5.04 & 25.35 & 16.30 \\
    PKS B2330-017       & 0.62 & 1.48 & 47.66 & 2.00 & 2.12 & 3.96 & 4.85 & 5.92 & 20.41 & 11.36 \\
    PKS 2332017         & 2.03 & 1.68 & 47.06 & 2.47 & 1.91 & 3.15 & 5.01 & 4.25 & 22.98 & 13.94 \\
    PKS B2224+006       & --- & --- & ---- & --- & --- & --- & --- & --- & --- & --- \\
    PKS 0420022         & 1.26 & 1.49 & 47.32 & 2.51 & 1.91 & 3.23 & 5.23 & 4.62 & 26.16 & 17.12 \\
    PMN J2118+0013      & 0.88 & 1.52 & 45.29 & 2.75 & 2.68 & 4.58 & 5.11 & 5.55 & 24.02 & 14.98 \\
    B2 1811+29          & 1.81 & 1.91 & 44.33 & 2.32 & 1.22 & 3.70 & 5.07 & 3.52 & 21.20 & 12.15 \\
    B2 0849+28          & 1.03 & 2.18 & 43.35 & 2.41 & 1.62 & 4.20 & 5.00 & 3.41 & 20.86 & 11.81 \\
    4C +31.51           & 1.01 & 1.69 & 45.24 & 2.38 & 1.62 & 4.19 & 5.00 & 4.75 & 21.27 & 12.23 \\
    TXS 1015+057        & 0.66 & 1.63 & 48.57 & 2.50 & 2.48 & 3.60 & 5.01 & 5.36 & 23.79 & 14.75 \\
    PKS 2058-297        & 1.66 & 1.52 & 47.74 & 2.73 & 2.13 & 3.59 & 5.22 & 5.73 & 25.34 & 16.29 \\
    PMN J0206-1150      & 1.51 & 1.66 & 46.90 & 2.72 & 2.31 & 3.79 & 5.13 & 5.25 & 25.68 & 16.63 \\
    OX 110              & 1.07 & 1.48 & 46.74 & 2.39 & 1.99 & 3.67 & 5.03 & 4.86 & 22.73 & 13.69 \\
    TXS 2210+065        & 0.87 & 1.45 & 47.68 & 2.23 & 2.27 & 3.60 & 5.05 & 5.13 & 22.23 & 13.19 \\
    NVSS J134240+094752 & 1.61 & 1.74 & 45.86 & 2.56 & 2.84 & 3.70 & 5.29 & 4.42 & 23.46 & 14.42 \\
    B2 1100+30B         & 1.72 & 1.35 & 43.41 & 2.67 & 2.17 & 4.60 & 5.00 & 4.50 & 25.81 & 16.76 \\
    \bottomrule
    \end{tabular}%
    \tabletext{Note: All log values are in base 10.}
\end{table*}

\noindent The values listed in Tables \ref{Table:results} and \ref{Table:Derived} as a consequence of the MCMC method are in the range of those reported in the literature \citep[e.g., see][]{2008ApJ...686..181F, 2022MNRAS.512.1557A, 2021PhRvD.104h3013A}. For example, the Doppler factor lies in $20 \ \le \delta_{D} \ \le 100 $ and the volume of the emission region lies in values of $\sim 10^{17}\,{\rm cm^{3}}$ which corresponds to a minimum variability time of $\sim1 \rm{days}$ \citep{2011ApJ...727..129A, 2011ApJ...736..131A, 2017ApJS..232....7F, 2019ApJS..245...18F}, which is lower than that recorded with the observation campaigns in other sources \citep{2001ApJ...560..659K, 2011ApJ...727..129A, 2003MNRAS.340.1095D}. Compared to the predicted Eddington luminosity $L_{\rm Edd} \sim 10^{48} \, {\rm erg\, s^{-1}}$, based on the evaluation of the supermassive black hole's mass $M_{\rm BH} \sim 10^9\, M_\odot$ \citep{2008MNRAS.385..119W, 2004MNRAS.352.1390M}, the calculated proton luminosity ($L_{p\rm }$) is proportional to a small amount. By analyzing the magnetic field, electron and proton densities, and their respective ratios $U_{\rm i}/U_{\rm j}$ derived and listed in Table \ref{Table:Derived}, we can observe that the potential existence of the principle of equipartition can be inferred.\\

Also, we have adopted an electron/proton population described by a power-law. This is a consequence of considering that the acceleration of particles lies in a First-Order Fermi acceleration, in which the acceleration and scape particles are in steady state \citep{1962SvA.....6..317K}. On the other hand, when the diffusive term is not negligible, then the solution becomes a log-Parabola shape which corresponds to a Second-Order Fermi acceleration regime. This scenario has been explored by \cite{2006A&A...448..861M, 2015ApJ...809..174D}, in which the blazar Mkn 501 spectra has been studied. They found that the broadband SED was described as successful assuming this paradigm, in which, the gamma ray band had an internal absorption rather than the EBL.\\

Recently, \cite{2020ApJ...898...48A} have studied a sample of the LAT Bright AGN Sample and compared with the values obtained in this work. Across the sample, the magnetic energy densities ($u_B$) are typically well below the combined particle energy densities $(u_{e} + u_{p})$, indicating particle-dominated jets, in agreement with previous blazar studies. The electron spectral indices ($p \sim 2.0 - 3.3$) and the Lorentz break factors ($\gamma_{br} \sim 10^{3-4.5}$) are consistent with the expectations of shock-acceleration processes and subsequent radiative cooling. Sources with the highest inferred $\gamma_{br}$ and $\gamma_{max}$ tend to display comparatively higher electron luminosities ($L_{e}$). Moreover, according to the theory of spectral curvature as discussed by \cite{2020ApJ...898...48A}, stronger cooling or weaker stochastic acceleration leads to increased curvature of the SED. In our sample, this is broadly supported by the trend of a lower $\gamma_{br}$ and a steeper p in objects with larger $u_{p}$ and lower $L_{e}$. These inter-comparisons reinforce the physical consistency of our fits and provide additional justification for our modeling choices.\\

It is also observed that in the IR-Optical bands, an excess is present that is not quite fitted by synchrotron emission; this effect also can be appreciated in the gamma ray band, where the SSC does not fit as well, this can be explained due to the SED being built with historical data and makes discrepancies with the expected behavior predicted by models due to the higher variability exhibit in this objects. For example \cite{2017MNRAS.469..255G} who revisited the blazar sequence detected by Fermi-LAT on four years of operations found that a significant proportion of FSRQs exhibit indications of thermal emission from the accretion disk, influencing the optical-UV spectrum. This point also has been discussed by \cite{2016Galax...4...36G}, when they normalized the entire SED to the peak of the disk emission, they found a reduced scatter in the other bands, especially in the radio and in the X-rays (in the $\gamma$-rays the variability amplitude is so large to hide the reduction of the scatter) and found that this can be taken as model independent evidence that the disk and the jet luminosity are related. This exhibits the importance of multi-wavelength campaigns of observations on this kind of objects, since blazars are highly variable on multiple timescales, the use of heterogeneous epochs can mis-represent the true emission state and lead to biased parameter inference in e.g. single-zone SSC or EC models. Accordingly, while our synchrotron peak is constrained and modeled, we refrain from attempting an overly strict fit of the high-energy component when relying on non-simultaneous data.\\ 

The comparison of our fitted parameters with the calculated jet energy indicates a consistent physical model in which relativistic collisions and stochastic processes collaboratively influence the observed emissions of FSRQs.  The predominance of particle energy densities over magnetic fields supports efficient acceleration and radiative cooling within particle-dominated jets.  Although limited by the non-simultaneous nature of the data, the overall agreement between our findings and previous investigations of blazars at multiple wavelengths corroborates the robustness of our modeling methodology. This also highlights the need for synchronized observations that integrate future multi-frequency observation campaigns in AGNs. Their data will be crucial for constraining time-dependent models and understanding the relationship between high-energy neutrino generation and jet energy in blazars.



\begin{table*}[htbp]
  \centering\small
  \caption{Derived quantities. \label{Table:Derived}} 
    \begin{tabular}{l|cccccccc}
    \toprule
    Name & \begin{tabular}[c]{@{}c@{}}$u_{B}$ \\ ($\times10^{-4}$) \end{tabular} & \begin{tabular}[c]{@{}c@{}}$u_{e}$ \\ ($\times10^{-4}$) \end{tabular}  &\begin{tabular}[c]{@{}c@{}}$u_{p}$ \\ ($\times10^{-4}$) \end{tabular} & $U_B/(U_e + U_p)$  & \begin{tabular}[c]{@{}c@{}}$L_B$ \\ ($\times10^{45}$) \end{tabular} & \begin{tabular}[c]{@{}c@{}}$L_e$ \\ ($\times10^{45}$) \end{tabular} & \begin{tabular}[c]{@{}c@{}}$L_p$ \\ ($\times10^{45}$) \end{tabular} & \begin{tabular}[c]{@{}c@{}}$L_j$ \\ ($\times10^{45}$) \end{tabular}\\

    \midrule

    PKS B2224+006       & 0.34  & 0.01  & 12.98 & 0.03 & 0.36 & 0.01 & 13.75 & 14.13 \\
 PKS 1741-03         & 0.40  & 0.12  & 2.96  & 1.30$\times10^{-1}$ & 0.09 & 0.03 & 0.66  & 0.78  \\
 OP 313              & 0.05  & 0.06  & 3.42$\times10^{4}$ & 1.41$\times10^{-6}$ & 0.03 & 0.04 & 2.10$\times10^{4}$ & 2.10$\times10^{4}$  \\
 RX J131058.8+323335 & 1.68 & 9.74 & 84.12 & 0.02 & 0.01 & 0.06 & 0.55 & 0.62  \\
 MG1 J120448+0408    & 1.42 & 218.31 & 1.46$\times10^{5}$ & 9.72$\times10^{-6}$ & 5.50$\times10^{-4}$ & 0.08 & 56.60 & 56.70  \\
 PKS B2330-017       & 0.01 & 0.01   & 6.10$\times10^{6}$ & 1.17$\times10^{-9}$ & 0.01 & 0.01 & 7.64$\times10^{6}$ & 7.64$\times10^{6}$  \\
 PKS 2332017         & 4.57 & 8.34   & 5.28$\times10^{3}$ & 8.64$\times10^{-4}$ & 0.01 & 0.03 & 17.20 & 17.20  \\
---                 & ---  & ---  & ----  & --- & --- & --- & --- & ---   \\
 PKS 0420022         & 6.31 & 2.18 & 6.59$\times10^{5}$ & 9.56$\times10^{-6}$ & 0.08 & 0.03 & 8.22$\times10^{3}$ & 8.22$\times10^{3}$   \\
 PMN J2118+0013      & 0.02 & 0.09 & 26.09 & 8.98$\times10^{-4}$ &  1.45$\times10^{-3}$ & 5.40$\times10^{-3}$ & 1.61 & 1.61  \\
 B2 1811+29          & 1.73 & 22.05 & 5.75$\times10^{3}$ & 2.99$\times10^{-4}$ & 0.01 & 0.18 & 45.90 & 46.00  \\
 B2 0849+28          & 0.05 & 0.05 & 223.15 & 2.08$\times10^{-4}$ & 0.27 & 0.29 & 1.32$\times10^{3}$ & 1.32$\times10^{3}$  \\
 4C +31.51           & 0.04 & 0.93 & 1.32$\times10^{3}$ & 3.21$\times10^{-5}$ & 2.76$\times10^{-3}$ & 0.06 & 86.00 & 86.00  \\
 TXS 1015+057        & 0.01 & 0.91 & 1.61$\times10^{4}$ & 5.27$\times10^{-7}$ & 1.49$\times10^{-3}$ & 0.16 & 2.83$\times10^{3}$ & 2.83$\times10^{3}$  \\
 PKS 2058-297        & 0.84 & 0.17 & 8.95$\times10^{3}$ & 9.33$\times10^{-5}$ & 0.17 & 0.03 & 1.79$\times10^{3}$ & 1.79$\times10^{3}$  \\
 PMN J0206-1150      & 0.42 & 0.25 & 157.71 & 2.69$\times10^{-3}$ & 0.05 & 0.03 & 17.60 & 17.70  \\
 OX 110              & 0.05 & 2.46 & 4.23$\times10^{5}$ & 1.30$\times10^{-7}$ & 5.71$\times10^{-3}$ & 0.26 & 4.40$\times10^{4}$ & 4.40$\times10^{4}$   \\
 TXS 2210+065        & 0.02 & 0.04 & 8.57$\times10^{5}$ & 2.63$\times10^{-8}$ & 0.02 & 0.04 & 7.69e$\times10^{5}$ & 7.69$\times10^{5}$  \\
 NVSS J134240+094752 & 0.68 & 2.80 & 0.35 & 0.22 & 2.35$\times10^{-4}$ & 9.65$\times10^{-4}$ & 1.21$\times10^{-4}$ & 1.32$\times10^{-3}$ \\
 B2 1100+30B         & 1.14 & 0.65 & 46.33 & 0.02 & 8.67$\times10^{-3}$ & 4.92$\times10^{-3}$ & 0.35 & 0.37  \\

 \bottomrule

    \end{tabular}%
  \label{tab:addlabel}%
\end{table*}


\section{FACILITIES}

We acknowledge the use of public data and analysis tools provided by the Fermi Science Support Center (FSSC) and the Fermi-LAT Collaboration, as well as data from the IceCube Neutrino Observatory available through the IceCube Public Data Portal. This research has made use of the NASA/IPAC Extragalactic Database (NED), operated by the Jet Propulsion Laboratory, California Institute of Technology, under contract with NASA, and data products from the Sloan Digital Sky Survey (SDSS). We also thank the Infrared Processing and Analysis Center (IPAC) for providing access to archival resources that supported this work.

\section{ACKNOWLEDGEMENTS}

This work was supported by UNAM Posdoctoral Program (POSDOC). NF acknowledges financial support from UNAM-DGAPA-PAPIIT through the grant IN112525. HLV and AG acknowledges financial support from UNAM-DGAPA-PAPIIT through the grant IN102223 and SECIHTI grant CF-2023-I-645. JD acknowledges financial support from UNAM-DGAPA-PAPIIT through the grant IN116325. Also, we thank the referee for his helpful comments on how to improve this work.

\renewcommand{\refname}{REFERENCES}
\bibliography{rmaa_main_v2}



    


\end{document}